\documentclass[aps,prx,showpacs,notitlepage,twocolumn, floatfix,superscriptaddress,nofootinbib]{revtex4-2}
\usepackage{amssymb}
\usepackage{amsmath}
\usepackage{amsfonts}
\usepackage{bbm}
\usepackage{braket}
\usepackage{ytableau}
\usepackage{graphicx}
\usepackage[colorlinks,linkcolor=blue,anchorcolor=blue,citecolor=blue,urlcolor=blue]{hyperref}
\usepackage[titletoc]{appendix}
\graphicspath{{Figures/}}

\def\bea{\begin{equation}\begin{aligned}}
\def\eea{\end{aligned}\end{equation}}
\def\bn{\bar{n}}
\def\F{\mathcal{F}}
\def\S{\mathcal{S}}
\def\H{\mathbb{H}}
\def\U{\mathbb{U}}
\def\Q{\mathcal{Q}}
\def\syk{\text{syk}}
\def\o{\text{out}}
\def\i{\text{in}}

\def\tr{\text{tr}}

\begin{document}
\title{Emergent Symmetry in Brownian SYK Models and Charge Dependent Scrambling }
\author{Lakshya Agarwal}
\affiliation{Department of Physics \& Astronomy, Texas A\&M University, College Station, Texas 77843, USA}
\author{Shenglong Xu}
\email{slxu@tamu.edu}
\affiliation{Department of Physics \& Astronomy, Texas A\&M University, College Station, Texas 77843, USA}    

\begin{abstract}
In this work, we introduce a symmetry-based approach to study the scrambling and operator dynamics of Brownian SYK models at large finite $N$ and in the infinite $N$ limit. We compute the out-of-time-ordered correlator (OTOC) in the Majorana model without charge conservation and the complex model with charge conservation, and demonstrate that in both models taking the random average of the couplings gives rise to emergent symmetry structures. The random averaging exactly maps the operator dynamics of the Majorana model and the complex model to the imaginary time dynamics of an SU(2) spin and an SU(4) spin respectively, which become solvable in the large $N$ limit. Furthermore, the symmetry structure drastically reduces the size of the Hilbert space required to calculate the OTOC from exponential to linear in $N$, providing full access to the operator dynamics at all times for large finite  $N$. In the case of the complex model with charge conservation, using this approach, we obtain the OTOC within each charge sector both numerically at finite $N$ and analytically in the large $N$ limit. We find that the time scale of the scrambling dynamics for all times and in each sector is characterized by the charge density. Furthermore, after proper rescaling, the OTOC corresponding to different finite charge densities collapses into a single curve at large finite $N$. In the large $N$ limit, the rescaled OTOCs at finite density are described by the same hydrodynamic equation as in the Majorana case.

\end{abstract}
\maketitle
\tableofcontents

\section{Introduction}
In an isolated quantum many-body system, a simple initial state, such as a product state, generally becomes complicated under the unitary time evolution. As time increases, the local density matrix of the state approaches equilibrium, which only depends on the macroscopic properties of the state~\cite{Deutsch1991,Srednicki1994,Polkovnikov2011,rigol2008thermalization}. Nevertheless, the local information of the initial state is not lost but flows to non-local degrees of freedom and cannot be recovered by local probes, a phenomenon dubbed as quantum information scrambling~\cite{Sekino_2008,Hayden_2007, shenker2014black, shenker2015stringy, Maldacena_2016, Hosur_2016}. Scrambling can be formulated in terms of operator dynamics and is related to the spreading of a local operator under Heisenberg time evolution
~\cite{shenker2014black, kitaev2015, Maldacena_2016, Hosur_2016, aleiner2016microscopic, roberts2016lieb}. The growth of the support of a Heisenberg operator $W(t)$ can be quantified by the out-of-time ordered correlator~(OTOC) between $W(t)$ and a local static probing operator $V$
\bea \label{eq:defineOTOC}
\F(W(t),V) = \frac{1}{\tr I} \tr (W^\dagger(t) V^\dagger W(t) V),
\eea
which was first introduced in the context of superconductors~\cite{larkin1969quasiclassical} and has received renewed interest because of its connection to scrambling. The OTOC is sensitive to whether the support of $W(t)$ overlaps with $V$, which is initially far away, and therefore is a good indicator for scrambling. 
In all-to-all interacting models with few-body interactions and a large number of degrees of freedom, such as the Sachdev-Ye-Kitaev model~\cite{Sachdev1992,kitaev2015,Sachdev_2015,Polchinski2016,Maldacena_2016_SYK}, $\F$ typically grows exponentially fast in the early-time regime, $\F\sim \frac{1}{N} e^{\lambda_L t}$ \cite{Sekino_2008, Lashkari_2013, Maldacena_2016, parker2019universal}, where $\lambda_L$ is the Lyapunov exponent. In a local extended system, the support of $W(t)$ typically grows ballistically, and $\F$ features a wavefront travelling at the butterfly velocity~\cite{shenker2014black,aleiner2016microscopic, roberts2016lieb, blake2016universal, Gu2017diffusion,luitz2017information,bohrdt2017scrambling,Nahum_2018,von_Keyserlingk_2018,lin2018out,Xu_2019_nature,khemani2018velocity, Xu_2019, Operatorhydrodynamics}.
More generally, the behavior of $\F$ depends on the interaction between the different degrees of freedom~\cite{bentsen2019treelike,Bentsen2019}. The OTOC has been experimentally measured in nuclear magnetic resonance quantum simulators~\cite{Li_2017,wei2018exploring,nie2019detecting,sanchez2020perturbation}, trapped ions~\cite{Garttner_2017,Landsman_2019,joshi2020quantum} and superconducting qubits~\cite{Geller_2018,braumuller2021probing, mi2021information, Blok_2021, zhao2021probing}.

There have been many fruitful discussions on the interplay between scrambling dynamics, conserved quantities such as energy~\cite{Maldacena_2016, blake2016universal,Gu2017diffusion,grozdanov2018poleskipping,blake2018poleskipping,choi2020pole,lucas2019operator,Qi_2019,han2019quantum,huang2019finite, sahu2020information, cheng2021scrambling}, charge~\cite{Patel_2017, Khemani_2018, Rakovszky_2018, Friedman_2019,chen2020quantum, Piroli_2020, Chen_2020, Kos_2021}, dipole~\cite{pai2019localization,Moudgalya_2021,feldmeier2021critically} and other symmetries~\cite{bao2021symmetry,kudlerflam2021information}.
It has been shown that the presence of conserved quantities bounds the operator growth~\cite{Maldacena_2016, Chen_2020} and also slows down the relaxation of OTOC in higher dimensions when the conserved quantities display diffusive transport~\cite{Khemani_2018, Rakovszky_2018, cheng2021scrambling}. From the opposite perspective, operator growth can also influence transport properties~\cite{hartnoll2015theory,blake2016universal,lucas2019operator}. 
In systems with conserved quantities, the operator dynamics contain contributions from different sectors of the Hilbert space, each labelled by the corresponding value of the conserved charge. A natural curiosity therefore arises, concerning the relation between scrambling dynamics and the density of the conserved quantities. In all-to-all interacting models, most studies related to this question focus on the early time behavior of $\F$ characterized by the Lyapunov exponent. In the SYK model, it is found that the Lyapunov exponent $\lambda_L=2\pi T$ at low energy~\cite{kitaev2015,Maldacena_2016_SYK}, saturating the conjectured chaos bound~\cite{Maldacena_2016}. More recently, the bound on $\lambda_L$ from the charge density $\rho$ has also been computed, showing $\lambda_L$ vanishes algebraically at low density~\cite{Chen_2020}. These results are consistent with the intuition that scrambling should slow down at low density of conserved quantities because of the restricted Hilbert space. However, most calculations of the Lyapunov exponent are in the large $N$ limit or at small $N$ up to $\sim 60$ in the Majorana case~\cite{kobrin2021many}.  Furthermore, precise results concerning the OTOC beyond the early time exponential regime are difficult to obtain, even in the large $N$ limit. Therefore, to further understand the interplay between conserved quantities and scrambling, exact large but finite $N$ calculations of the OTOC at a given energy or charge are required.

Brownian many-body models~\cite{Lashkari_2013, shenker2015stringy, saad2018semiclassical, Xu_2019, Zhou_2019, Sunderhauf2019, Jian_2021}, in which the couplings are random variables uncorrelated in the time direction, are useful for understanding scrambling dynamics. In the absence of conserved quantities, the operator dynamics in Brownian models can be mapped to a Markovian process~\cite{Xu_2019,Zhou_2019,kuo2020markovian} or imaginary time dynamics of bosonic models~\cite{Sunderhauf2019} post disorder average, in which case the OTOC can be calculated efficiently for all time at large finite $N$. The analytical expression of the OTOC in the large $N$ limit can also be derived, which obeys a logistic type differential equation~\cite{Zhou_2019} for all-to-all connected models and reaction-diffusion equation~\cite{aleiner2016microscopic, Xu_2019} in higher dimensions. Related to Brownian models, scrambling for random Hamiltonians~\cite{you2018entanglement} and noisy spin systems~\cite{knap2018entanglement,rowlands2018noisy} has also been studied. 
Although the Brownian model does not conserve energy because of the time-dependent couplings, one can design such a model to conserve U(1) charge. These properties make such models appealing for understanding the interplay between conserved quantities and scrambling.
One of the simplest Brownian models that conserves charge is the complex Brownian SYK model. However, the technique used to solve Brownian models previously does not directly apply to this case, because of the charge conservation. So far, only the charge dependence of the Lyapunov exponent is available in the large $N$ limit based on a standard field theory calculation~\cite{Chen_2020}.

In this work, we study operator dynamics of the Brownian SYK model, focusing on the complex case with charge conservation. Building on previous work on the Majorana Brownian SYK model~\cite{Sunderhauf2019}, we demonstrate that Brownian SYK models have a simple symmetry structure for arbitrary $N$ after taking the random disorder average, which makes the numerical calculation of the OTOC at large finite $N$ possible even in the case with charge conservation. The operator dynamics can be organized into various irreducible representations~(irreps) of the symmetry group. We show that in the case of the Majorana Brownian SYK model, this approach maps the operator dynamics to that of an SU(2) spin in imaginary time, with the angular momentum being related to $N$. In the case of the complex Brownian SYK model, which is the primary focus of this work, this approach maps the operator dynamics to an SU(4) spin with conserved weights and the particular irrep is related to $N$. In both cases, the largest Hilbert space dimension required to compute the OTOC scales linearly with $N$, drastically reduced from the original Hilbert space that scales exponentially with $N$. This allows for numerical computation of the OTOC for large but finite $N$ and also makes the derivation of the OTOC possible for all times in the large $N$ limit. We emphasize that the original Brownian SYK model does not have the specified symmetry structure, which only appears after taking the random average. This work is also related to recent studies on the emergent discrete symmetry resulting from the disorder average over replicas in random circuit models~\cite{Nahum_2018, vasseur2019entanglement, zhou2020entanglement, jian2020measurement, bao2021symmetry, nahum2021measurement, bao2020theory} and large $N$ field theory calculations of the static SYK model~\cite{jian2021measurement, jian2021quantum}. These discrete symmetries can be intuitively understood as the interplay between permutation among replicas and the physical symmetries of the model. This work demonstrates that the effective model emerging from the Brownian SYK model at any $N$ is not only invariant under these discrete symmetries, but is closed within a larger continuous symmetry group, SU(2) in the case without charge conservation, or SU(4) with charge conservation, for which the discrete symmetry group is a subgroup. Furthermore, when the model is non-interacting, i.e., quadratic in the fermionic operators, it is invariant under the continuous symmetry group \cite{ winer2020exponential,zhang2021syk}.  

We begin this work with a discussion of the map between operator spreading and scrambling in systems with Majorana fermions, followed by complex fermions. To this end, we provide a picture of operator dynamics when they are restricted by the U(1) symmetry in the complex model. We utilize this understanding to analytically compute the late time values of the OTOC in each charge sector, for an arbitrary complex fermionic model with charge conservation.

Next, using the new approach based on the emergent symmetry structures, to connect to the previously known result, we compute the OTOC of the Brownian Majorana SYK model for finite $N$ and obtain an analytical equation that is exact in the large $N$ limit for all times. From the formalism used it becomes clear that two previously known methods to solve the model, the approach which maps the Brownian model to a stochastic or bosonic model, and the Hamiltonian approach, are simply related by a similarity transformation. 
Following this, we apply the method to study the charge dependent scrambling in the complex Brownian SYK model for various operators $W$ and $V$. We obtain numerically exact results of the OTOC for different charge sectors and for $N$ up to 500. We mainly focus on the model with four-fermion terms~($q_{\syk}=4$), and demonstrate that the charge dependent Lyapunov exponent $\lambda_L^{\rho}$, as well as the late time relaxation rate $\lambda_{late}^{\rho}$, are proportional to $\rho(1-\rho)$ where $\rho$ is the charge density. Furthermore, by taking the large $N$ limit in our approach, we obtain the hydrodynamic equation and analytical expression of the charge dependent OTOC for all time scales. By appropriately transforming the time variable to be functionally dependant on $\rho$ and $N$, we also find that all the different kinds of OTOCs considered in this work collapse into a single function for various charge sectors and values of $N$, thereby elucidating the primary functional dependence of the OTOCs on $\rho$ and $N$.

The rest of the paper is organized as follows. In Sec.~\ref{sec:operator}, we provide a physical description of the information scrambling in systems with Majoranas and complex fermions. Sec.~\ref{sec:preparation} is devoted to understanding the essential components of the formalism, namely the mapping of operators to states and the imaginary-time evolution that emerges post disorder averaging in Brownian models. In Sec.~\ref{sec:BrownianSYK}, to illustrate our approach based on the special unitary group, we first study the Brownian SYK model of Majoranas, where the relevant group is SU(2). We demonstrate that in this case, the OTOC can be exactly mapped to an SU(2) spin problem with angular momentum  $L=N/2$. In Sec.~\ref{sec:ComplexBrownianSYK} we use a similar approach to study the Brownian SYK model of complex fermions with U(1) symmetry. In this case, the relevant group is enhanced to SU(4) $\otimes$ U(1) and the operator dynamics is organized into irreps of SU(4). Furthermore, the charge conservation in the original model manifests as conservation of weights in the SU(4) irrep. As a result, the OTOC is exactly mapped to SU(4) spin dynamics with weight conservation, where the largest Hilbert space dimension is linear in $N$. Sec.~\ref{sec:chargedscrambling} discusses the charge-resolved OTOCs in the complex model and analyzes the density dependence of the correlators for large finite $N$ numerically, and large-$N$ limit analytically. Sec.~\ref{sec:discussion} is devoted to discussion and conclusion.

\section{Scrambling in fermionic systems with and without charge conservation}
\label{sec:operator}
In this section, we provide a general physical understanding of quantum information scrambling in terms of operator dynamics in systems of Majorana and complex fermions. The system of complex fermions conserves the total charge. In this case, we also discuss how the initial and late-time values of OTOC depend on the different charge sectors.

The key ingredient to understand operator dynamics in both the Majorana system and the complex fermionic system, is the basis of the operators, denoted by $\mathcal{S}$, which is similar to Pauli strings in the spin system~\cite{Nahum_2018, von_Keyserlingk_2018}. Before discussing each case individually, we first discuss the general properties of the operator basis $\{\mathcal{S}\}$ and set the convention used throughout the rest of the paper. We construct the operator basis $\mathcal{S}$ as a product of local operators, which is sometimes also called an operator string. The number of operator strings is the dimension of the Hilbert space squared. To be consistent with the Pauli string operator basis used in spin systems, we demand that the operator string $\mathcal{S}$ satisfy the following orthogonal and completeness relations
\bea
\frac{1}{\tr I} \tr (\S^\dagger \S') =  \delta(\S',\S), \ \ \frac{1}{\tr I} \sum\limits_{\S} \S^{\dagger}_{mn} \S_{pq} = \delta_{mq}\delta_{np}.
\label{eq:orthcomp}
\eea

\subsection{Operator dynamics of Majorana fermions}
We begin with the analysis of the operator dynamics of Majorana fermions, which is relevant in the case of the regular and Brownian SYK models. We consider a system of $N$ Majoranas with the Hilbert space dimension $\tr(I)=2^{N/2}$, where the Hamiltonian is a function of the Majorana operators $\chi_i$, and the subscript $i$ goes from 1 to $N$. The operators obey the anti-commutation relation $\{\chi_i, \chi_j\}=2\delta_{ij}$. 
A good basis for the operator dynamics are the Majorana strings~\cite{roberts2018operator}, which are products of local operators, either $\chi_i$ or $I_i$. The Majorana strings take the form
\bea
\S =  s_1s_2\cdots s_N
\eea
Where each $s_i$ in the string is either the identity ($I$) or the Majorana operator ($\chi$) at that site. We define the size of the Majorana string as $\text{size}(\S)$, which counts the number of $\chi$'s in the string.
The Majorana strings satisfy the orthogonality and completeness relations in Eq.~\eqref{eq:orthcomp}.
Therefore, any Heisenberg operator at arbitrary time can be expanded in this basis with coefficients $c(\S,t)$:
\bea
W(t) = \sum_{\S} c(\S, t) \S.
\eea
We choose the operator $W$ to have the normalization $\tr(W^\dagger W) =\tr I$ and this leads to the constraint $\sum_{\S} |c(\S,t)|^2 = 1$. Thus the coefficients $|c(\S,t)|^2$ have the interpretation of a probability distribution over the different strings of operators. Quantum information scrambling is tied to the fact that a simple initial operator becomes as complicated as possible under Heisenberg time evolution. This suggests that the operator probability distribution, starting with one localized at a single operator-string, would approach uniform distribution in the late time regime where every operator is equally probable and the system becomes fully scrambled. An important caveat to keep in mind is that the operator cannot spread to the identity or parity operator at late times if it has a null overlap with the specified steady operators at zero time.

Physical systems usually conserve the fermionic parity, since the Hamiltonian only contains an even number of Majorana operators and commutes with the parity operator $\prod \chi$. As a result, the parity of an operator, whether it starts with an even or odd number of Majorana operators, remains invariant under the unitary time evolution. Only the $\S$ with even (odd) lengths appear in the expansion of the operator $W(t)$ with even (odd) parity. In the late time regime, the operator probability becomes uniform in the parity sector determined by the initial operator but remains zero in the opposite sector. The simplest quantity to characterize scrambling of an initially simple operator, such as $\chi(t)$, is the average size of the Majorana strings $\sum_\S |c(\S,t)|^2 \text{size}(\S)$. This average size is precisely measured by the OTOC. Using the operator expansion and the anti-commutation relation of Majorana operators, one can show that
\bea\label{eq:F_majorana}
\sum\limits_i \F(W(t),\chi_i) &= \frac{1}{2^{N/2}} \sum_{i} \tr (W^\dagger(t)\chi_i W(t) \chi_i) \\
&= \pm \left( N-2 \, \overline{\text{size}(\S)}\right).
\eea
The plus or minus sign depends on whether $W(t)$ is parity even or odd. Each OTOC in the sum $\sum_{i} \F(W(t),\chi_i)$ is related to the probability of the operator $\chi_i$ appearing in the operator string. For the simple OTOC  $\F(\chi_i(t), \chi_j)$ that will be considered in Sec.~\ref{sec:BrownianSYK}, the initial and final values are
\bea \label{eq:MajoranaScrmablingExpectation}
\F(\chi_i(0),\chi_j) = -1+2\delta_{ij}, \ \ \F(\chi_i(\infty),\chi_j) =0. 
\eea
We will compute the time evolution of $\F$ from $t=0$ to $t=\infty$ in the Brownian Majorana SYK model and also show that the general expectation is violated when the system becomes non-interacting.
Here we emphasize that the late-time value goes to zero because as intuitively expected, starting from an operator in the parity odd sector will result in the operator spreading uniformly to all the operator-strings in the odd sector, therefore odd operator-sizes will be binomially distributed and the average operator size will be $N/2$. On the other hand, if we consider a bosonic initial operator, say $\chi_i \chi_{i'}$, the late-time operator distribution will be uniform over the operator-strings in the even sector, with the exception of the identity and parity operator which are static. The binomial distribution of all operators with even sizes excluding the identity and the parity operator also leads to average operator size $N/2$. Therefore, in the majorana system where the only symmetry is the fermionic parity, the late-time value of the OTOC in Eq.~\eqref{eq:F_majorana} approaches zero. 
This is in contrast with general spin models, where the only operator excluded from the late-time distribution would be the identity. In this case, the late-time value of the OTOC contains a finite-size correction exponentially small as a function of the system size~\cite{roberts2017chaos}.

\subsection{Operator dynamics of complex fermions} \label{subsec:complexotoc}
In this section, we discuss the operator dynamics of complex fermions in systems with charge conservation. In general, we consider a system of $N$ fermions, and a Hamiltonian that is a function of the creation and annihilation operators $\chi^\dagger_i$ and $\chi_i$, where the subscript $i$ goes from 1 to $N$. These operators obey the standard anti-commutation relation $\{\chi_i^\dagger, \chi_j\} = \delta_{ij}$. We define the operator $n_i=\chi_i^\dagger \chi_i$, which measures the local charge. We also define $\bar{n}_i=I-n_i$ for later convenience. The Hamiltonian conserves the total charge, meaning that $[H, \sum_i n_i]=0$. As a result, given an initial state with a fixed charge, its dynamics is always restricted to the corresponding charge-sector of the Hamiltonian. 

One can also define the conserved charge for an operator in such systems. Unlike the state, the operator has two conserved quantities, resulting from measuring the total charge on the left or on the right. An eigen-operator $W$ of two U(1) symmetries is defined as
\bea
\left(\sum_i n_i\right) W= m_a W, \quad  W \left(\sum_i n_i\right) = m_b W.
\eea
In general, $m_a$ and $m_b$ are independent and the tuple $(m_a,m_b)$ is labelled as the charge-profile of the operator. Because the Hamiltonian conserves the total charge $\sum n_i$, the Heisenberg operator $W(t)$ remains an eigen-operator with the same conserved quantities $m_a$ and $m_b$. The appropriate local eigen-operators have the charge profile :
\bea \label{eq:singleoperatorcharge}
\chi^\dagger:(1, 0) \ \ \chi:(0,1) \ \ n:(1,1) \ \ \bar{n}:(0, 0).
\eea
Note that the identity operator is not an eigen-operator of the two U(1) symmetries.

Using the operators in Eq.~\eqref{eq:singleoperatorcharge}, one can construct a complete operator basis for $N$ fermions that fully respects the two U(1) symmetries of the operator dynamics
\bea
\mathcal{S} = 2^{N/2}s_1 s_2 \cdots s_N, \quad s_i \in \{ \chi^\dagger, \chi, n , \bar{n} \}.
\eea
One can also show that $m_a(\S)$ counts the number of $\chi^\dagger$ plus the number of $n$ in the string, while $m_b(\S)$ counts the number of $\chi$ plus $n$. Under unitary time evolution driven by the charge-conserving Hamiltonian, the only permissible building block for the dynamics of the operator-states of the form $\mathcal{S}$ is the move 
\bea
\bn n \longleftrightarrow \chi^{\dagger} \chi
\eea
which preserves the charge-profile. Following this, one can immediately detect the set of operators that may have time-dependence but display no operator spreading, i.e. $W(t) = e^{ict}W$ where $c$ is a real constant. For example, it is evident that the operators of the form $\prod_{i=1}^{N} \chi_i^{\dagger}$ and $\prod_{i=1}^{N} \chi_i$ are the unique operators with the charge profiles $(N,0)$ and $(0,N)$ respectively, therefore they simply gain a phase under the dynamics. The identity operator is more special because it always commutes with the unitary and therefore does not have dynamics. We can further expand $I$ in the basis $\mathcal{S}$ as
\bea \label{Eq:IdentitySplit}
&I = \prod_{i=1}^{N} (n_i + \bn_i) =  \sum_{m=0}^{N} I_{m, N}\\
&I_{m,N} = \sum_{i_1<...<i_m} n_{i_1}...n_{i_m}\bn_{i_{m+1}}...\bn_{i_{N-m}}.
\eea
Here the operator $I_{m, N}$ is a component of the identity over $N$ fermions in the U(1) basis, with the charge profile $(m,m)$. Since the identity does not have time dependence, each component with a different $m$ is also static under the charge-conserving dynamics. 

We can determine the dynamics of a local operator $W_i I$ after expanding the identity over $N-1$ fermions in the basis $\S$
\bea \label{eq:localdynamics}
W_i I = W_i \prod_{j \neq i} (n_j + \bn_j) =  \sum_{m=0}^{N-1}W_iI_{m, N-1}.\\
\eea
Each component $I_{m,N-1}$ has profile $(m,m)$. Thus it is apparent that when a local operator chosen from the set $\{ \chi_i^\dagger, \chi_i, n_i , \bar{n}_i \}$ is expanded in such a basis, each operator-string in the sum has a fixed charge profile $(m_a,m_b)$, where $m_a-m_b$ is invariant across all the components and takes the values $\{1, -1, 0 , 0\}$ corresponding to the choice of operator from the set respectively.
Once the expansion is obtained, the dynamics, in the form of the charge conserving move, take place independently within each charge sector (labelled by $m$).

Given an initially simple operator $W_i$, one way to track its complexity under Heisenberg time evolution is using the OTOC
\bea \label{eq:OTOCVW}
\F(W_i(t), V_j) = \frac{1}{2^N}\tr\left( W_i^\dagger(t) V_j^\dagger W_i(t) V_j \right)
\eea
where $V_j$ is a local probing operator. Since $W_i(t)$ is a local operator, it does not respect the two U(1) symmetries and has the expansion shown in Eq.~\eqref{eq:localdynamics}. Therefore the OTOC $\F$ contains contributions from different charge sectors,
\bea \label{eq:chargeOTOCVW}
\F(W_i(t), V_j) = &\sum\limits_m \frac{\tr{(P_m)}}{2^N} \F^{m}(W_i(t), V_j)\\ 
\F^m(W_i(t), V_j) =& \frac{1}{\tr (P_m)}\tr(P_m W^\dagger_i(t) V_j^\dagger W_i(t) V_j )\\
\eea
where $P_m = \sum \limits_m \ket{\psi_m}\bra{\psi_m}$ is the projection operator for the subspace of the Hamiltonian with charge $m$ and dimension $\binom{N}{m}$. We denote $\F^m$ as the charge-resolved OTOC, contributing to the overall $\F$ based on the binomial distribution.
To illustrate this, let us consider $\F(\chi_i(t), \chi_j^\dagger)$ where one can start with $\chi_i$ and use $\chi_j^{\dagger}$ to probe its growth. 
In $\F^m(\chi_i(t), \chi_j^{\dagger})$ given by
\bea
\F^m(\chi_i(t), \chi^{\dagger}_j) =& \frac{1}{\tr (P_m)}\tr(P_m \chi_i^{\dagger}(t) \chi_j \chi_i(t) \chi_j^{\dagger} ).
\eea
The charge profiles of the operator $\chi_i^\dagger(t)$ (notice the dagger) and $\chi_i(t)$ are fixed to be $(m,m-1)$ and $(m,m+1)$ respectively for all time as a result of the charge conserving dynamics. Therefore, $\F^m(\chi_i(t), \chi_j^{\dagger})$ probes the correlation between the components of the operator $\chi_i^\dagger(t)$ in two different charge sectors $(m,m-1)$ and $(m+1,m)$. It is generally true that $\F^m(W_i(t), \chi_j)$ probes the correlation between different charge sectors of $W_i$. 
To probe the operator growth within a charge sector, one can also use $n_j$ instead as the probing operator and study the charge resolved OTOC $\F^m(W_i(t), n_j)$. For example, in $\F^m(\chi_i(t), n_j)$ the charge profiles of $\chi_i^\dagger(t)$ and $\chi_i(t)$ are fixed to be $(m,m-1)$ and $(m-1,m)$. Therefore, $\F^m(\chi_i(t),n_j)$ probes the operator dynamics of $\chi_i(t)$ within one charge sector $(m-1,m)$.
Using combinatorics and the assumptions that the Heisenberg operator $W(t)$ becomes as complicated as possible at late times, we obtain the initial and late-time values of the charge resolved OTOC $\F^m(\chi_i(t),\chi_j^\dagger)$ and $\F^m(\chi_i(t),n_j)$, which are summarized in Table~\ref{tab:earlylatevalue}. 

We also consider the Heisenberg dynamics of the operator $n_i(t)$. The operator growth can be probed by the charge resolved OTOC $\F^m(n_i(t), \chi_j^{\dagger})$ and $\F^m(n_i(t), n_j)$. When the local probing operator is $\chi_j^{\dagger}$, $\F^m$ probes the correlation between the charge sectors $(m, m)$ and $(m+1, m+1)$ of $n_i$. When the local probing operator is $n_j$, $\F^m$ probes the operator growth of $n_i$ within the charge sector $(m,m)$. In the case of the OTOCs $\F^m (n_i(t), \chi^{\dagger}_j)$ and $\F^m (n_i(t),n_j)$, the late-time values are more nontrivial to compute than when the initial operator is chosen to be $\chi_i (t)$, because $n_i(t)$ is not traceless. Within each charge sector, we have $\tr(P_m n(t))/\tr(P_m)=m/N$, and this will put constraints on the coefficients of the operators present in the component of the identity in the charge sector (i.e., all the different operators that make up $I_{m,N}$ in Eq.~\eqref{Eq:IdentitySplit}). These constraints will in turn lead to non-uniform operator spreading. To remedy this, one can investigate the OTOC through the operator-spreading of a modified operator $n(t)-\Delta_m I$ within each charge sector, which is related to the OTOC of $n(t)$ in a simple way. The constant $\Delta_m$ can be chosen to precisely guarantee uniform operator spreading at late times, through the equation $\tr{(P_m (n_i - \Delta_m I))} = \sqrt{(m/N) (1-2\Delta_m+\Delta_m^2 N/m)}$. The initial and late-time values for $\F^m(n_i(t),\chi_j^\dagger)$ and $\F^m(n_i(t),n_j)$ are also listed in Table~\ref{tab:earlylatevalue},
and one can use them to compute the late time values of the overall OTOCs $\F$ from the weighted average $\F=\sum_m \tr(P_m) \F^m/\tr(I)$
\bea\label{eq:overalllatevalue}
&\F(\chi_i(t \rightarrow \infty),\chi_j^{\dagger}) = 0\\
&\F(\chi_i(t \rightarrow \infty), n_j)= \frac{(N-1)(N+2)}{8N^2} \overset{N \rightarrow \infty}{=} \frac{1}{8}\\
&\F(n_i(t \rightarrow \infty),\chi_j^{\dagger}) = \frac{(N-1)(N+2)}{8N^2} \overset{N \rightarrow \infty}{=} \frac{1}{8}
\eea
For the OTOC $\F(n_i(t),n_j)$, it is difficult to obtain a closed form expression for the overall OTOC. However, from the charge-resolved value, one can compute the large-$N$ expansion and find that the leading order piece at late times is $\F(n_i(t \rightarrow \infty),n_j)\overset{N \rightarrow \infty}{=} 3/16.$
This OTOC will also have finite sized effects in its late time value, similar to both the other OTOCs computed where one of the operators is chosen to be $n$.
In Sec.~\ref{sec:chargedscrambling}, we verify these late-time values in the case of the complex Brownian SYK model (The values for the charge-resolved case are verified in the appendix). Furthermore, we also provide an exact formalism to track the time evolution of various OTOCs from the initial value to the late-time value in different charge sectors.

Using the result in Table.~\ref{tab:earlylatevalue}, one can also compute the late-time value of the overall OTOC between traceless operators by summing over the contribution from each charge sector and compare the result without charge conservation. For example, we can consider the overall OTOC between $\chi_i$ and $n_j-I/2$. Importantly, within each symmetry sector, the operator $n_j-I/2$ is not traceless, and the charge resolved OTOC approaches a finite value that depends on $m$. Summing over contributions from each charge sector leads to a late-time value $(N-2)/8N^2$, scaling as $\sim 1/N$ in sharp contrast with $1/\exp(\alpha N)$ that is found in systems without symmetry~\cite{roberts2017chaos}. The $1/N$ corrections to the late-time value are present for all pairs of operators in Table.~\ref{tab:earlylatevalue}, except $\F(\chi_i(t), \chi_j^\dagger)$ because the operators involved are traceless in each charge sector. 
Related to this, the $1/\text{poly}(N)$ correction to the late-time value of the OTOC is found in energy conserving systems as well~\cite{huang2019finite}. 

We also note that some subtlety arises for the late value of $\F^m(W_i(t), V_j)$ when the Hamiltonian is $q_{\syk}$-uniform and only contains terms of the form $\chi_{i_1}^\dagger\cdots \chi_{i_{q_{\syk}/2}}^\dagger \chi_{j_1} \cdots \chi_{j_{q_{\syk}/2}} $. In this case, some operators in sectors of dilute charge ($m \sim \mathcal{O}(1)$) have restricted dynamics owing to the symmetries of the $q_{\syk}$-uniform Hamiltonian, and the late time value of the OTOC can be different for the cases $i=j$ and $i \neq j$, due to imperfect scrambling. 

\begin{table}
\renewcommand{\arraystretch}{2}
 \begin{tabular}{||c| c |c |c||} 
 \hline
 $W_i$ & $V_j$ & $\F^m(W_i(t=0), V_j)$ & $\F^m(W_i(t \rightarrow \infty), V_j)$ \\ 
 \hline\hline
 $\chi_i$ & $\chi_j^{\dagger}$ & $(1-\delta_{ij})\frac{m(m-N)}{N(N-1)}$ & 0 \\
 \hline
 $\chi_i$ & $n_j$ & $(1-\delta_{ij})\frac{m(m-1)}{N(N-1)}$ & $\frac{m^2(m-1)}{N^3}$ \\
 \hline
 $n_i$ & $\chi_j^{\dagger}$ & $(1-\delta_{ij})\frac{m(m-N)}{N(N-1)}$ & $\frac{m(m+1)(N-m)}{N^3}$ \\
 \hline
 $n_i$ & $n_j$ & $\frac{m(m(1-\delta_{ij})-1) }{N(N(1-\delta_{ij})-1)}$ & $\frac{m^3}{N^3} + (\frac{m}{N}-(\frac{m}{N})^2)\bigg(\frac{\binom{N-1}{m-1}^2-m/N}{\binom{N}{m}^2-1} \bigg)$ \\[5pt]
 \hline
\end{tabular}
\caption{\label{tab:earlylatevalue} The early and late time value of the charge resolved OTOC in Eq.~\eqref{eq:chargeOTOCVW}  for each charge sector labelled by $m$ and for different choices of the operators $W$ and $V$.}
\end{table}

\section{General formalism} \label{sec:preparation}
\subsection{Operator to state mapping on 4 copies of the Hilbert Space}\label{sec:OperatorstateOTOC}
To compute the OTOC, we map it to the overlap between two quantum states in four replicas of the original Hilbert space, labelled by $a$, $b$, $c$, $d$. The OTOC in Eq.~\eqref{eq:OTOCVW} can be written as
\bea \label{eq:LRotoc}
\F(W(t), V) = \tr{I}\bra{\o} \mathbb{U} \ket{\i}
\eea
where $\ket{\i}$ and $\ket{\o}$ are the input and output states defined as
\bea
&\ket{\i}=\frac{1}{\tr I}\sum W^\dagger_{m n} W_{pq}\ket{m\otimes n \otimes p \otimes q} \\
&\ket{\o} = \frac{1}{\tr I}\sum V^\dagger_{m q} V_{p n }\ket{m\otimes n \otimes p \otimes q}
\eea
and $m$, $n$, $p$, $q$ are the computational basis states spanning the Hilbert space of each replica.
The time evolution operator $\mathbb U$ is given by
\bea
\mathbb U = U\otimes U^* \otimes U \otimes U^*.
\eea
The time evolved state $\ket{\i (t)}$ is
\bea
\ket{\i (t)}=\mathbb U \ket{\i}=\frac{1}{\tr I}\sum W^\dagger(t)_{m n} W(t)_{pq}\ket{m\otimes n \otimes p \otimes q}
\eea
One can verify that $\braket{\o|\i (t)}$ does indeed lead to $\F(W(t),V)$. To simplify the notation, we rewrite $\ket{\i}$ as
\bea
\ket{\i} = \frac{1}{\tr I}\ket{W^\dagger\otimes W}.
\eea
On the other hand, using the completeness relation of the operator basis $\mathcal{S}$, we rewrite the state $\ket{\o}$ as
\bea
\ket{\o} &= \frac{1}{\tr^2 I}\sum V^\dagger_{mm'}\S^\dagger_{m'n'}V_{n'n}\S_{pq}\ket{m\otimes n\otimes p\otimes q} \\
&=\frac{1}{\tr^2 I} \sum\limits_{\mathcal{S}} \ket{ V^\dagger  \S^\dagger V \otimes \S}.
\eea
Therefore, the OTOC is written as
\bea  \label{eq:otoc_4}
\F(W_i(t),V_j)&= \frac{1}{\tr^2(I)} \sum\limits_{\mathcal{S}} \bra{ V_j^\dagger   \S^\dagger V_j \otimes \S}\mathbb U \ket{ W_i^\dagger \otimes W_i}.
\eea
This is equivalent to Eq.~\eqref{eq:LRotoc} and does not reduce the computational complexity in general. However, as we will show in the following sections, for a special class of chaotic quantum many-body models called Brownian models, the symmetry and algebraic structure of $\mathbb{U}$ after the random averaging significantly reduces the computational complexity. This allows us to calculate the OTOC exactly for all time scales, including the early time growth and late-time saturation, for large system size. 

\begin{figure}
\includegraphics[width=1\columnwidth]{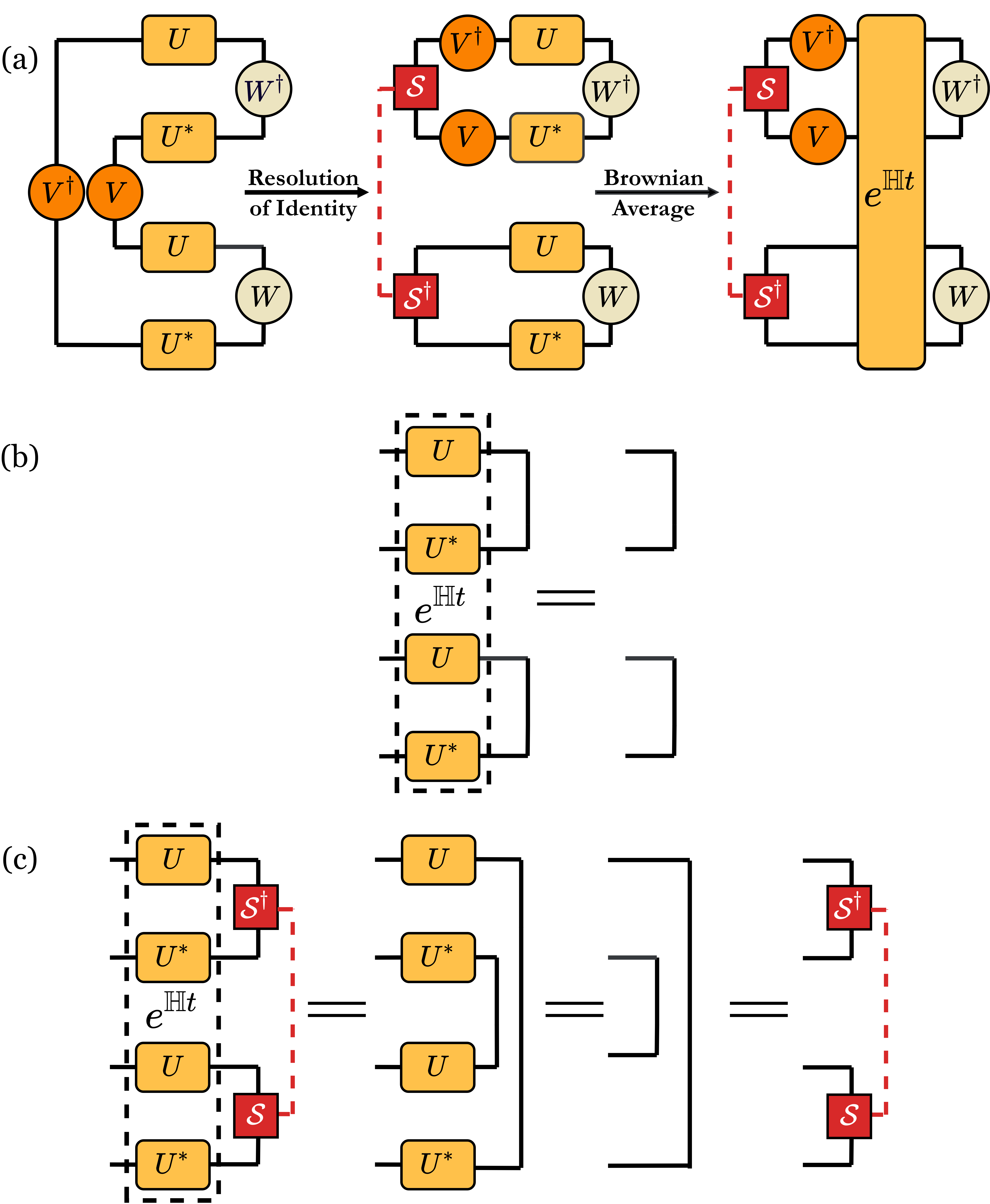}
\caption{(a) Resolution of the identity in the OTOC, followed by the Brownian average. (b) The invariance of the Identity operator-state $\ket{I \otimes I}$ under the effective imaginary-time evolution. (c) The invariance of the complete set of operator-states $\sum_{\S} \ket{\S^{\dagger} \otimes \S}$ under the effective imaginary-time evolution. Both operator-states considered in (b) and (c) remain eigenstates of the effective Hamiltonian $\H$ because they are invariants of the circuit even before the random averaging.}
\label{Fig:Tensor}
\end{figure}

\subsection{General structure of Brownian models}
The Hamiltonian of Brownian models takes the following general form,
\bea\label{eq:H_brownian}
H(t) = \sum_A J_A(t) \hat{X}_A + h.c.
\eea
where $\hat{h}_A$ describes few body interactions, i.e. spin-spin interactions or 4-body fermionic interactions between the different degrees of freedom, and $A$ is a generic index labeling the degrees of freedom this term acts on.  The model is Brownian because the coupling strength $J_A(t)$ is a Gaussian white noise uncorrelated in time obeying
\bea
\overline{J_A(t)}=0, \quad \overline{J_A(t) J^{*}_{A'}(t')}= J \delta_{AA'}\delta(t-t').
\eea
Because the disorder is uncorrelated in time, the disorder average can be computed independently at different times. At each time slice, we have
\bea
\overline{d\U}=\overline{dU^a dU^{b,*} dU^{c} dU^{d,*}}=(1+\H dt)
\eea
where
\bea
&\H = -\frac{J}{2}\sum_A\left( \mathbb{X}_A \mathbb{X}^\dagger_A + \mathbb{X}^\dagger_A \mathbb{X}_A \right),\\ 
&\mathbb{X}_A =X_A^a-{X_A^{b,*}}^{\dagger}+X_A^c-{X_A^{d,*}}^{\dagger}.
\eea
The operator $\H$ is Hermitian and semi-negative, namely all the eigenvalues are either zero or negative. 
As is evident from the expression above, there is a discrete symmetry under the exchange of replicas $a \leftrightarrow c$ and $b \leftrightarrow d$, along with a conjugation symmetry between the pairs $(a,c)$ and $(b,d)$. These discrete symmetries have been studied in detail recently~\cite{bao2021symmetry}.

In addition $\H$ has two zero energy eigenstates that are independent of the details of $\hat X_A$. We have 
\bea
e^{\H t} \ket{I\otimes I}=\ket{I\otimes I}, \ \ e^{\H t} \sum\limits_\S \ket{\S^\dagger\otimes S}=\sum\limits_\S \ket{\S^\dagger\otimes S}.
\eea
This is because the identity operator stays the same under unitary time evolution (Fig.~\ref{Fig:Tensor}(b) and (c)). This condition is useful for consistency checks when we build $\H$ for specific models. If the Hamiltonian has additional symmetries, the corresponding operator-states are also zero-energy eigenstates with respect to the emergent Hamiltonian. As an example, the parity operator-state $\ket{\prod \chi \otimes \prod \chi}$ has null energy with respect to the emergent Hamiltonian in the case of the Brownian SYK model. 

The unitary time evolution operator on the replicated Hilbert space after Brownian average $\overline{\U}$ becomes $\exp (\H t)$. In other words, the Brownian disorder average converts the unitary real-time evolution to the imaginary time evolution governed by the Hamiltonian $\H$ acting on four copies of the original Hilbert space. This property follows from the observation that the term corresponding to first order time in $dU$ has to be paired with another first order term to be nonzero after disorder average. Then the OTOC in Eq.~\eqref{eq:otoc_4} after the disorder average can be written as
\bea \label{eq:OTOCgeneralform}
\F(W_i(t),V_j)&=\\
\frac{1}{\tr^2(I)}& \sum\limits_{\S} \bra{ V_j^\dagger   \S^\dagger V_j \otimes \S} \exp(\H t) \ket{ W_i^\dagger \otimes W_i}
\eea
which measures the overlap between the input state and output state after quenched \textit{imaginary} time evolution. This is valid for any Brownian model described by Eq.~\eqref{eq:H_brownian}. In general, with this formalism, it is still very challenging to obtain the OTOC $\F$ because one needs to diagonalize the Hamiltonian $\H$ numerically, which is limited to small system sizes. 

In the following sections, we will show that for a certain class of Brownian models, namely the Brownian-SYK models, the Hamiltonian $\H$ exhibits elegant symmetry structures that only appear after the disorder average. By exploiting the symmetry structures, we show that the largest Hilbert space involved in the computation of $\F$ scales linearly with the number of Majoranas or complex fermions in the systems, making results for large but finite $N$ accessible. More specifically, we find that
\begin{itemize}
    \item For the Brownian Majorana SYK model without charge conservation, the operator dynamics can be mapped to imaginary time dynamics of an SU(2) spin with angular momentum $\sim N/2$.
    \item For the complex Brownian SYK model with charge conservation, the operator dynamics can be mapped to imaginary time dynamics of an SU(4) spin with fixed weight.
\end{itemize}
In what follows, we discuss each model individually. For the Brownian Majorana SYK model, our approach, inspired by~\cite{Sunderhauf2019}, simplifies the method used therein and unifies the approach used in~\cite{Sunderhauf2019} and~\cite{Xu_2019, Jian_2021, Zhou_2019}. More importantly, our approach can be generalized to the complex Brownian SYK model, which is the primary focus of the current work.

\section{The Brownian SYK model} \label{sec:BrownianSYK}

In this section we will review the work done on the Brownian SYK model for the purpose of completeness and notational clarity. We start with the Hamiltonian 
\bea \label{eq:MajoranaHamiltonian}
H(t)=i^{\frac{q_{\syk}}{2}} \sum_{ i_{1}<\ldots<i_{q_{\syk}}} J_{i_{1}, \ldots, i_{q_{\syk}}}(t) \chi_{i_{1}} \chi_{i_{2}} \ldots \chi_{i_{q_{\syk}}}
\eea
where the generalised index $i_{j}$ can take values between $1$ and $N$. The couplings for general $q_{\syk}$ are distributed according to
\bea
&\overline{J_{i_{1} \ldots, i_{q_{\syk}}}(t) J_{i_{1}^{\prime} \ldots, i_{q_{\syk}}^{\prime}}\left(t^{\prime}\right)}\\
&\quad =\delta_{i_{1} i_{1}^{\prime}} \cdots \delta_{i_{q_{\syk}} i_{q_{\syk}}^{\prime}} \delta\left(t-t^{\prime}\right) \frac{(q_{\syk}-1)!}{2N^{(q_{\syk}-1)}}.
\eea
We are interested in computing the OTOC, which can be rewritten in terms of four copies of the Hilbert space, as shown in Sec.~\ref{sec:OperatorstateOTOC}.
The four copies of the unitary operator $d\U$ that encode the time evolution in the OTOC are built using operators of the form 
\bea
\chi_{j}^{a}:=\chi_{j} \otimes I \otimes I \otimes I && \chi_{j}^{b}:=I \otimes \chi_{j}^{*} \otimes I \otimes I\\
\chi_{j}^{c}:=I \otimes I \otimes \chi_{j} \otimes I && \chi_{j}^{d}:=I \otimes I \otimes I \otimes \chi_{j}^{*}
\eea
These operators satisfy the (anti-)commutation relations $\left[\chi_{j}^{\alpha}, \chi_{k}^{\beta}\right]=0$ for $\alpha \neq \beta $
and $\left\{\chi_{j}^{\alpha}, \chi_{k}^{\alpha}\right\}=2 \delta_{j, k}$.
We can use the parity operator $\Q$ on each copy 
\bea
\mathcal{Q}^{\alpha}=\prod_{k=1}^{N} \chi_{k}^{\alpha}, \quad \alpha=a, b, c, d 
\eea
to turn $\chi_i^\alpha$ into purely anti-commuting operators as follows~\cite{Sunderhauf2019}
\bea
&\psi_{j}^{a}=i \mathcal{Q}^{a}\chi_{j}^{a}, && \psi_{j}^{b}=\mathcal{Q}^{a} \chi_{j}^{b}\\ & \psi_{j}^{c}=i \mathcal{Q}^{a} \mathcal{Q}^{b} \mathcal{Q}^{c} \chi_{j}^{c}, &&
\psi_{j}^{d}=\mathcal{Q}^{a} \mathcal{Q}^{b} \mathcal{Q}^{c} \chi_{j}^{d}.
\eea
These new operators obey the relation $\left\{\psi_{j}^{\alpha}, \psi_{k}^{\beta}\right\}=2 \delta_{\alpha, \beta} \delta_{j, k}$, and since $(Q^{\alpha})^2 = 1$ (for $N \equiv 0$ mod $4$), we can exploit the identity $\prod_{k=1}^{M} \chi_{j_{k}}^{\alpha}=\prod_{k=1}^{M} \psi_{j_{k}}^{\alpha}$ to rewrite $d\U$ and thus the effective Hamiltonian $\H$ in terms of these new operators. We introduce the bilinear operators $S^{\alpha\beta} = \sum\limits_i \psi_i^\alpha \psi_i^\beta$, and remark that the effective Hamiltonian takes the following general form :
\bea \label{eq:MajoranaHamiltonianStructure}
\H = \H\left(S^{\alpha\beta}\right), \quad \alpha, \beta \in a, b, c, d.
\eea
The explicit expression, which depends on $q_{\syk}$, is provided in the appendix. The main observation is that the Hamiltonian $\H$ is always a function of the six operators $S^{\alpha\beta}$  for $\alpha\neq \beta$~(the term with $\alpha=\beta$ contributes a constant term to the Hamiltonian).

\subsection{The emergent SU(2) $\otimes$ SU(2) algebra}

The emergent Hamiltonian $\H$ acts on a Hilbert space of dimension $4^N$. The system contains $N$ sites, each one hosting 4 local states, which can be thought of as the vacuum state ($\ket{I \otimes I}$), doubly-occupied state ($\ket{ \Q \chi_i \otimes \Q \chi_i}$), and two singly occupied states ($\ket{ \Q \chi_i \otimes I}, \ket{I \otimes \Q \chi_i}$), much like those in the Fermi-Hubbard model. Here $\Q$ is the parity operator $\prod_{i} \chi_i$. The vacuum and doubly occupied state are even parity states whereas the singly occupied states have odd parity. Now we explore the symmetry of $\H$ to block-diagonalize the Hamiltonian and reduce the effective dimension. First, $S^{\alpha\beta}$ and $\H$ commute with the onsite parity operator $\psi^a_i\psi^b_i\psi^c_i\psi^d_i$. As a result, the number of fermions per site stays even or odd. One can show that for the input state of the form $2^{N/2}\ket{\i}=\ket{W^\dagger \otimes W}$, the parity on each site is even, either the empty state or doubly occupied state. This reduces the total Hilbert space dimension to $2^N$. To this end, we map the system to $N$ two-level systems. 

In order to further reduce the Hilbert space dimension, we will rely on additional symmetries of the emergent Hamiltonian. 
Let us define the following quantities:
\bea \label{eq:MajoranLJ}
L_x &= \frac{1}{4i}(S^{bc} + S^{ad}), \quad J_x = \frac{1}{4i}(S^{bc} - S^{ad}) \\
L_y &= \frac{1}{4i}(S^{ca} + S^{bd}), \quad J_y = \frac{1}{4i}(S^{ca} - S^{bd}) \\
L_z &= \frac{1}{4i}(S^{ab} + S^{cd}),\quad  J_z = \frac{1}{4i}(S^{ab} - S^{cd}).
\eea
These operators are the generators of the  $\text{SU}(2)\otimes \text{SU}(2)$ algebra, which can by checked by verifying the commutation relations 
\bea
\left[L_i , L_j\right] = i \epsilon_{ijk} L_k, \   [J_i , J_j] = i \epsilon_{ijk} J_k, \  [L_i, J_j] = 0.
\eea
The full emergent Hamiltonian can now be written as a function of both the $L$ and the $J$ operators (Eq.~\eqref{eq:MajoranaHamiltonianStructure}). Identifying the SU(2) algebras in the Hamiltonian has reduced the maximum dimensionality of the dynamical subspace to be of order $N^2$ since the input state will split into the irreps of the algebras, and the Hamiltonian evolves separate irreps independently. Consider the four states on a site. They split into a $\textbf{2} \oplus \textbf{2}$ representation of the SU(2) $\otimes$ SU(2) group. The empty and doubly occupied states form a doublet of the $L$ algebra and a singlet of the $J$ algebra, while the two singly occupied states form a singlet of the $L$ algebra and a doublet of the $J$ algebra. Since the input state $\ket{\i}$ has an even number of fermions per site, it is a singlet state of the $J$ algebra and we have $J_\alpha \ket{\i}=0$ ($\alpha = x,y,z$). 
Thus the emergent Hamiltonian will now have only one copy of the SU(2) algebra, i.e.\ the $L$ algebra. The effective Hamiltonian in this form does not depend on the details of the system but is fully determined by the irrep of the SU(2) algebra. More explicitly, for $q_{\syk}=2$, the emergent Hamiltonian $\H$ is
\bea
\H_{q_{\syk}=2}=\frac{1}{2N}\left(-2\left(\begin{array}{c}
N \\
2
\end{array}\right) -3N +4L^2 \right).
\eea
The Hamiltonian is SU(2) invariant as it only contains the total angular momentum $L^2$, and is a c number once the irrep is fixed. 
On the other hand, for $q_{\syk}=4$, the Hamiltonian is 
\bea \label{eq:q=4MajoranaHamiltonian}
\H_{q_{\syk}=4}=\frac{3}{N^{3}}\left(-2\left(\begin{array}{c}
N \\
4
\end{array}\right)+\frac{1}{4!} \bigg(H_x +H_z - H_y\bigg) \right),
\eea
where
\bea
H_{\alpha} = 32 L_{\alpha}^4 + 8(8-6N)L_{\alpha}^2 + 6N(N-2); && \alpha = x,y,z.
\eea
The full SU(2) symmetry in the $q_{\syk}=2$ case is reduced to a discrete rotation symmetry in the $xz$ plane. For general $q_{\syk}$, the part of the emergent Hamiltonian which depends on the angular momentum takes the form $H_x + H_z -(-1)^{q_{\syk}/2} H_y$, and hence the symmetry is either discrete $\pi/2$ rotations within the $xz$ plane or between the $x,y,z$ axes, depending on whether $q_{\syk}/2$ is even or odd respectively. The additional symmetry in the case where $q_{\syk}/2$ is odd occurs as a result of the  time-reversal symmetry operator commuting with the unitary time evolution operator~\cite{saad2018semiclassical}.

The emergent Hamiltonian is always a function of the angular momentum $\vec{L}$ for arbitrary $q_{\syk}$. When $q_{\syk}=2$, the Hamiltonian is also an SU(2) invariant, making it analytically tractable. For $q_{\syk}>2$, the Hamiltonian only has a square or cubic symmetry. The enhanced symmetry in the $q_{\syk}=2$ case makes the operator dynamics non-scrambling and distinct from general $q_{\syk}$. This is expected since the original Brownian SYK is quadratic at $q_{\syk}=2$.

Although we still need to diagonalize $\H$ to obtain the dynamics for general $q_\syk$, the largest Hilbert space dimension, which is determined by the angular momentum, is $N+1$. This is drastically reduced from the original Hilbert space size of $4^N$ and enables the exploration of operator scrambling dynamics for large but finite $N$ and arbitrary time scales. Furthermore, as we will show in Sec.~\ref{subsec:Majorana-hydrodynamics}, this formalism also makes it possible to derive an analytical expression for the OTOC in the large $N$ limit for arbitrary time scales and is naturally connected to the previously known approach which consists of mapping to a stochastic model.

\begin{figure}
\includegraphics[width=1\columnwidth]{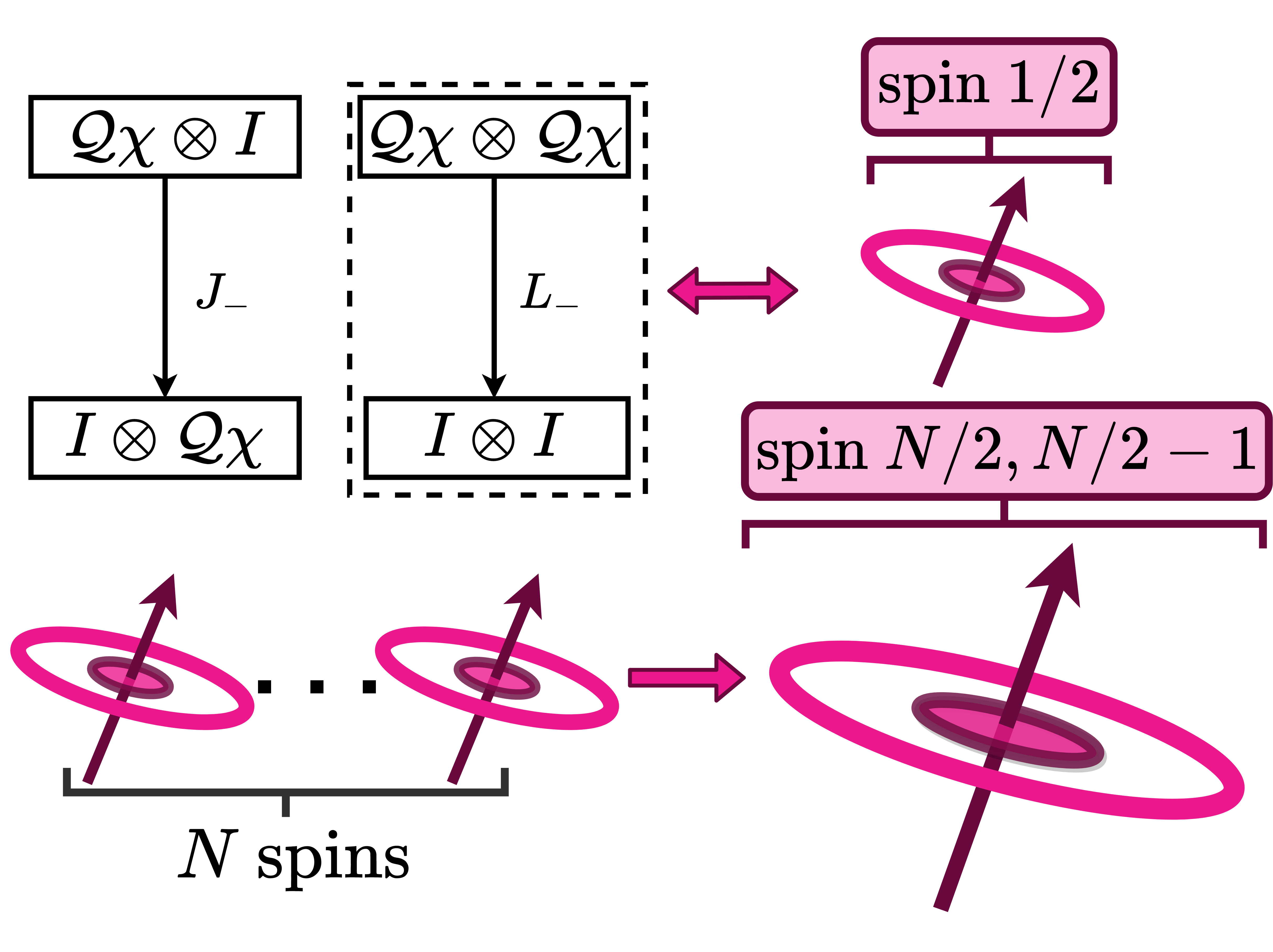}
\caption{The four local operator-states form two doublets of the emergent SU(2) $\otimes$ SU(2) algebra. The $J$ doublets do not contribute to the OTOC because they have odd local parity. The operator dynamics of $N$ Majoranas are mapped to the imaginary time dynamics of $N$ 1/2-spins from the $L$ doublets, which compose to form $N/2$ and $N/2-1$ spins 
of the global SU(2) algebra generated by $\vec L$ in Eq.~\eqref{eq:MajoranLJ}. The largest Hilbert space dimension is $N+1$.}
\label{Fig:MajoranaSpin}
\end{figure}

\subsection{Classifying the states in the SU(2) algebra} \label{sec:MajoranaOTOC}
We are interested in studying the operator dynamics of the Majorana fermions $\chi(t)$ through the OTOC $\F(\chi_i(t), \chi_j)$, which, based on Eq.~\eqref{eq:OTOCgeneralform}, takes the following form:
\bea
&\F(\chi_i(t),\chi_j) = 2^{N/2} \bra{\o}\exp(\H t) \ket{\i} \\
&\ket{\i} =\frac{1}{2^{N/2}} \ket{\chi_i \otimes \chi_i}, \ \  \ket{\o}=\frac{1}{2^{N}} \sum\limits_\S \ket{\chi_j \S^\dagger \chi_j\otimes S}.
\eea
Now that we have described the structure of the emergent Hamiltonian, the next task is to decompose the input state $\ket{\i}$ and the output state $\ket{\o}$. The problem is very similar to decomposing states of $N$ $1/2$-spins into various irreps of the total angular momentum. In general, we have
\bea \label{eq:MajoranaYoungComposition}
2\otimes 2 & \cdots 2\otimes 2 
= (N+1) \oplus \underbrace{(N-1)}_{N-1 \text{ copies}} \oplus \cdots 
\eea
There is a single irrep with largest angular momentum $L=N/2$, $(N-1)$ irreps with $L=N/2-1$, etc.  However, in calculating the OTOC $F(\chi_i(t),\chi_j)$, only the irreps with $L=N/2$ and $L=N/2-1$ appear in the decomposition (As shown in Fig.~\ref{Fig:MajoranaSpin}). Now we construct the input and the output state explicitly from the SU(2) algebra.

\textit{Input State}--We first notice that the operator states corresponding to the identity operator $\ket{I\otimes I}$ and the parity operator $\ket{\Q \otimes \Q}$ are the two fully polarized states in the $z$ direction and thus belong to the unique $L=N/2$ irrep. This can be checked explicitly as
\bea
L_z |I \otimes I \rangle = -\frac{N}{2} |I \otimes I \rangle, \quad 
L_z |\Q \otimes \Q \rangle = \frac{N}{2} |\Q \otimes \Q \rangle .
\eea
Therefore, the parity operator state and the identity state can be regarded as the $N$ up spin state and $N$ down spin state, respectively. Generally, the eigenstates of the $L_z$ operator are classified as follows :
\bea
L_z & |\Q \chi_{i_{1}}...\Q \chi_{i_{j}} \otimes \Q \chi_{i_{1}}...\Q \chi_{i_{j}}\rangle \\ 
&=  \bigg(j - \frac{N}{2}\bigg) |\Q \chi_{i_{1}}...\Q \chi_{i_{j}} \otimes \Q \chi_{i_{1}}...\Q \chi_{i_{j}}\rangle.
\eea
In the same spirit, the input state is obtained by flipping a local spin at site $i$ from the fully polarized state $\ket{\Q \otimes \Q}$. In the picture of $N$ 1/2-spins, the input state is
\bea
\ket{\i} = -\ket{\uparrow_1...\uparrow_{i-1} \downarrow_i \uparrow_{i+1}...\uparrow_{N}}.
\eea
This state splits into two irreps of the total angular momentum $L$ as 
\bea
&|\chi_i \otimes \chi_i\rangle \\
&= \bigg(\frac{1}{N}\sum_j |\chi_j \otimes \chi_j\rangle\bigg) + \bigg(|\chi_i \otimes \chi_i\rangle - \frac{1}{N}\sum_j |\chi_j \otimes \chi_j\rangle\bigg)\\
 &= -\sqrt{\frac{1}{N}}\bigg|\frac{N}{2}, \frac{N}{2}-1 \bigg\rangle_z + \sqrt{\frac{N-1}{N}} \bigg|\frac{N}{2}-1, \frac{N}{2}-1 \bigg\rangle_{z,i}
\eea
in the $\ket{l,m}$ notation. 

\textit{Output State}--The output state is more non-trivial to interpret in the spin formalism because it requires us to insert a complete set of operators and resolve the identity, as shown in Fig.~\ref{Fig:Tensor}(a)
\bea \label{eq:ouputstate}
\ket{\o}= \frac{1}{2^N} \sum_{\mathcal{S}} \ket{\chi_j \mathcal{S}^{\dagger} \chi_j \otimes  \mathcal{S} }.
\eea
For an intuitive understanding of expressing this state in the SU(2) language, one can start with the state corresponding to the complete set of operators and its respective spin representation 
\bea 
\frac{1}{2^N}\sum \limits_{\S} \ket{ \S^\dagger  \otimes S} =\frac{1}{2^{N/2}} \prod\limits_i (\uparrow - \downarrow)_i =\ket{\leftarrow\cdots\leftarrow}.
\eea
This is just the lowest weight state polarised along the x-direction, which is a steady state with respect to the emergent Hamiltonian because of the discrete rotational square~(or cubic) symmetry. For the output state, the term in the summation gains a relative minus sign when the Majorana string $\S$ contains $\chi_j$. As a result, the $j$th spin is flipped from $\leftarrow$ to $\rightarrow$, and  we have
\bea
&\ket{ \o}
=-\ket{ \leftarrow_1...\leftarrow_{j-1}\rightarrow_j\leftarrow_{j+1}...\leftarrow_N }.
\eea
This output state splits into two irreps of SU(2) as well, similar to the input state, but in the $x$ direction
\bea
\ket{\o} &= \\
-&\sqrt{\frac{1}{N}}\bigg|\frac{N}{2}, 1-\frac{N}{2} \bigg\rangle_x + \sqrt{\frac{N-1}{N}} \bigg|\frac{N}{2}-1, 1-\frac{N}{2} \bigg\rangle_{x,j} .
\eea

\subsection{The OTOC}
The problem reduces to the time evolution of the input state followed by the computation of the overlap with the output state. Because the Hamiltonian only depends on the total angular momentum $\vec{L}$, the two irreps in the input state do not mix during the time evolution. As a result, the OTOC can be can be succinctly written as the contribution from the two irreps 
\bea\label{eq:OverallMajorana}
\F(\chi_i, \chi_j) = \F_{N/2}(t) +  \left(\frac{N \delta_{i j}-1}{N-1}\right)\F_{N/2-1}(t)
\eea
where
\bea\label{eq:Majorana_irrep}
&\F_{N/2}(t)= 
 \frac{2^{N/2}}{N} \, {}_x\langle l, 1-l| e^{\H t}|l, l-1 \rangle_z 
 \\
  &\F_{N/2-1}(t)= 2^{N/2}\bigg(\frac{N - 1}{N}\bigg)\,{}_{x}\langle l-1, 1-l| e^{\H t}|l-1, l-1 \rangle_{z} 
\eea
and $2l = N$. Therefore the operator dynamics have been exactly mapped to the imaginary time dynamics of SU(2) spins with angular momenta $L=N/2, N/2-1$.

\subsubsection{Analytical results for the non-interacting model $(q_{\syk}=2)$}
We first discuss the non-interacting case, i.e. $q_{\syk}=2$ in Eq.~\eqref{eq:MajoranaHamiltonian}. This special case is manifest in the effective Hamiltonian $\H$, since it only depends on the total angular momentum $L^2$ and the SU(2) algebra is promoted to an exact symmetry. Recall that the effective Hamiltonian for $q_{\syk}=2$ takes the form
\bea
\H_{q_{\syk}=2}=\frac{1}{2N}\left(-2\left(\begin{array}{c}
N \\
2
\end{array}\right) -3N +4L^2  \right). \nonumber
\eea
Where $L^2 = L^2_x+L^2_y+L^2_z$ is the total angular momentum squared, which is the Casimir of the SU(2) group, and hence the Super-Hamiltonian becomes a constant within a given irrep. For the irreps relevant to the computation of the OTOC,
\bea
\H_{l,q_{\syk}=2} = 0;\quad \H_{l-1,q_{\syk}=2} = -2.
\eea
Hence the OTOC for $q_{\syk}=2$ becomes:
\bea
\F_{q_{\syk}=2}(\chi_i (t), \chi_j) = \left(-1 + \frac{2}{N}\right) + 2\left(\frac{N \delta_{i j}-1}{N}\right)e^{-2t}.
\eea
It exponentially decays to the late-time value $(-1+2/N)$, which is nonzero and in contrast with the expectation from scrambling (Eq.~\eqref{eq:MajoranaScrmablingExpectation}), because the model for $q_\syk=2$ is not interacting.

\subsubsection{Scrambling dynamics for $q_{\syk}=4$}
For $q_{\syk}>2$, the model becomes interacting and the effective $\H$ is not only a function of the total angular momentum $L^2$ but depends on $L_x$, $L_y$ and $L_z$. An example (for $q_{\syk}=4$) is shown in Eq.~\eqref{eq:q=4MajoranaHamiltonian}. Therefore, to obtain the OTOC one needs to diagonalize $\H$ for the two irreps $L=N/2$ and $L=N/2-1$. The Hilbert space dimension scales linearly with $N$, permitting calculation from small to large but finite $N$.

\begin{figure}
\includegraphics[scale=0.5]{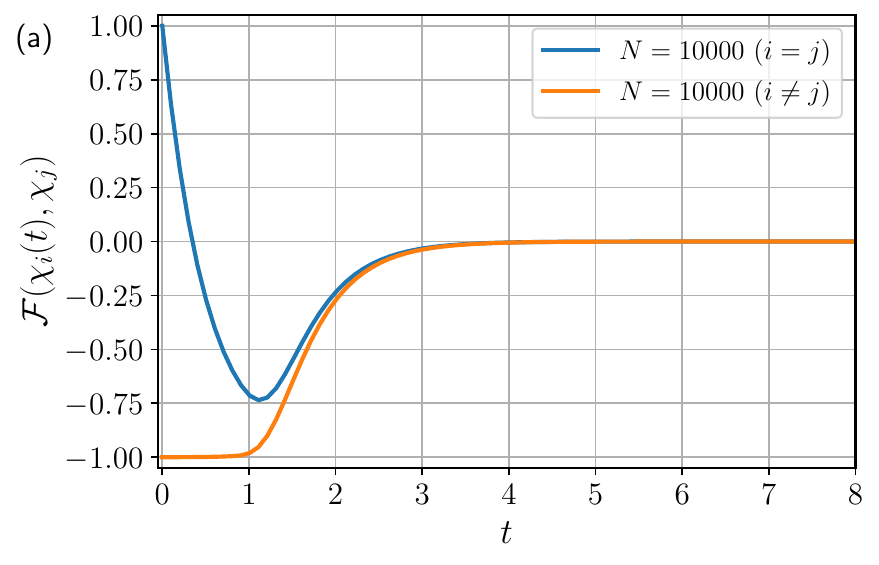}
\includegraphics[scale=0.5]{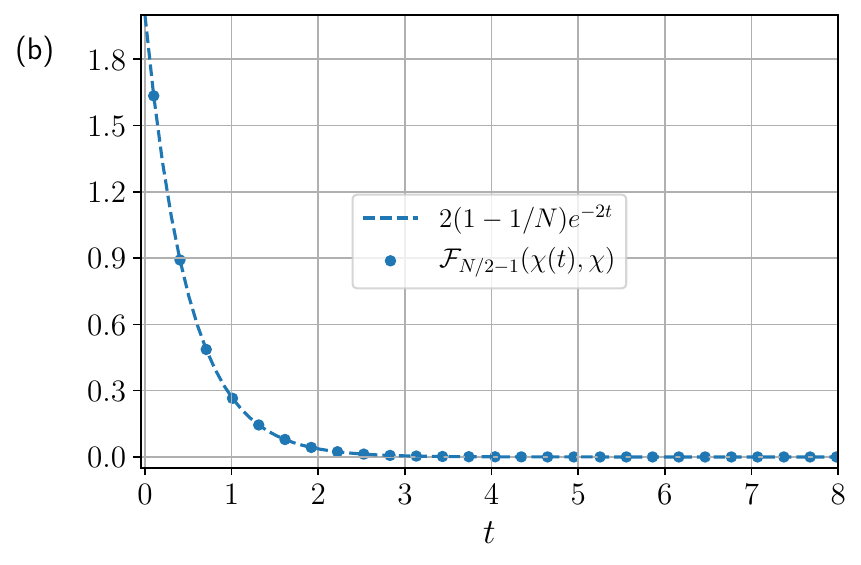}
\includegraphics[scale=0.5]{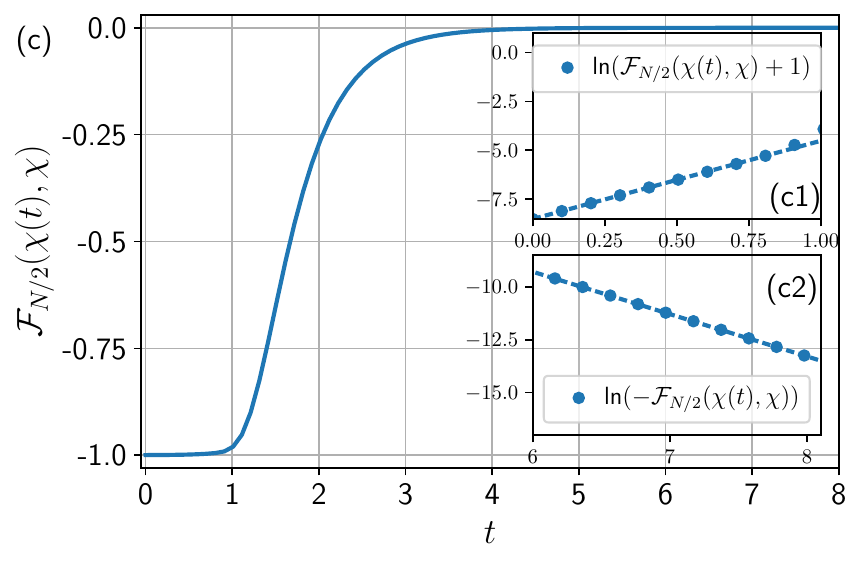}
\includegraphics[scale=0.5]{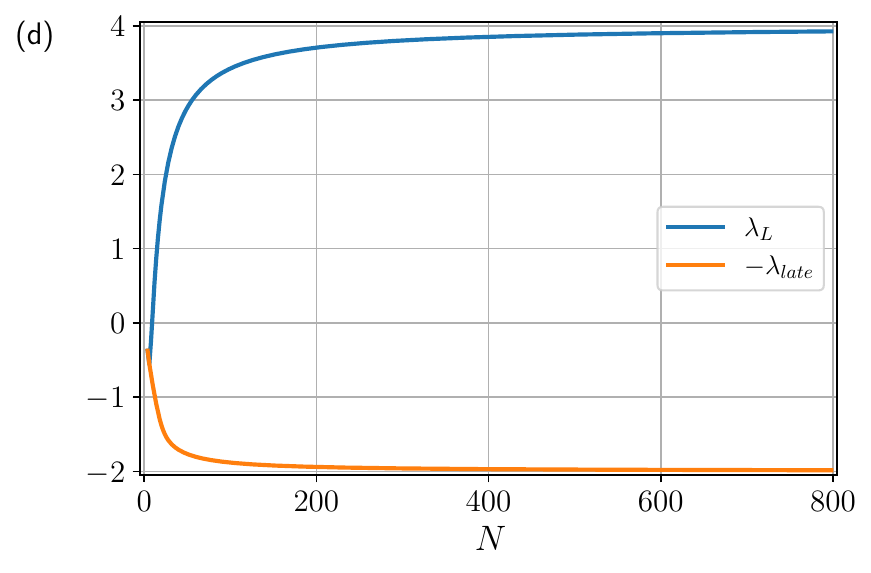}
\caption{(a) The OTOC $\F(\chi_{i}(t), \chi_j)$ for $i=j$ and $i \neq j$, computed according to Eq.~\eqref{eq:OverallMajorana}, for $N=10000$ fermions. (b) The contribution to the OTOC from the irrep $L=N/2-1$ exponentially decays with time. The curve is fitted against $e^{-2t}$. (c) The contribution to the OTOC from the irrep $L=N/2$, vs time for $N=10000$ fermions, displaying  scrambling behavior. The insets (c1) and (c2) show exponential Lyapunov growth in the early time regime and exponential decay in the late time regime, fitted against the lines $4t-$ln$(N)$ and $-2t+$ ln$(15)$ respectively. Thus the curve follows the ansatz in Eq.~\eqref{eq:Majoranaearlylate}. (d) The plot of $\lambda_L$ and $-\lambda_{late}$ as a function of $N$.}
\label{Fig:Majoranairrep}
\end{figure}

In Fig.~\ref{Fig:Majoranairrep}(a) we plot the overall OTOC for the cases $i=j$ and $i \neq j$, for $N=10000$, which agrees with the results in~\cite{Sunderhauf2019} up to an overall time scale due to the different convention of $J$ used in this work. The results for smaller $N$ are also benchmarked with exact diagonalization on the original model in Eq.~\eqref{eq:MajoranaHamiltonian}, averaged over 200 noise realizations in the appendix. This demonstrates the validity of our method.
The two curves start with different values and both relax to the late time value $0$, in contrast with $q_{\syk}=2$ and agreeing with the general expectation for scrambling dynamics.
The difference between $i=j$ and $i\neq j$ drastically decreases as time increases, the latter characterized by the Lyapunov growth in the early time. 

From Eq.~\eqref{eq:OverallMajorana}, the difference between $i=j$ and $i\neq j$ is proportional to $\F_{N/2-1}$, the contribution from the smaller irrep. In this irrep, one can show that the initial state $\ket{l-1,l-1}_z$ displays an exponential decay, as shown in Fig.~\ref{Fig:Majoranairrep}(b), which also fits the ansatz $\F_{N/2-1}(\chi (t), \chi) \simeq 2\left( 1- 1/N \right) e^{-2t}$.
 As a result, the difference in $\F(\chi_i(t), \chi_j)$ between $i=j$ and $i\neq j$ vanishes at a short time scale. On the other hand, the scrambling dynamics is contained in $\F_{N/2}(t)$, the contribution from the largest irrep which corresponds to the angular momentum $L=N/2$. This irrep shows early time Lyapunov growth and late time exponential decay and follows the ansatz
\bea \label{eq:Majoranaearlylate}
\F_{N/2}(t) \sim \left\{\begin{array}{ll}
-1 + \frac{2}{N} e^{4 t}  \quad & t\ll \frac{1}{4}\operatorname{ln} N \\
 e^{-2t} \quad & t\gg\frac{1}{4}\operatorname{ln} N
\end{array}\right.
\eea

As discussed in Eq.~\eqref{eq:Majorana_irrep}, 
\bea
\F_{N/2}(t)=\frac{2^{N/2}}{N}\langle N/2, 1-N/2 |e^{\H t}| N/2,N/2-1\rangle \nonumber
\eea
where $\H$ given in Eq.~\eqref{eq:q=4MajoranaHamiltonian} is a negative $N+1$ dimensional Hermitian matrix with all eigenvalues smaller or equal to 0. As a result, $\F_{N/2}(t)$ in general is a sum of $N+1$ exponentially decaying terms. Therefore, it is  natural to expect that $\F_{N/2}(t)$ decays exponentially in the late time, where the exponents are given by the largest nonzero eigenvalue of $\H$. On the other hand, the early-time Lyapunov exponential growth emerges from the interplay of sufficiently many exponentially decaying terms. For example, when $N=4$, $\H$ only has five eigenvalues, including four zeros modes ($\H$ is identically zero for $N=2,q_{\syk}=4$). As a result, $F_{N/2}(t)$ would display an exponential decay for all time scales, instead of exponential growth at early time. Therefore, the Lyapunov growth only occurs for sufficiently large $N$, the time scale for which increases logarithmically with $N$.

To investigate how the early time behavior changes from the exponential decay to exponential growth as $N$ increases, we use a more general ansatz $\F_{N/2}(t \sim 0) \sim a + be^{\lambda_L t}$
 for the early time regime of $\F_{N/2}$,
and study how $\lambda_L$ change as $N$ increases. By Taylor expanding the ansatz and the definition of $\F_{N/2}$ in Eq.~\eqref{eq:Majorana_irrep} and comparing coefficients, we get
\bea
\lambda_L =  \frac{{}_x\langle l, 1-l| \H^2|l, l-1 \rangle_z }{ {}_x\langle l, 1-l|\H|l, l-1 \rangle_z }.
\eea
This is a more accurate approach to extract $\lambda_L$ than curve fitting, especially for relatively small $N$ where the Lyapunov growth can be short-lived.

We plot the $\lambda_L$ obtained using this approach as a function $N$ in Fig.~\ref{Fig:Majoranairrep}(d), and we also include the largest non-zero eigenvalue $\lambda_{late}$ that controls the late time relaxation for comparison ($\F_{N/2}(t \rightarrow \infty) \sim c e^{-\lambda_{late} t}$). At $N=4$, because there is only decaying mode from $\H$, both $\lambda_L$  and $-\lambda_{late}$ start with the same value. As $N$ increases, $\lambda_L$ also increases, changes sign at $N=10$ and asymptotes to $4$, while $\lambda_{late}$ increases and asymptotes to $2$.

To understand why $\lambda_L=2\lambda_{late}$ in the large $N$ limit and provide an analytical understanding of $\F_{N/2}(t)$'s behavior for all time scales, in the next subsection, we will derive $\F_{N/2}(t)$ directly in the infinite $N$ limit (It should be noted that the ratio of $\lambda_L/\lambda_{late}$ is a function of $q_{\syk}$ and the behavior for general $q_{\syk}$ is provided in the appendix. The results obtained there agree with the analysis in~\cite{stanford2021subleading}). In particular, we will see precisely how the early time Lyapunov growth emerges from many exponentially decaying modes.

\subsection{Hydrodynamic equation for the Brownian SYK model} \label{subsec:Majorana-hydrodynamics}
We focus on the sector given by $L=N/2$, which contains the scrambling behavior. In this sector, the OTOC is
\bea
\F_{N/2}(t) = 2^{N/2}\braket{\o|\i(t)}=\sum\limits_m \psi_{\o}(m) \psi_{\i}(m,t)
\eea
where
\bea
\psi_{\o}(m)=-\frac{2m}{N}\binom{N}{N/2-m}^{1/2}.
\eea
The coefficient of the input state satisfies the imaginary time Schrodinger equation
\bea
\partial_t \psi_{\i}(m,t) =\sum\limits_{m'}\H_{m,m'}\psi_{\i}(m',t).
\eea
To proceed to derive $\F$ in the large $N$ limit, we exploit a similarity transformation to remove the non-uniform $N$ dependence of $\psi_{\o}(m)$ such that,
\bea
\psi_{\o}(m)\rightarrow \tilde{\psi}_{\o}(m)=-\frac{2m}{N}.
\eea
The OTOC therefore becomes $\F_{N/2}(t)=-2/N \sum\limits_m m \tilde\psi_{\i}(m,t)$. After the similarity transformation, the ladder operator takes the following simple form
\bea
(\tilde L_+)_{m+1,m} = \frac{N}{2}-m, \quad (\tilde L_-)_{m-1,m} = \frac{N}{2}+m
\eea
while $\tilde L_z$ is the same as $L_z$. From these angular momentum operators, The transformed effective Hamiltonian $\tilde \H$ can be constructed from Eq.~\eqref{eq:q=4MajoranaHamiltonian} and becomes non-hermitian. 

The transformed input state now obeys,
\bea
\partial_t \tilde{\psi}_{\i}(m,t)=\sum_{m'}\tilde \H_{mm'}\tilde{\psi}_{\i}(m',t).
\label{eq:Majorana_master}
\eea
Remarkably, $\tilde \H$ satisfies the property that $\sum_m \tilde \H_{m m'} =0$. This is because one of the eigenstates of $\H$, the fully polarized state in the $x$ direction becomes uniform after the transformation, and is a left eigenvector of $\tilde \H$. As a result, Eq.~\eqref{eq:Majorana_master} is a master equation and $\tilde \psi_{\i}(m,t)$ has the interpretation of a probability because $\sum_m \tilde \psi_{\i}(m,t)$ is conserved for all time.

Following this, we take the large $N$ limit of the master equation by using the continuous variable $\xi = 2m/N$. To the leading order of $1/N$, the master equation becomes
\bea
\label{eq:Majorana_infiniteN}
\partial_t \tilde \psi_{\i}(\xi, t)= -2 \partial_\xi \left(\xi (\xi^2-1)\tilde \psi_{\i}(\xi,t) \right)
\eea
which can be solved analytically. In particular, if $\tilde\psi_{\i}$ starts with a Delta distribution , it remains a Delta distribution $\tilde\psi_{\i}(\xi,t)=\delta(\xi-\xi(t))$ for all time. As a result, $\F_{N/2}(t)=-\xi(t)$.
The peak value $\xi(t)$ obeys a logistic differential equation
\bea \label{eq:MajoranaHydrodynamics}
\xi'(t) = 2 \xi(t) \big(\xi^2(t)-1).
\eea
There are 3 static solutions; $\xi=\pm 1$ are the unstable solutions corresponding to the states $\ket{I\otimes I}$ and $\ket{\Q \otimes \Q}$, while $\xi=0$ is the stable solution corresponding to complete scrambling. When $\xi(0)=1-2\delta$, i.e., slightly deviates away from the unstable solution, we obtain the OTOC $\F$ as
\bea
\label{eq:Majorana_F_infiniteN}
\F_{N/2}(t) =-\xi(t)=-\frac{1}{\sqrt{1+4 e^{4t} \delta}}.
\eea
It demonstrates the characteristic early-time Lyapunov exponential growth and late-time exponential relaxation,
\bea
\F_{N/2}(t) \sim \left\{\begin{array}{ll}
 -1 + 2 e^{4 t} \delta \quad & t\ll-\frac{1}{4}\operatorname{ln} \delta \\
 e^{-2t} \quad & t\gg-\frac{1}{4}\operatorname{ln} \delta
\end{array}\right.
\eea
Which is in agreement with the numerical results in Sec.~\ref{sec:MajoranaOTOC} and \cite{Sunderhauf2019}.
With the analytical expression in hand, one can expand $\F_{N/2}(t)$ as 
\bea \label{eq:Majoranaghostmodes}
\F_{N/2}(t)=\sum\limits_{n=0}^\infty \frac{(-1)^n(2n)!}{2^{4n+1}(n!)^2\delta^{n+1/2}}\exp \left(-2(2n+1)t\right).
\eea
This demonstrates the emergence of the Lyapunov growth from many decaying modes with alternating sign.

Several remarks are in order. First, $\tilde\psi_{\i}(\xi,t)$ remains a Delta probability distribution over time only in the infinite $N$ limit. One can include the $1/N$ term in Eq.~\eqref{eq:Majorana_infiniteN} when expanding the master equation and will obtain a Fokker-Planck equation. The $1/N$ term would broaden the distribution, which is a result of quantum fluctuations. Such terms can lead to observable effects, such as wavefront broadening in higher dimensions~\cite{Nahum_2018, von_Keyserlingk_2018, Xu_2019}. Second, the analytical form of $\F$ in Eq.~\eqref{eq:Majorana_F_infiniteN} does not have to precisely match the numerics in Sec.~\ref{sec:MajoranaOTOC}. This is because in the numerics, we always use the simplest initial operator state corresponding to $\delta=1/N$ while the analytical form is valid when $\delta$ is kept a constant as one approaches the infinite $N$ limit. 
Finally, the master equation we obtained in Eq.~\eqref{eq:Majorana_master} precisely matches that obtained by solving the model using the stochastic method~\cite{Jian_2021}. Thus our approach, by taking advantage of the emergent symmetry structure after the random disorder average, reveals that the stochastic approach and the Hamiltonian approach used to solve the model are simply connected by a similarity transformation. The logistic equation we obtain in Eq.~\eqref{eq:MajoranaHydrodynamics} is slightly different from that obtained in the spin Brownian model~\cite{Xu_2019,Zhou_2019}, where there are only two steady solutions because of the absence of the fermionic parity operator. In the spin Brownian model, $\lambda_L$ and $\lambda_{late}$ are the same. In the Majorana case, the relation between $\lambda_L$ and $\lambda_{late}$ depends on $q_{\syk}$, as discussed in the appendix. 

Our approach, exactly mapping the operator dynamics to the imaginary-time dynamics of a spin, is readily generalized to the complex Brownian SYK model with charge conservation, which we will present next.

\section{The complex Brownian SYK model} \label{sec:ComplexBrownianSYK}
We start with the Brownian version of the  complex SYK model~\cite{Sachdev_2015} with $N$ complex fermionic pairs $(\chi, \chi^{\dagger})$ and ($q_{\syk}=4$)-body interactions with complex time-dependent couplings $J$
\bea \label{eq:ComplexBrownianHamiltonian}
H(t) =  \sum_{j_{1},j_{2},k_{1},k_{2}} J_{j_{1},j_{2}, k_{1},k_{2}}(t)  \chi_{j_{1}}^{\dagger} \chi_{j_{2}}^{\dagger} \chi_{k_{1}} \chi_{k_{2}} + h.c.
\eea
These fermions satisfy the usual anti-commutation relations 
\bea
\left\{\chi_{j},\chi_{k}\right\}=0 \quad \left\{\chi_{j}^{\dagger},\chi_{k}\right\}= \delta_{jk}.
\eea
The couplings $J$ are sourced independently from a Gaussian distribution with zero mean and variance
\bea
\overline{J^{\text{}}_{j_{1} ,j_{2}, k_{1}, k_{2}}(t) J^*_{j_{1}^{\prime},j_{2}^{\prime},k_{1}^{\prime},k_{2}^{\prime}}\left(t^{\prime}\right)}=\, \delta^{j_{1} }_{j_{1}^{\prime}} \delta^{j_2}_{j_{2}^{\prime}}  \delta^{k_{1} }_{k_{1}^{\prime}} \delta^{k_2}_{k_{2}^{\prime}} \, \delta\left(t-t^{\prime}\right) \frac{1}{2 N^{3}}
\eea
This relation can be generalized for other $q_{\syk}$ as well, although in this work we primarily focus on $q_{\syk}=4$. In the main text, as an example, we will primarily be focusing on computing the OTOC $\F(\chi_i(t), \chi_j^\dagger)$ and its charge resolved version $\F^m(\chi_i(t),\chi_j^\dagger)$
\bea
\F(\chi_i(t), \chi_j^\dagger) = &\sum\limits_m \frac{\tr{(P_m)}}{2^N} \F^{m}(\chi_i(t), \chi_j^\dagger)\\ 
\F^m(\chi_i(t), \chi_j^\dagger) =& \frac{1}{\tr (P_m)}\tr(P_m \chi_i^\dagger(t) \chi_j \chi_i(t) \chi_j^{\dagger}).\\
\eea
We will quote the result for the other OTOCs but leave the details of the calculation in the appendix. 

One can rewrite the OTOC as shown in Sec.~\ref{sec:OperatorstateOTOC}.
This gives us the idea to work with four copies of the Hilbert space, occupied by four ``replica" fermions labelled by the indices $(a,b,c,d)$. This larger Hilbert space is spanned by the basis vectors 
\bea \label{eq:particle-hole}
\chi_{j}^{a}:=\chi_{j} \otimes I \otimes I \otimes I \qquad \chi_{j}^{b}:=I \otimes \chi^{\intercal}_{j} \otimes I \otimes I \\
\chi_{j}^{c}:=I \otimes I \otimes \chi_{j} \otimes I \qquad \chi_{j}^{d}:=I \otimes I \otimes I \otimes \chi^{\intercal}_{j}
\eea
and their Hermitian conjugates $({\chi_{j}^{a}}^{\dagger}, {\chi_{j}^{b}}^{\dagger}, {\chi_{j}^{c}}^{\dagger}, {\chi_{j}^{d}}^{\dagger})$. Here we use the notation $\chi_{j}^{\intercal} = {\chi_{j}^{*}}^{\dagger}$, which implies that we have performed a particle-hole transformation on copies $b,d$. This is just a convention which makes things easier when defining the operators of the $\text{SU}(4)$ algebra in the next section. The replica fermions with different indices commute with each other. 
This `mixed' species of particles populating the Hilbert space is rather inconvenient to work with, and we convert them to fermions that anti-commute with each other by using the parity operator similar to the Majorana case
\bea
\mathcal{Q}^{\alpha}= \prod_{k=1}^{N} \exp(i\pi n^\alpha_k), \quad \alpha=a, b, c, d.
\eea
One can check that this operator satisfies the following relations :
$\left\{\mathcal{Q}^{\alpha},\chi_{k}^{\alpha}\right\}= 0, \, \left\{\mathcal{Q}^{\alpha},{\chi_{k}^{\alpha}}^{\dagger}\right\}= 0, \, \left(\mathcal{Q}^{\alpha}\right)^2 = 1$.
Following this, we define 
\begin{equation}
\begin{array}{ll}
\psi_{j}^{a}= \mathcal{Q}^{a} \chi_{j}^{a}, & \psi_{j}^{b}=\mathcal{Q}^{a} \chi_{j}^{b} \\
\psi_{j}^{c}= \mathcal{Q}^{a} \mathcal{Q}^{b} \mathcal{Q}^{c} \chi_{j}^{c}, & \psi_{j}^{d}=\mathcal{Q}^{a} \mathcal{Q}^{b} \mathcal{Q}^{c} \chi_{j}^{d}
\end{array}
\end{equation}
These operators are purely fermionic, i.e.\ they anti-commute with themselves and fermions from other replicas, satisfying :
$\left\{{\psi_{j}^{\alpha}}^{\dagger}, \psi_{k}^{\beta}\right\}= \delta_{\alpha, \beta} \delta_{j, k} , \, \left\{\psi_{j}^{\alpha}, \psi_{k}^{\beta}\right\}=0$.
One can also confirm that 
\bea
{\psi_{j_{1}}^{\alpha}}^{\dagger} \ldots {\psi_{j_{q}}^{\alpha}}^{\dagger} \psi_{k_{1}}^{\alpha} \ldots \psi_{k_{q}}^{\alpha} = {\chi_{j_{1}}^{\alpha}}^{\dagger} \ldots {\chi_{j_{q}}^{\alpha}}^{\dagger} \chi_{k_{1}}^{\alpha} \ldots \chi_{k_{q}}^{\alpha} 
\eea
and hence we can replace the operators in the original Hamiltonian with these new purely fermionic operators.

We can now take the disorder average of each time step in the discretized time-evoluton independently using the approach outlined in Sec.~\ref{sec:preparation} and arrive at the averaged time evolution operator, as well as the effective Hamiltonian $\overline\U(t)=\exp( \H t)$.
Although the full expression of $\H$ is quite complicated (which is given in the Appendix), it takes a general simple form that results from the general construction as well as the charge conservation on each replica $a\sim d$. 

We introduce the bilinear operators
\bea
S^{\alpha\beta}=\sum\limits_i \psi_i^{\alpha\dagger} \psi_i^{\beta}.
\label{eq:complex_bilinear}
\eea
The total charges  $\sum_i \chi_i^\dagger \chi_i$ on each replica are given by $(S^{aa}, N-S^{bb}, S^{cc}, N- S^{dd})$, which measures the charge profile of an operator state.  Note the particle-hole transformation on replicas $b$ and $d$.
In terms of these bilinear operators, the Hamiltonian takes the following general form:
 \bea \label{eq:Hamiltonianstructure}
 \H = \H(S^{\alpha\beta}S^{\beta\alpha}, S^{\alpha\alpha}), \quad \alpha, \beta \in a, b, c, d.
 \eea
Which results from the independent charge conservation on the four replicas. One can explicitly verify that $\H$ commutes with all four operators $S^{\alpha\alpha}$ in this functional form.  
\subsection{Emergence of the SU(4) $\otimes$ U(1) algebra}
Now we analyze the full symmetry structure of $\H$.
The dimension of the total Hilbert space of the effective Hamiltonian $\H$ is $16^N$ since it contains 4 copies of the original system. The dimension scales exponentially with $N$ even with the charge conservation. In this section, we exploit the additional symmetry structure of $\H$ in Eq.~\eqref{eq:Hamiltonianstructure} to further reduce the dimension. 

From the bilinear operator introduced in Eq.~\eqref{eq:complex_bilinear}, we define $\tilde{S}^{\alpha\beta}= S^{\alpha \beta}-\frac{1}{4} \delta^{\alpha \beta} S^{\sigma \sigma}$.
The operators satisfy the commutation relations 
\begin{equation}
[\tilde{S}^{\alpha \beta} , \tilde{S}^{\gamma \sigma}] = \delta^{\beta \gamma} \tilde{S}^{\alpha \sigma} - \delta^{\alpha \sigma} \tilde{S}^{\gamma \beta}.
\end{equation}
There are 15 independent operators because $\sum_\alpha \tilde{S}^{\alpha\alpha}=0$ and they are generators of the SU(4) algebra.  Along with this, the operator $Q=\sum_{\alpha} S^{\alpha \alpha}$ commutes with all the operators of the SU(4) algebra and defines the U(1) charge to make the overall algebra SU(4)$\otimes$U(1).

Since the emergent Hamiltonian commutes with the four operators $S^{\alpha \alpha}$, in addition to commuting with the total charge operator, it also commutes with the three operators that form the Cartan-subalgebra of the SU(4) algebra. Each subsector of $\H$ can therefore be labeled by the irrep of the SU(4) algebra, the three weights in the weight diagram of the irrep,  and the total U(1) charge. This fully resolves the symmetry structure of $\H$. It is well known that there are multiple states corresponding to the same weights in SU($n$) irreps for $n>2$. In our case, the dimensions of the irreps of the SU(4) algebra scales as $\mathcal{O}(N^4)$ and fixing the weights decreases the scaling to $\mathcal{O}(N)$. In other words, the largest dimension of subsectors of $\H$ scales linearly with $N$, drastically reducing the computational cost.

Let us compare the structure of $\H$ between the Majorana case and the complex case. In the Majorana case, $\H$ can be written as a function of SU(2) generators, and the dimension of largest subsector scales linearly with $N$. Here in the complex case, $\H$ is a function of SU(4) generators. In addition, it commutes with the weight of the SU(4) algebra, and as a result, the largest Hilbert space dimension also scales linearly with $N$.

The strategy for calculating the OTOC in the complex case is similar to the Majorana case. We need to decompose the input and the output state into different subsectors of $\H$, let the different components of the input state evolve in the imaginary time given by $\H$, and then take the overlap with the different components of the output state. Thus the OTOC contains the contribution from the different irreps of SU(4) as well as different charge sectors~(weights). This method therefore naturally provides us the OTOC in the charge resolved manner.

Before proceeding to discuss the decomposition of the initial and the final states into irreps of SU(4), we note that the non-interacting nature of the quadratic model $(q_{\syk}=2)$ manifests itself in the effective Hamiltonian, similar to the Majorana case.  A straightforward calculation reveals that the emergent Hamiltonian $\H_{q_{\syk}=2}$ takes the following simple form,
\bea \label{eq:q=2Hamiltonian}
\H_{q_{\syk}=2} = \frac{1}{N} (C_2 +\frac{Q^2}{8}-2N -\frac{QN}{2})
\eea
where $Q$ is the total charge and $C_2$ is the quadratic Casimir of SU(4)
\bea
C_2 = \frac{1}{2} \sum_{\alpha, \beta} S^{\alpha \beta} S^{\beta \alpha} - \frac{Q^2}{8}.
\eea
In this case, $\H_{q_{\syk}=2}$ commutes with all $S^{\alpha\beta}$ and becomes a constant within a given irrep, i.e. the algebra structure is enhanced to an exact symmetry. These properties make it possible to solve the OTOC analytically for $q_{\syk}=2$, which will be presented in Sec.~\ref{sec:OTOC}. Other than $q_{\syk}=2$, the emergent Hamiltonian $\H_{q_{\syk}}$ only conserves the weights within the SU(4) irrep. This remarkable difference between $q_{\syk}=2$ and $q_{\syk}\neq 2$ leads to the distinct operator dynamics.

The full emergent Hamiltonian for $q_{\syk}=4$ is present in the appendix. The general structure of the Hamiltonian for all $q_{\syk}$ takes the form in Eq.~\eqref{eq:Hamiltonianstructure},
which manifestly preserves the charges for each of the four replicas. This is a powerful property that will be exploited to `chop up' the input state in the OTOC into states with different weights within the SU(4) irreps so that each piece will only have dynamics within its own fixed-weight~(charge) subspace.

\subsection{Classification of states in the SU(4) algebra}
The emergent Hamiltonian acts on four replicas of the original Hilbert space. In other words, the input states correspond to two copies of operators, in total $16^N$ independent states. 
In the last section, we demonstrated that the emergent Hamiltonian is closed within the irreps of the SU(4) algebra. In this section, we organize the input states and the output states into various irreps of SU(4), from which the operator dynamics and OTOC can be efficiently computed exactly using Hilbert space of size $\mathcal{O}(N)$. 

\subsubsection{$N=1$ representation}
\begin{figure}
\includegraphics[width=1\columnwidth]{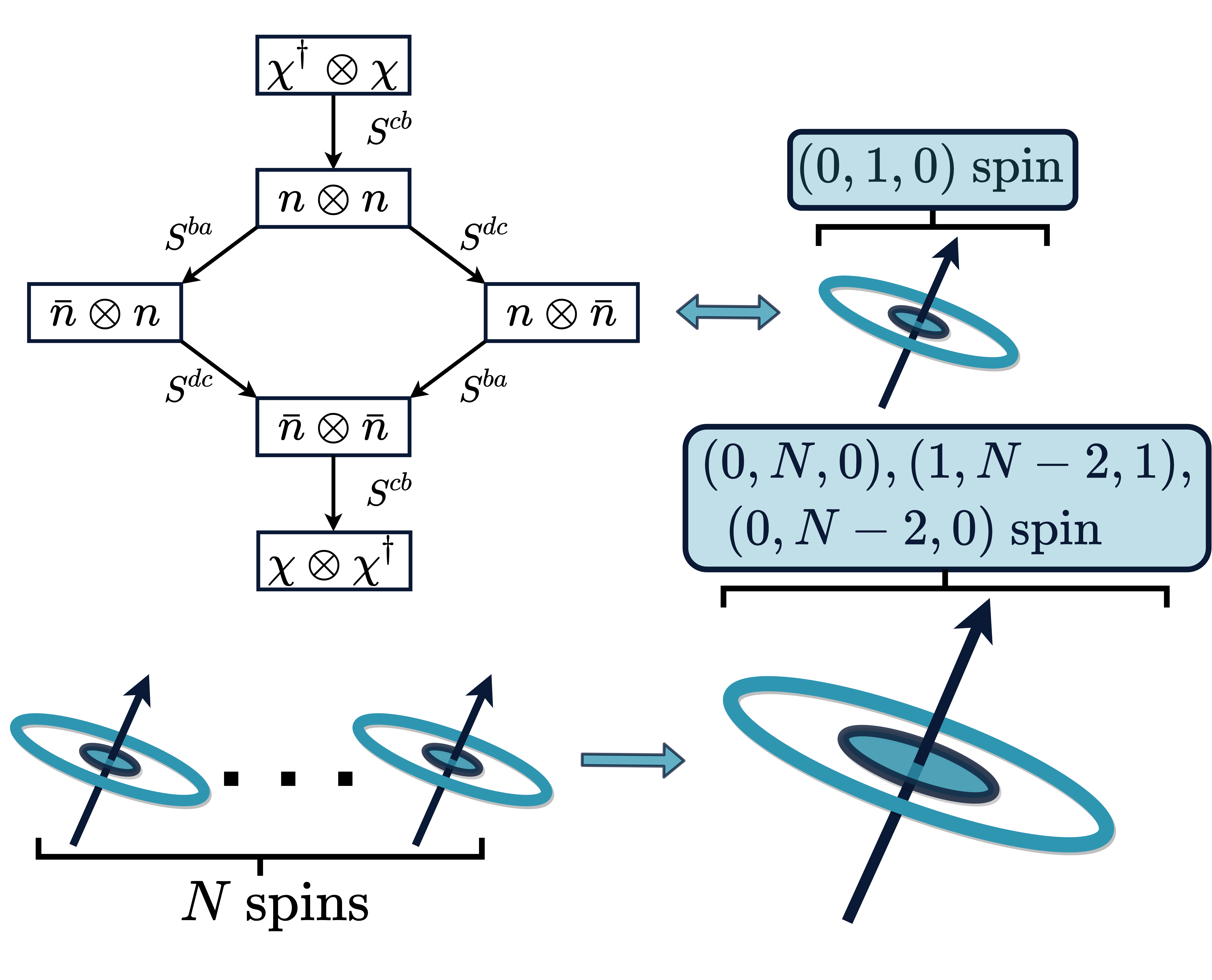}
\caption{The six onsite operator-states for the complex Brownian SYK model form the SU(4) irrep $(0, 1, 0)$. Here $\bar{n} = 1- n$ and the construction of the generators is provided in Eq.~\eqref{eq:complex_bilinear}. The operator dynamics of $N$ complex fermions with charge conservation is mapped to the imaginary time dynamics of $N$ SU(4) spins in the $(0, 1, 0)$ irrep, which compose to form $(0,N,0)$, $(1,N-2,1)$ and $(0,N-2,0)$ irreps of the global SU(4) algebra as shown in Eq.~\eqref{eq:ComplexYoungComposition}. In addition, the weights of the global SU(4) spin are also conserved because of the charge conservation. The largest Hilbert space dimension for a fixed weight sector within a given global SU(4) irrep scales linearly with $N$. }
\label{fig:6su4}
\end{figure}

We first consider the operators acting on the same fermionic index, i.e., $N=1$. There are four independent operators per site, $\chi^\dagger$, $\chi$, $n=\chi^\dagger \chi$ and $\bar{n}=\chi \chi^\dagger$. As discussed in Sec.~\ref{sec:operator}, this basis fully utilizes the U(1) symmetry of the complex Brownian model. The identity operator $I$ is $n+\bar{n}$. Combinations from the 4 operators on each copy leads to 16 independent operator states. The 16 operator states can be grouped into irreps of SU(4) in the following way,
\bea
\textbf{16} = \textbf{1} \oplus \textbf{4} \oplus \textbf{6} \oplus \bar{\textbf{4}} \oplus \textbf{1}.
\eea
 The initial operator state of interest, $\ket{\chi_i^\dagger I \otimes \chi_i I}$, contains the operator state $\chi^\dagger \otimes \chi $ and $I\otimes I$ acting on different fermions, which belong to the six dimensional irrep $(0, 1, 0)$.  In other words, the initial states are only made using states per fermion within the $(0, 1 , 0)$ irrep. The operator states in the $(0, 1, 0)$ irrep and their transformation under SU(4) generators are shown in Fig.~\ref{fig:6su4}. 
Since the emergent Hamiltonian only contains the generators of SU(4), the other single fermion irreps do not contribute to the dynamics. To this end, the operator dynamics is mapped to the dynamics of $N$ six-dimensional SU(4) spins. This is in the same spirit with the operator dynamics of the Majorana model mapping to $N$ spin-$1/2$ SU(2) spins. Crucially, unlike the Majorana case, the operator on each copy does not have to match during the dynamics, and the configuration $n\otimes \bar{n}$, for example, can be generated in the dynamics. This leads to the rich charge-dependent operator dynamics of the complex model. 

\subsubsection{Representation for general $N$}
Taking into account the SU(4) irreps of operator states at each site reduces the Hilbert space dimension from $16^N$ to $6^N$, a significant reduction which however is still difficult to work with. Since the Hamiltonian only depends on the SU(4) $\otimes$ U(1) generators, the dynamics must be closed within the SU(4) irreps of $N$ total fermions. Hence the next step is to build SU(4) irreps of $N$ fermions from $N$ copies of the irrep $(0, 1, 0)$ corresponding to a single fermion, based on the composition rule of SU(4) irreps,
\begin{equation}
    \underbrace{(0,1,0) \otimes...\otimes (0,1,0)}_{N \text{ times}} = (0,N,0) \oplus \underbrace{(1,N-2,1)}_{N-1 \text{ copies}} \oplus ...,
    \label{eq:su4comp_general}
\end{equation}
and then decompose the initial operators into the various irreps, which evolve \textit{independently} under the emergent Hamiltonian. The dimension of each irrep scales polynomially with $N$, drastically reduced from $6^N$. The dimension of the Hilbert space can be further reduced to linear scaling because the emergent Hamiltonian also conserves the weight of the states within the irrep. 

This is a well-defined but tedious procedure for a general initial operator state $\ket{W^\dagger \otimes W}$ since many SU(4) irreps can appear in the composition. However, for simple initial operator states of interest, this procedure is significantly simplified, and at most three irreps appear in the composition. The initial operator states we consider are local operators of the form $\ket{W^\dagger_1 I \otimes W_1 I}$, made from the identity operators except for one fermion, which is located at site $1$ without loss of generality. We notice that the operator state $\ket{I\otimes I}$ on the remaining $N-1$ fermions belongs to a single SU(4) irrep $(0, N-1, 0)$. Therefore the composition in Eq.~\eqref{eq:su4comp_general} is reduced to the composition of two irreps,
\bea \label{eq:ComplexYoungComposition}
(0, 1, 0)&\otimes  (0, N-1, 0) \\
&= (0, N, 0)\oplus (1, N-2,1) \oplus (0, N-2, 0).
\eea

Now we present the explicit decomposition of the initial operators into the three irreps, which will then be used to calculate OTOC later. We will assume that $N$ is even throughout this work for simplicity. The notation we use will assume that unless an index is specified, it is implied that the operator has support on all sites which are not populated by other operators in the given operator string, i.e.: \bea
O \equiv \Pi_{i=1}^{N} O_i ;\quad \tilde{O}_j O \equiv \tilde{O}_j \Pi_{i \neq j} O_i.
\eea
Our strategy is to build the states from the highest weight state within each of the three irreps, just like building the state from the fully polarized state using $L_{\pm}$ in the Majorana case. The highest weight state is defined so that it is annihilated by $S^{ab}$, $S^{bc}$ and $S^{cd}$, using the following convention
\bea \label{eq:highestweight}
&\ket{W_{(0,N,0)}} = \frac{1}{N} \ket{ \chi^{\dagger}  \otimes \chi}\\
&\ket{W_{(1,N-2,1)}} = - \ket{n_1 \chi^{\dagger} \otimes n_1 \chi} + \frac{1}{N} \sum_{i=1}^{N} \ket{ n_i \chi^{\dagger} \otimes n_i \chi}.
\eea

\textit{Input state} -- We will first demonstrate the procedure to build all the states required to compute the OTOC, using the OTOC $\F(\chi_i(t),\chi_j^{\dagger})$ as an example. We start with the operator state $\ket{\chi_1^\dagger I \otimes \chi_1 I} $, which corresponds to the input state in the OTOC. Using the $\text{SU}(2) \otimes \text{SU}(2)$ sub-algebra of SU(4) generated by $S^{ba}$ and $S^{dc}$, one can show that,
\bea \label{eq:inputweights}
\ket{\chi_1^\dagger I \otimes \chi_1 I} =  \sum\limits_{k,l=0}^{N-1} (-1)^{k+l} \frac{(S^{ba})^k}{k!} \frac{ (S^{dc})^l}{l!} \ket{\chi_1^\dagger n \otimes \chi_1  n} .
\eea
Each term in the summation has the fixed charge profile $(N-k, N-1-k, N-1-l, N-l)$. The operator state $\ket{\chi_1^\dagger n \otimes \chi_1 n}$ can be built from the highest weight states using the operator $S^{cb}$ from the following simple relation:
\bea
|\chi^\dagger_1 n  \otimes &\chi_1 n\rangle = (-1)^{N/2+1} \\ 
 &\bigg(\frac{(S^{cb})^{N-1}}{(N-1)!}\ket{W_{(0,N,0)}}  + \frac{ (S^{cb})^{N-2}}{(N-2)!}\ket{W_{(1,N-2,1)}}\bigg).
 \eea
Thus the initial operator state $\ket{\chi^\dagger_1 I \otimes \chi_1 I}$ is completely decomposed into the two irreps and built from the highest weight states. This state does not have a component in the $(0, N-2, 0)$ irrep. Each term in the summation can therefore be restricted to one irrep and has fixed charges, thus evolving independently under the emergent Hamiltonian. The dimension of each subspace is given in Eq.~\eqref{eq:weightmultiplicity}, which scales linearly with $N$. 
The other initial state of interest $\ket{n_1 I\otimes n_1 I}$ can be built from the highest weight state in a similar fashion, but has components in all three irreps. The details can be found in the appendix. 

\textit{Output state} -- Now we discuss building the output state using SU(4) irreps. Similar to the case of the Majorana model, building the output state requires inserting the resolution of the identity operator to ensure a proper overlap with the input state. We again consider the operator state $\ket{\chi^\dagger_1 I \otimes \chi_1 I}$. After inserting the resolution of the identity, the state becomes
\bea
\frac{1}{4^N} \sum\limits_{\S} \ket{\chi_1 \S^\dagger\chi^\dagger_1 \otimes  \S} =  \sum\limits_{\S^{c}} \frac{P(S_c)}{2^{2N-1}} \ket{\bar{n}_1 {\S^c}^\dagger \otimes n_1 \S^c}
\eea
where $P(S^c)$ is the parity of the string $S^c$,  equaling $1$ if the total number of $\chi$ and $ \chi^{\dagger}$ in the string $S^c$ is even, and $-1$ otherwise. The string $S^c$ represents all complex fermionic strings over the fermions except site 1. Using the SU(2)$\otimes$SU(2) sub-algebra of SU(4) generated by $S^{cb}$ and $S^{da}$, one can show that
\bea \label{eq:outputweights}
&\sum\limits_{\S^c} \frac{P(S^c)}{2^{N-1}} \ket{\bar{n}_1 {\S^c}^\dagger \otimes n_1 \S^c}  \\
&=\sum\limits_{k,l=0}^{N-1} (-1)^{N/2+k}  \frac{(S^{cb})^k}{k!} \frac{(S^{da})^l}{l!}(- S^{ba}) \ket{n_1 \chi^\dagger \otimes n_1 \chi}.
\eea
where the operator state $\ket{n_1\chi^\dagger \otimes n_1 \chi}$ is built from the highest weight states as
\bea
\ket{n_1 \chi^\dagger \otimes n_1 \chi} = S^{cb} \ket{W_{(0,N,0)}} - \ket{W_{(1,N-2,1)}}.
\eea
Thus this completes the prescription of constructing the output state from the highest weight states.
\subsection{The OTOC} \label{sec:OTOC}
As is evident from the sum in Eqs.~\eqref{eq:inputweights} and \eqref{eq:outputweights}, both the input and output states are composed of $N^2$ different weights. Since the Hamiltonian preserves weights and states with different weights are orthogonal, it is important to check how many of the weights are shared between the input and output states. One can deduce that there are $N-1$ such weights, and they can be labelled by a single integer $m$ that also labels the charge in each corresponding sector, and ranges between $\left[ 1,N-1 \right]$
\bea \label{eq:chargedstates}
&|\i_m\rangle =  -\frac{1}{2^N} \frac{(S^{dc})^{N-m-1}(S^{ba})^{N-m}}{(N-m-1)! (N-m)!} \ket{\chi_1^\dagger n \otimes \chi_1  n}  \\ 
&|\o_m\rangle =  \frac{(-1)^{N/2-m-1}}{2^N} \frac{(S^{da})^{N-m-1}(S^{cb})^{m-1}}{(N-m-1)! (m-1)!} \ket{\bar{n}_1 \chi^\dagger \otimes n_1 \chi} \\
\eea
 This is exactly equivalent to restricting the OTOC to a specific charge sector, as shown in Sec.~\ref{subsec:complexotoc}. The states $\ket{\chi_1^\dagger n \otimes \chi_1  n}$ and $\ket{\bar{n}_1 \chi^\dagger \otimes n_1 \chi}$ have been built in the previous section. One can check that the states defined above have equal weight for the corresponding $m$, hence the problem reduces to dynamics within $N-1$ subspaces, each of maximum size $\mathcal{O}(N)$. Following this, the OTOC is governed by the equation 
\bea \label{eq:OverallOTOC} 
&\F(\chi_i(t),\chi_j^{\dagger}) = \F_{(0,N,0)}(t) + \bigg( \frac{N \delta_{ij}-1}{N-1} \bigg) \F_{(1,N-2,1)}( t )\\
&\F_{(0,N,0)}(t) = 2^N \sum_{m=1}^{N-1} \langle \o_{m} |e^{\mathbb{H}_m t}|\i_{m}\rangle_{(0,N,0)} \\ & \F_{(1,N-2,1)}(t) = 2^N \sum_{m=1}^{N-1} \langle \o_{m} |e^{\mathbb{H}_{m}t}|\i_{m}\rangle_{(1,N-2,1)}.
\eea
From here on out, we will use the notation $\F_{\text{irrep}}(t)$ to refer to the OTOC $\F(\chi_i(t),\chi_j^{\dagger})$ and the notation $\F^m_{\text{irrep}}(t)$ to discuss its charge resolved version  restricted to the particular irrep, unless other operators are explicitly specified. As we will see in the upcoming sections, the scrambling dynamics are present solely in the contribution from the symmetric $(0,N,0)$ irrep (for $q_{\text{syk}} \geq 4$) and all other irrep contributions are marked by exponential decays at all times. 

\subsubsection{Analytical results for the non-interacting model $(q_{\syk}=2)$} \label{subsec:q=2}

One can compute the OTOC for the free case ($q_{\syk}=2$) using the emergent Hamiltonian which in the case of the non-interacting model depends only on the quadratic Casimir $C_2$ and the total charge $Q$, as shown below
\bea \nonumber
\H_{q_{\syk}=2} = \frac{1}{N} (C_2 +\frac{Q^2}{8}-2N -\frac{QN}{2}).
\eea
In this case the dynamics are analytically accessible and the two different irreps involved in the OTOC are a constant. These are given by
\bea
C_2^{(0,N,0)} = \frac{N^2}{2} + 2N ;&& C_2^{(1,N-2,1)} = \frac{N^2}{2} + N.
\eea
Utilizing this knowledge and the fact that the total charges of both irreps are $Q=2N$, one can compute the emergent Hamiltonian corresponding to both irreps
\bea
\H_{q_{\syk}=2}^{(0,N,0)} = 0; && \H_{q_{\syk}=2}^{(1,N-2,1)} = -1.
\eea
This implies that the dynamics within each irrep takes the following simple form
$\F_{(0,N,0)}(t)=a, \, \F_{(1,N-2,1)}=be^{-t}$.
The constant parameters $a$ and $b$ can be conveniently obtained from the initial value of the overall OTOC $\F_{q_{\syk}=2}(\chi_i(t), \chi_j^\dagger)$
\bea
\F_{q_{\syk}=2}(\chi_i(t=0),\chi_j^{\dagger}) =\left\{\begin{array}{ll} 0 &(i=j) \\ -\frac{1}{4}  &(i \neq j) \end{array}\right.
\eea
The OTOC is then written as
\bea \label{eq:freecomplexotoc}
\F_{q_{\syk}=2}(\chi_i(t),\chi_j^{\dagger}) = \bigg(\frac{N-1}{4N}\bigg)\bigg[-1 + \bigg(\frac{N \delta_{ij} - 1}{N-1}\bigg) e^{-t} \bigg].
\eea
Hence the OTOC starts from a value dependent on $\delta_{ij}$ and $N$, and exponentially decays to $-(N-1)/4N$ instead of $0$ as in the case of scrambling dynamics.

\subsubsection{Scrambling dynamics for $q_{\syk}=4$}
The effective Hamiltonian is not a constant anymore for $q_{\syk}>2$. One needs to diagonalize $\H$ for each irrep and every charge sector to compute the OTOC in Eq.~\eqref{eq:OverallOTOC}. This can be done for large but finite $N$ because the Hilbert space for each sector is at most of size $\mathcal{O}(N)$, given the symmetry structure of $\H$. 
Constructing the effective Hamiltonian for each sector requires us to build the matrix representation for the bilinear operators $S^{\alpha\beta}$ for both irreps $(0,N,0)$ and $(1, N-2, 1)$. Unlike the SU(2) case in which all the states within an irrep can be uniquely labelled by $L_z$, or equivalently the weight, for SU($n$) and $n>2$, there are multiple states within each irrep that are labelled by the same weight. In fact, these subspaces labelled by the weights are the ones leading to the operator dynamics within each charge sector in the complex Brownian SYK model. Fortunately, there are well-established schemes to uniquely label the states of arbitrary irreps for SU($n$) groups, called Gelfand-Tsetlin patterns~\cite{Alex_2011}, a brief introduction to which is provided in the appendix. Based on GT patterns, the matrix representation of $S^{\alpha\beta}S^{\beta\alpha}$ for each irrep and fixed weight subsectors and thus the effective Hamiltonian $\H$ can be constructed efficiently, which is local in this basis. 

\begin{figure}
\includegraphics[scale=0.5]{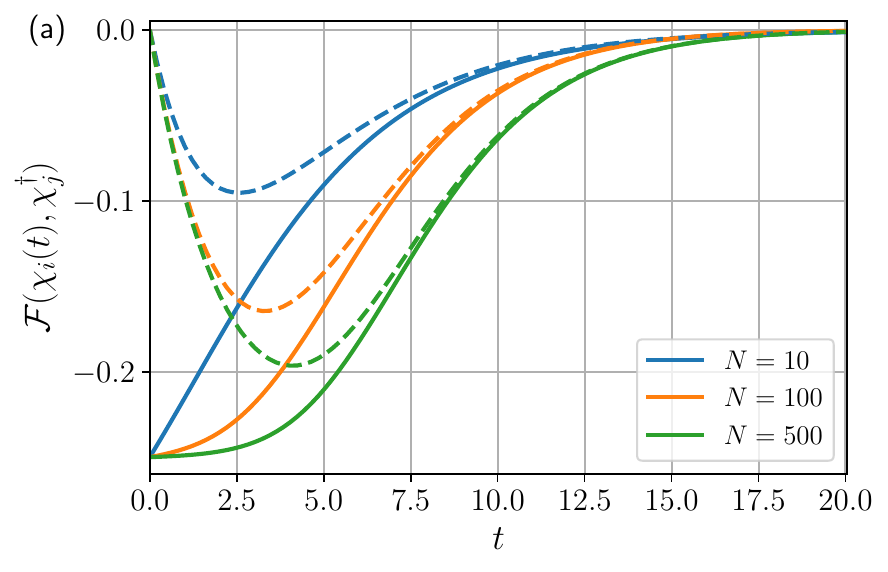}
\includegraphics[scale=0.5]{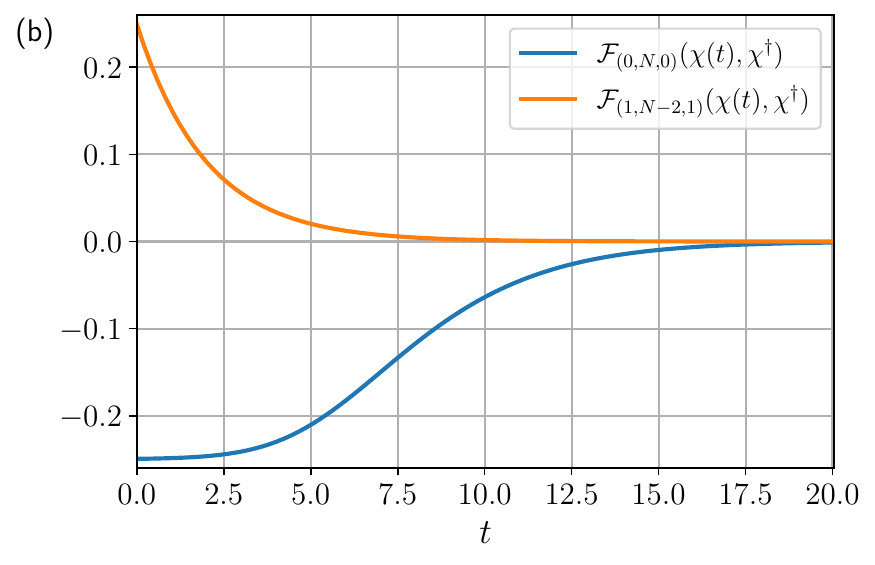}
\caption{(a) The overall OTOC $\F(\chi_i(t),\chi_j^{\dagger})$ vs time for $i=j$ (dashed lines) and $i \neq j$ (solid lines) for different values of $N$. The OTOC decays to zero at late times as expected from scrambling,  which is in contrast with the non-interacting $(q_{\syk}=2)$ case given in Eq.~\eqref{eq:freecomplexotoc}. (b) The OTOC resolved into the contributing irreps, according to Eq.~\eqref{eq:OverallOTOC}, for $N=500$ particles. The scrambling behavior is present in the $(0,N,0)$ irrep while the $(1,N-2,1)$ irrep displays an exponential decay at all times.}
\label{Fig:OTOCccdagger}
\end{figure}

Using this approach, we compute the overall OTOC $\F(\chi_i(t), \chi_j^\dagger)$ for $N$ up to 500 and plot the results in Fig.~\ref{Fig:OTOCccdagger}(a) and (b) for $i=j$ and $i\neq j$, respectively. Similar to the Majorana model, the two cases start with distinct initial values but quickly approach the same behavior that exponentially decays to zero. Furthermore, the case with $i \neq j$ develops the characteristic early time Lyapunov growth as $N$ increases. As shown in Eq.~\eqref{eq:OverallOTOC}, the difference between the two cases is from $\F_{(1,N-2,1)}$, the contribution from the $(1, N-2,1)$ irrep. We plot $\F_{(0,N,0)}(t)$ and $\F_{(1,N-2,1)}(t)$ in Fig.~\ref{Fig:OTOCccdagger}(b). Evidently, $\F_{(1,N-2,1)}(t)$ shows purely exponential decay, explaining the early time difference between $i=j$ and $i\neq j$ in $\F(\chi_i (t), \chi_j^\dagger)$. On the other hand, the early time Lyapunov growth is from $\F_{(0,N,0)}(t)$. This demonstrates scrambling in the interacting complex Brownian SYK model for local operators of the type $\ket{\chi^{\dagger} \otimes \chi}$.

\begin{figure}
    \includegraphics[scale = 0.5]{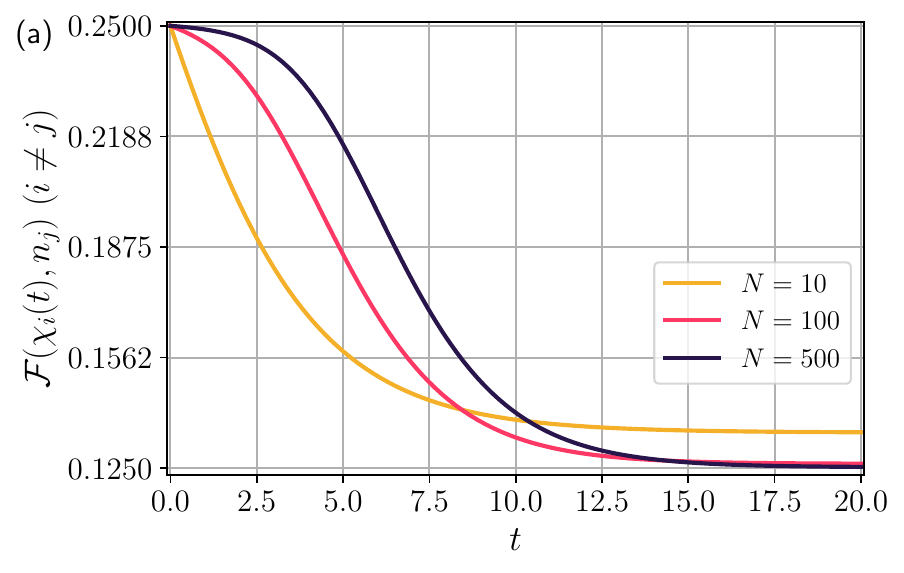}
    \includegraphics[scale = 0.5]{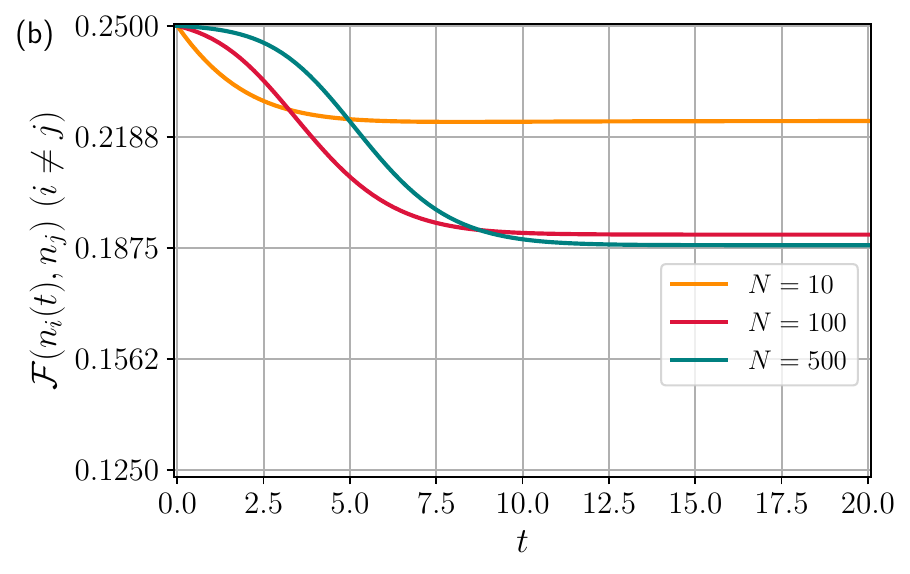}
    \caption{The overall OTOC $\mathcal{F}(\chi_i(t),n_j)$ in (a) and $\mathcal{F}(n_i(t),n_j)$ in (b), for $i \neq j$ and different values of $N$. Both the OTOCs display pronounced finite-sized effects in their late-time values, which remain finite as $N\rightarrow\infty$, in contrast with that of $\mathcal{F}(\chi_i(t),\chi^{\dagger}_j)$. }
\label{Fig:OtherOTOCs}
\end{figure}

OTOCs $\F(W(t),V)$ for other operators, even non local ones and other $q_{\syk}$ can be in principle calculated following the same procedure, which we summarize below
\begin{enumerate}
    \item Find the form of $\H$ as a function of the operators $S^{\alpha\beta}$
    \item Decompose the input and output states, $\ket{\i}$ and $\ket{\o}$, which depend on the operators $W$ and $V$ respectively, into different irreps and weight sectors of SU(4).
    \item Construct the matrix representation of $\H$ using GT patterns, for the sectors in which the component of the input and output states have a non-zero overlap. 
    \item Evolve the input state in imaginary time, using the Hamiltonian constructed in the last step, for each sector, and take the overlap with the output state.
\end{enumerate}
Among all these steps, step~2 is the most tedious. In what follows, we discuss the other OTOCs, $\F(\chi_i(t), n_j)$ and $\F(n_i(t), n_j)$. Note that $\F(\chi_i(t),n_j)$ and $\F(n_j(-t),\chi_i)$ are identical. The calculation involves the same two irreps $(0, N, 0)$ and $(1, N-2,1)$ for $\F(\chi_i(t), n_j)$ but involves the third irrep $(0, N-2, 0)$ for $\F(n_i(t),n_j)$. The details of decomposing the input state and the output state can be found in Appendix~\ref{sec:Operatorstatemap}. Similar to the previous example, the scrambling behavior results from the dynamics in the irrep $(0, N, 0)$, while the contributions from the other irreps are purely exponential decay. In Fig.~\ref{Fig:OtherOTOCs}, we plot both $\F(\chi_i(j), n_j)$ and $\F(n_i(t), n_j)$ for $i\neq j$ both $N$ up to 500, the behavior of which is dominated by the contribution from the irrep $(0, N, 0)$. Evidently, they develop the characteristic early time Lyapunov growth as $N$ increases. Unlike $\F(\chi_i(t), \chi_j)$, the late time value has a strong finite-size effect and asymtotes to $1/8$ and $3/16$ for large $N$, in agreement with Eq.~\eqref{eq:overalllatevalue} \footnote{Since we are working with a $q_{\syk}$-uniform model, sectors corresponding to dilute charge will display imperfect scrambling and there will be small corrections to the predicted values of the overall OTOC}.

This concludes our discussion on the behavior of the overall OTOC in the complex Brownian SYK model. The overall OTOC contains contributions from each charge sector that we label $\F^m$ in Sec.~\ref{sec:chargedscrambling}, where $m$ denotes the charge. Remarkably, our approach naturally provides full access to $\F^m$ from each charge sector and can be exploited to understand charge dependent scrambling for finite but large $N$ and all time scales, which we discuss in the next section.

\section{Charge-dependent scrambling} \label{sec:chargedscrambling}
As discussed in Sec.~\ref{subsec:complexotoc}, in systems with charge conservation the dynamics of a generic operator splits into sectors with different charge profiles. We expect that dynamics with restricted access to states within the full Hilbert space will display slower operator growth as the restrictions are made more stringent. In other words, OTOCs involving operators confined to specific sectors of the Hilbert space should experience Lyapunov growth as a function of the size of the sector. In the case of U(1) conservation, these sectors can be labelled using the charge. The overall OTOC is a weighted sum of the charge resolved OTOC defined in Sec.~\ref{subsec:complexotoc}
\bea
\F^m(W, V)=\frac{1}{\tr (P_m)} \tr (P_m W^\dagger(t) V^\dagger W(t) V).
\eea
We also define $\rho=m/N$ as the charge density.
Our approach, which is based on the SU(4)$\otimes$U(1) symmetry structure of the emergent Hamiltonian in the complex Brownian SYK model, can be naturally used to compute $\F^m(W,V)$ for each $m$. 

Take the charge resolved OTOC $\F^m(\chi_i(t), \chi_j)$ for example. After resolving the symmetry structure of $\H$, the common subsectors for the input state and the output state are labeled by conserved quantities $S^{\alpha\alpha}$ for $\alpha$ in $a\sim d$, which are directly related to $m$ as $(m, N-m+1, m, N-m-1)$. Furthermore each charge sector splits into two irreps $(0, N, 0)$ and $(1, N-2, 1)$. Then from Eq.~\eqref{eq:OverallOTOC}, we can directly obtain the charge resolved OTOC from each term in the summation. They are
\bea \label{eq:chargeOTOC} 
&\F^m(\chi_i(t),\chi_j^{\dagger}) = \F^m_{(0,N,0)}(t) + \left( \frac{N \delta_{ij}-1}{N-1} \right) \F^m_{(1,N-2,1)}( t )\\
&\F^m_{(0,N,0)}(t) = \frac{4^N}{\tr(P_m)} \langle \o_{m} |e^{\mathbb{H}_{m}t}|\i_{m}\rangle_{(0,N,0)} \\ 
& \F^m_{(1,N-2,1)}(t) = \frac{4^N}{\tr(P_m)} \langle \o_{m} |e^{\mathbb{H}_{m}t}|\i_{m}\rangle_{(1,N-2,1)}.
\eea
where the states $\ket{\i_m}$ and $\ket{\o_m}$ for each charge sector are given in Eq.~\eqref{eq:chargedstates}.
The other charge resolved OTOCs share a similar structure, and more irreps might be involved in their computation, for example, in the case of $\F^m(n_i(t), n_j)$.

As discussed in the last section, the irrep $(0, N, 0)$ contributes to scrambling dynamics while the contribution from other irreps exponentially decays to zero. Therefore we focus on $\F^m_{(0,N,0)}(t)$ and study how the scrambling dynamics depend on the charge $m$ both numerically for large finite $N$, and analytically directly in the infinite $N$ limit.

\subsection{Numerical result at large finite $N$}
\begin{figure}
\includegraphics[scale = 0.5]{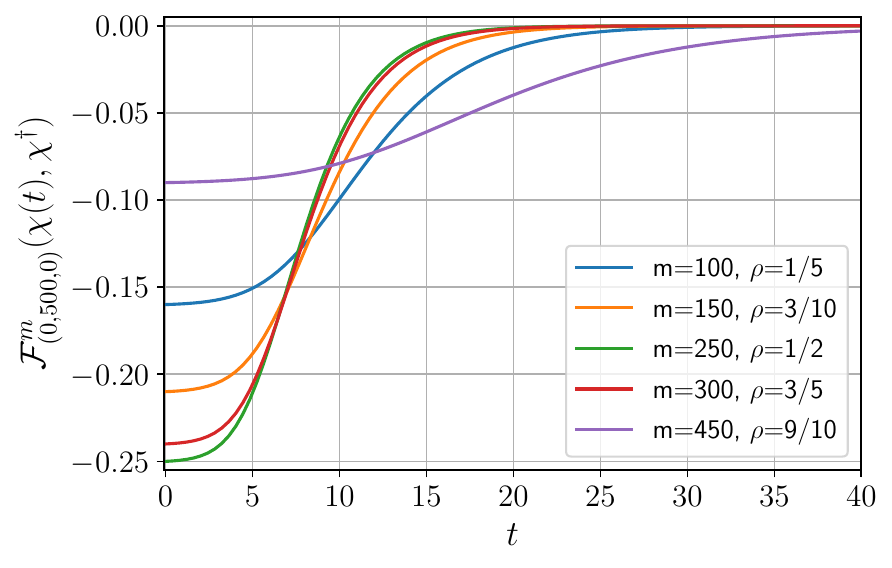}
\caption{Contributions to the OTOC from different charge densities $\rho=m/N$, for the OTOC $\F^{m}(\chi_i(t),\chi_j^{\dagger})$. The OTOC has been restricted to the $(0,N,0)$ irrep with $N=500$. This can be compared with the overall OTOC for the case $i \neq j$ for large $N$. The charge resolved OTOC $\F^m$ shows different initial values and time scales.
}
\label{fig:OTOCcartan}
\end{figure}

Using our approach, we compute $\F^m_{(0,N,0)}(\chi(t),\chi^\dagger)$ and plot the results for different $m$ in Fig.~\ref{fig:OTOCcartan} for $N=500$. They start with different initial values and relax to zero, consistent with Table.~\ref{tab:earlylatevalue}. Since $\F^m_{0,N,0}(\chi(t), \chi^\dagger)$ is related to the overall one as $\sum\limits_j \F^m(\chi_i(t), \chi_j^\dagger)/N$, we obtain the exact initial values as $-\rho(1-\rho)$. 
The early time behavior is characterized by the Lyapunov growth. Remarkably, in addition to the different initial values, the time scale in which $\F^m$ relaxes to zero only strongly depends on $m$ or the charge density $\rho$, which we analyze in full detail below.

\begin{figure}
\includegraphics[scale = 0.5]{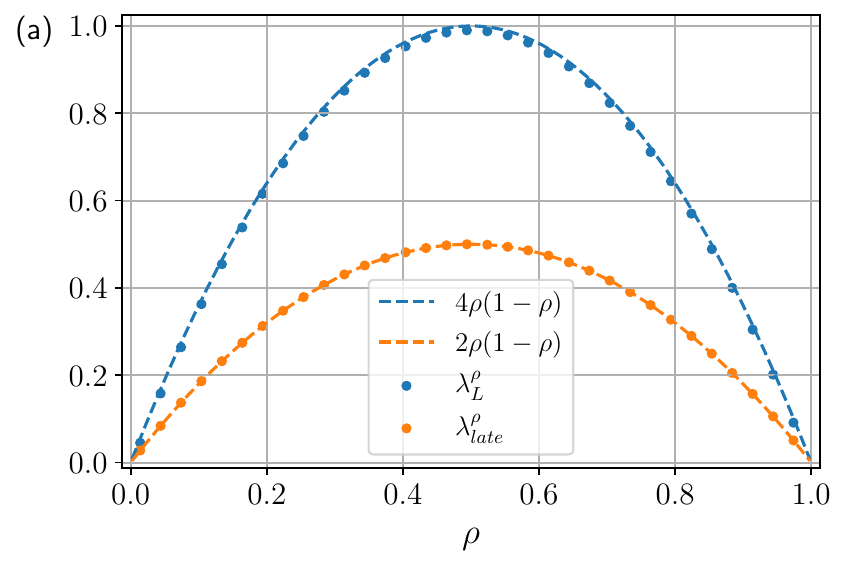}
\includegraphics[scale = 0.5]{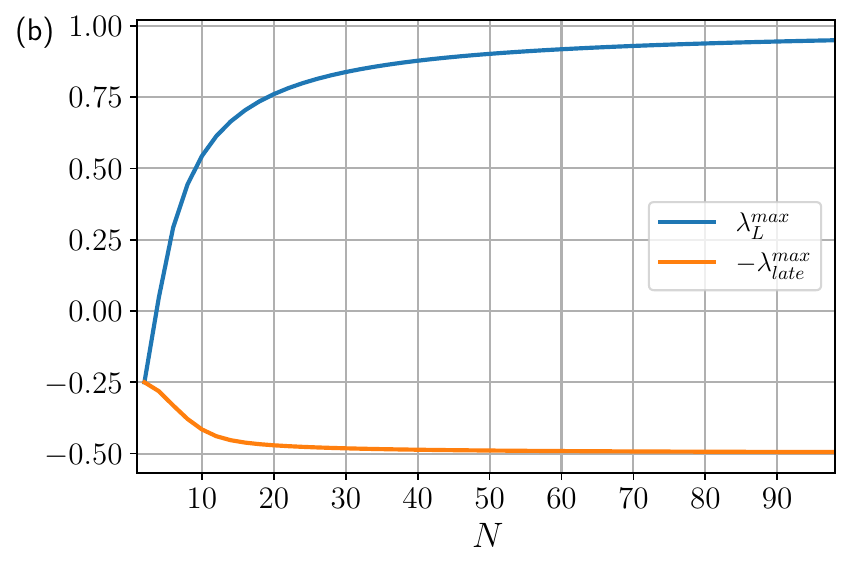}
\caption{(a) The plot of the Lyapunov exponent $\lambda_L^{\rho}$ and the late-time exponent $\lambda_{late}^{\rho}$ vs $\rho$ for $N=500$.  (b) The maximal Lyapunov exponent  $\lambda_L^{\rho}$ and late-time exponent corresponding to the half filled state $\rho= 1/2$ plotted as a function of $N$.
The exponents in both (a) and (b) have been extracted from a given charge sector within the irrep $(0,N,0)$ for $\F(\chi_i(t),\chi_j^{\dagger})$. The exponents of OTOCs involving other local operators, restricted to the $(0, N, 0)$ irrep,  show similar behavior.}

\label{fig:lyapcartan}
\end{figure}

To examine the early time growth and extract the Lyapunov coefficients for each charge sector, we will begin by assuming an early time ansatz of the form 
\bea \label{eq:earlytimeansatz}
\F^{m}_{(0,N,0)}(\chi(t \sim 0),\chi^\dagger) \sim -\rho(1-\rho) + \frac{a_\rho}{N} (e^{\lambda_L^{\rho} t}-1).
\eea
By Taylor expanding both the ansatz and Eq.~\eqref{eq:chargeOTOC} and matching coefficients, we can relate $\lambda_L^\rho$ to the moments of the effective Hamiltonian
\bea
\lambda_L^{\rho} = \frac{\braket{\o_m|\H_m^2|\i_m}_{(0,N,0)}}{\braket{\o_m|\H_m|\i_m}_{(0,N,0)}}.
\eea
This is a more accurate method to extract $\lambda^\rho_L$ than curve fitting. We plot the Lyapunov coefficients obtained in Fig.~\ref{fig:lyapcartan}(a) for $N=500$, via the method outlined above. We see that the Lyapunov exponent grows with the density till it reaches $\rho = 1/2$ after which it starts decreasing again, and the behavior is exactly described by an inverted parabola. The curve has been fitted against the function
\bea
\lambda_L^{\rho}(\rho) = 4\rho(1-\rho)
\eea
which agrees with \cite{Chen_2020}. We also plot the magnitude of the late-time relaxation exponent $\lambda^\rho_{late}$, which is given by the largest nonzero eigenvalues for each sector of $\H$ labelled by $m$,  as a function of the charge density $\rho$. It follows the same behavior as $\lambda^\rho_L$ described by the inverted parabola, but half in magnitude, $\lambda_{late}^\rho=2\rho(1-\rho)=\lambda_L^\rho/2$. This relation is reminiscent of the Majorana case shown in Fig.~\ref{Fig:Majoranairrep}(c). 

In Fig.~\ref{fig:lyapcartan}(b) we plot the maximal Lyapunov exponent and relaxation exponent at half filling as a function of $N$. Similar to the Majorana case, $\lambda_L^{max}$, starts with a negative value $-1/4$, changes sign and asymptotes to the large $N$ value 1 as $N$ increases, while $-\lambda_{late}^{max}$, starts with the same value, remains negative, and asymptotes to $-1/2$. The positive Lyapunov exponent is an effective behavior that emerges from a pool of a large number of eigenvalues, all of which are negative. For small Hilbert spaces (corresponding to low $N$), this pool is not large enough to produce a positive exponent. It should also be noted that although we obtain positive exponents for some small $N$, this behavior is short-lived and therefore difficult to obtain using standard curve-fitting techniques. In the large $N$ limit, this $\lambda_L^{max}$ is also the overall Lyapunov coefficient of the contribution from the $(0,N,0)$ irrep since it corresponds to the sector (half-filled) which dominates, and of the overall OTOC in the case $i \neq j$ since the contribution from the $(0,N,0)$ irrep dominates in that case.

\begin{figure}
\includegraphics[scale = 0.5]{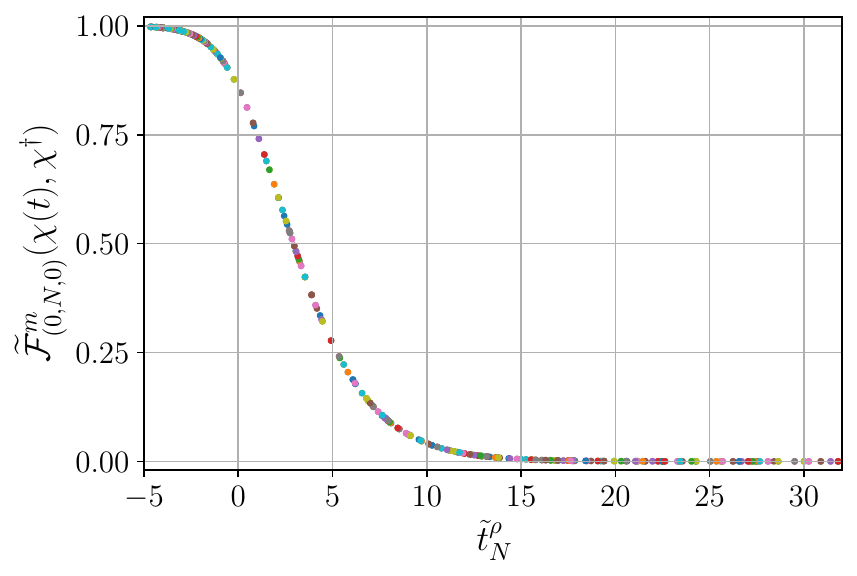}
\caption{The rescaled OTOC $\widetilde{\F}^m_{(0,N,0)}(\chi(t),\chi^{\dagger})$ for different charge sectors and two values of $N$ : $m= \{50,60,...,440,450\}$ for $N=500$, and $m=\{50,60,...,140,150\}$ for $N=200$, plotted against the transformed time variable $\tilde{t}^{\rho}_N$. The OTOCs for all different $m$ corresponding to both the values of $N$ collapse into a single function.}
\label{fig:chargecollapse}
\end{figure}

To capture the charge dependence of $\F^m_{(0,N,0)}$ beyond early time, we consider the rescaled OTOC $\widetilde{\F}^m=\F^m/(-\rho(1-\rho))$, which starts from 1 and relaxes to 0 for each $m$. We define a new parameter
\bea\label{eq:tNrho}
\tilde{t}_N^\rho = \text{ln} \left(\frac{e^ {\lambda_L^\rho t}-1}{N\rho(1-\rho)}\right).
\eea
Then the early time behavior of $\tilde{\F}^m_{(0,N,0)}$ is given by $1-a_\rho e^{\tilde{t}_N^\rho}$.  Using the numerical data, one can fix $a_\rho= 1$. Remarkably, we find that $\tilde{\F}^m_{(0,N,0)}$ for $N=200$ and $N=500$,  and various $m$ collapses to a single curve as a function of $\tilde{t}_N^\rho$ for all time scales, as shown in Fig.~\ref{fig:chargecollapse}. This indicates that the leading charge dependence and also $N$ dependence of $\F^m_{(0,N,0)}$ is captured by the following simple form
\bea
\F^m_{(0,N,0)}(\chi(t),\chi^\dagger)=-\rho(1-\rho)f(\tilde{t}^\rho_N).
\eea
It must be emphasized that the collapsing behavior is only observed for charge sectors of finite charge density $\rho$,  which have Hilbert-spaces that are large enough to produce a positive Lyapunov exponent.

\subsection{Charge dependent hydrodynamic equation}
To gain analytical understanding of the charge dependent scrambling and its relation to the Majorana case, we derive the charge resolved OTOC in the large $N$ limit analytically in this section.
Similar to the procedure in the Majorana case, we focus on the irrep $(0,N,0)$ that gives rise to the scrambling behavior, and write the OTOC $\F^m(\chi_i(t),\chi_j^{\dagger})$ as follows:
\bea
\F^{m}_{(0,N,0)}(t) = \frac{4^{N}}{\tr(P_m)} \braket{\o_{m}|\i_{m}(t)} = \sum_{k} \psi^m_{\o}(k) \psi^{m}_{\i}(k,t).
\eea
Since all the states within a fixed-$m$ subspace have equal weights in the eyes of the algebra, we use GT-patterns to distinguish them, labelled by the integer $k$. In this basis $k$ takes integer values ranging from $0$ to $\min(m-1,N-m-1)$ and the output state takes the form:
\bea
&\psi^m_{\o}(k) = -g(N,k)\sqrt{(N-m-k)(m-k)}/N \\ 
&g(N,k)=\bigg(\binom{N+1}{k+1} \binom{N+1}{k} \frac{N-2k}{N+1}\bigg)^{1/2}.
\eea
We define a similarity transformation to make the $N$-dependence of $\psi_{\o}^{m}(k)$ uniform, just as shown in Sec.~\ref{subsec:Majorana-hydrodynamics}, and the output state is transformed as
\bea
\psi^m_{\o}(k)\rightarrow \tilde{\psi}^m_{\o}(k)=-\sqrt{(N-m-k)(m-k)}/N.
\eea
After the transformation, the Hamiltonian and the input state are
\bea
\tilde H_{kk'}=g(N,k)H_{kk'}g(N,k)^{-1}; \, \tilde{\psi}^m_\i(k)=g(N,k)^{-1}\psi^m_\i(k).
\eea
The dynamics of $\tilde{\psi}_\i^m$ is governed by the equation $\partial_t \tilde{\psi}_\i^m(t) = \tilde{H}\tilde\psi_\i^m$, which now is ready to be expanded in the large-$N$ continuum limit.

The strategy is similar to that in the Majorana case. We use the continuous variables $x=k/N$ and $\rho=m/N$, where $0\leq x\leq \min(\rho,1-\rho)$. In the large $N$ limit, the building blocks of the Hamiltonian  $S^{\alpha\beta}S^{\beta\alpha}$ can be written as differential operators in terms of $x$ after the similarity transformation, from which the Hamiltonian $\tilde H$ as a differential operator acting on $\tilde \psi^m_{\i}$ can be obtained. In the infinite $N$ limit, $\tilde\psi_{\i}^m$ obeys
\bea \label{eq:Complexinputdynamics}
\partial_t \tilde\psi_{\i}^m =\partial_x\left( -\frac{4 (x-1) x (x^2-x+\rho(1-\rho))}{2 x-1} \tilde\psi_{\i}^m\right).
\eea
Similar to the Majorana case,
this equation predicts that $\int dx \, \psi^m_\i(x)$ is a constant and that $\psi^m_\i(x)$ can be interpreted as a probability distribution function. 
If $\tilde \psi^m_\i$ starts with a delta function, it will remain a delta function $\sqrt{\rho(1-\rho)}\delta(x-x(t))$, with the peak value $x(t)$ obeying an ordinary differential equation
\bea
\label{eq:complex_inf_N_ode}
\partial_t x(t) =\frac{4 (x-1) x (x^2-x+\rho(1-\rho))}{2 x-1}.
\eea
Then the charge resolved OTOC is given by 
\bea
\F^m_{(0,N,0)}&\sim \int \psi^m_\o(x) \psi^m_\i(x) dx\\
&=-\sqrt{(1-\rho)\rho(1-\rho-x(t))(\rho-x(t))}.
\eea
It should be noted that the transformed effective Hamiltonian $\tilde \H$ at finite $N$ is not stochastic due to the lack of a steady state for the OTOC we are considering, unlike the Majorana case. As a result, the interpretation of $\psi^m_\i(x)$ as a probability distribution is only valid in the infinite $N$ limit and breaks by a $1/N$ effect. We will discuss some interesting $1/N$ effects in the end of this section.

To solve Eq.~\eqref{eq:complex_inf_N_ode}, we introduce a new variable $\xi$ as a function of $x$
\bea
\xi=\left(\frac{(1-\rho-x)(\rho-x)}{(1-\rho)\rho}\right)^{1/2}
\eea
which ranges from $0$ to $1$. In the transformed variables
Eq.~\eqref{eq:complex_inf_N_ode} becomes
\bea
\partial_t \xi(t)=2\rho(1-\rho)\xi(\xi^2-1),
\eea
which is the same as the logistic differential equation derived for the Majorana case in Eq.~\eqref{eq:MajoranaHydrodynamics} up to a $\rho$ dependent factor that can be absorbed into $t$. This factor clearly demonstrates the characteristic time scale associated with the sector of charge density $\rho$. The differential equation has two unstable steady solutions $\xi=\pm 1$ and one stable steady solution $\xi=0$. Initially $\xi$ starts with a value close to $1$, corresponding to $x$ near 0, and relaxes to the stable steady solution over time.

In the new variable $\xi$, the output state is $-\sqrt{\rho(1-\rho)}\xi$, and OTOC is proportional to $\xi(t)$. Solving the logistic equation, we obtain
\bea
\F^m_{(0,N,0)}(t) =A_\rho \xi(t)=\frac{A_\rho}{\sqrt{e^{4 t \rho(1-\rho)}\delta_\rho+1}},
\eea
where $\delta_\rho$ and $A_\rho$ are determined by the input operator. We have assumed that $\delta_\rho \ll 1$.
The early and late time behavior are given by
\bea
\F^{m}_{(0,N,0)}(t) \sim \left\{\begin{array}{ll}
 A_\rho - \frac{1}{2} e^{4 \rho(1-\rho) t}  A_\rho \delta_{\rho} \,\, & t\ll-\frac{\operatorname{ln} \delta}{4 \rho(1-\rho)} \\
 e^{-2 \rho(1-\rho) t} \,\, & t\gg-\frac{\operatorname{ln} \delta}{4 \rho(1-\rho)}
\end{array}\right.
\eea
which results from the competition between $e^{4 t \rho(1-\rho)}$ and $\delta_\rho$.  The exponents $\lambda_L^{\rho}$ and $\lambda_{late}^{\rho}$ agree with the leading order of the numerical results. The same comments that apply to the Majorana case apply here as well, namely the analytical expression is not expected to match the numerical data as the initial state corresponding to local operators implies an initial condition of the form $x(0) \propto 1/N$, and the analytical technique is only applicable in cases where the input state is independent of $N$.

This result illustrates that the majority of the density dependence in the large-$N$ limit is contained in the initial value and the Lyapunov exponent, and scaling them appropriately will result in all sectors of finite charge density displaying the same behavior. Given the analytical expression, one can also expand the solution in a manner very similar to Eq.~\eqref{eq:Majoranaghostmodes} to observe the emergence of positive Lyapunov exponents from the negative eigenvalues of the emergent Hamiltonian.

There are also some interesting $1/N$ effects that we briefly mention here. Similar to the Majorana case, adding $1/N$ corrections will result in the dynamical equation for $\tilde{\psi}^m_\i$ (Eq.~\eqref{eq:Complexinputdynamics}) transforming into a second-order differential equation in $x$, which will lead to the initial delta function broadening under the time-evolution. However, there are two $1/N$ effects that are absent in the Majorana case. Firstly, the new logistic equation obtained will have zeros which when corrected for the $1/N$ effect, will lead to steady solutions being shifted outside of the physical Hilbert space. Secondly, since there will be no steady-state solutions within the physical Hilbert space, $\int dx \, \tilde{\psi}^m_\i$ will no longer be conserved and the interpretation of $\tilde{\psi}^m_\i$ as a probability distribution will no longer be valid. Hence we will see that this quantity will show a late-time decay away from its constant value, which can be numerically verified as well. 

\subsection{Other OTOCs}
\begin{figure}
\includegraphics[scale = 0.5]{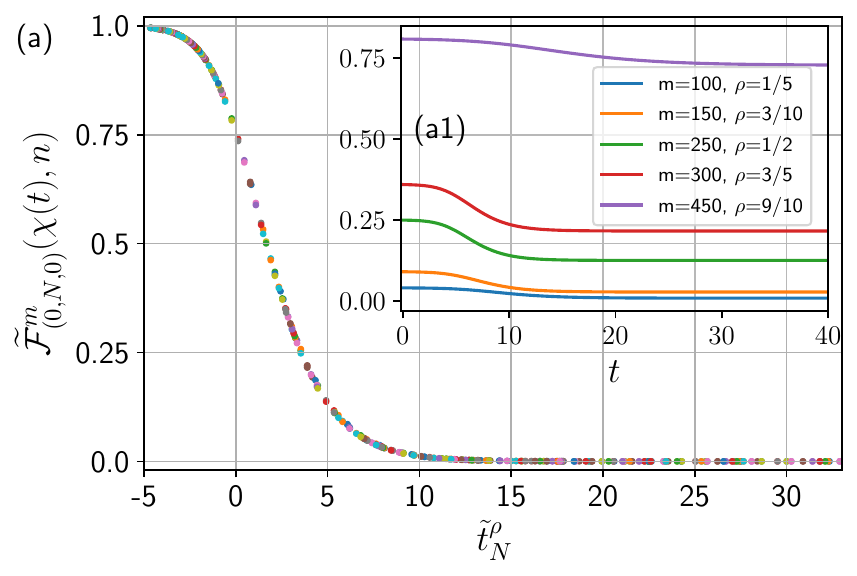}
\includegraphics[scale = 0.5]{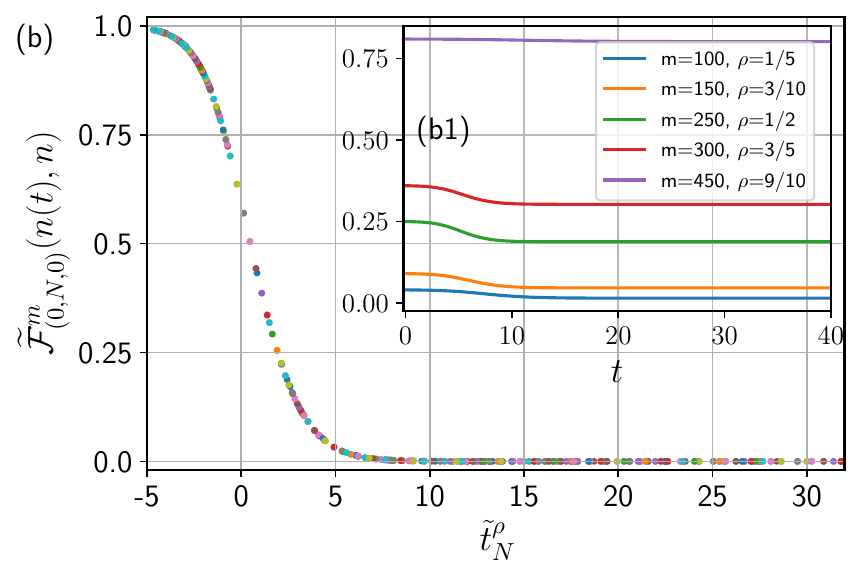}
\caption{The rescaled charge resolved OTOCs $\widetilde{\F}^m_{(0,N,0)}(\chi(t),n)$ in~(a) and $\widetilde{\F}^m_{(0,N,0)}(n(t),n)$ in~(b) for different $m$ and $N$ plotted against the transformed time variable $\tilde{t}^{\rho}_N$. We choose  $m= \{50,60,...,440,450\}$ for $N=500$, and $m=\{50,60,...,140,150\}$ for $N=200$.  The OTOCs for different $m$ and both the values of $N$ collapse into a single functional form. Insets (a1) and (b1) show the corresponding unrescaled charge-resolved OTOCs $\F^m_{(0,N,0)}(\chi(t),n)$ and $\F^m_{(0,N,0)}(n(t),n)$ for different $m$ at $N=500$ as a function of $t$.}
\label{Fig:OTOCcartan1}
\end{figure}

We now discuss two other kinds of OTOCs, namely $\F^m_{(0,N,0)}(\chi(t),n)$ and $\F^m_{(0,N,0)}(n(t),n)$. We are again restricting the discussion to the irrep $(0,N,0)$ because it contains all of the interesting scrambling dynamics. The relevant charge sector labelled by $S^{\alpha\alpha}$ are $(m,N-m+1,m-1,N-m)$ and $(m,N-m,m,N-m)$ for  $\F^m_{(0,N,0)}(\chi(t),n)$ and $\F^m_{(0,N,0)}(n(t),n)$, respectively. The main difference from the previous OTOC $\F^m_{(0,N,0)}(\chi(t),\chi^\dagger)$ is that there reside steady states within these charge sectors. In other words, the effective Hamiltonian $\H$ has a zero eigenvalue in addition to other negative eigenvalues. As a result, when the input state and the output state both have a finite overlap with the steady state, the OTOC develops a finite late-time value. This indeed is what is observed here. In Fig.~\ref{Fig:OTOCcartan1} (insets (a1) and (b1)), we plot both OTOCs for different $m$, which relax to different final values consistent with those given in Table~\ref{tab:earlylatevalue}.

To study the charge dependence, we rescale $\F^m_{(0,N,0)}$ as
\bea
\widetilde \F^m_{(0,N,0)}(t) = \frac{\F^m_{(0,N,0)}(t)-\F^m_{(0,N,0)}(\infty)}{\F^m_{(0,N,0)}(0)-\F^m_{(0,N,0)}(\infty)}.
\eea
The rescaling removes the dependence of the initial and the late-time values, and the rescaled OTOC monotonically decreases from 1 to 0 as time increases, for all $m$. Remarkably, as we show in Fig.~\ref{Fig:OTOCcartan1}, the rescaled OTOC for different $m$ and two values of $N$, $200 \text{ and } 500$, also collapses to a function of the variable $\tilde{t}_N^\rho$ defined in Eq.~\eqref{eq:tNrho}. This indicates that the leading charge dependence of all OTOCs considered in this work, apart from the initial and the late-time values, can be captured by $\tilde t_N^\rho$.

 As is evident from the discussions in this section, all OTOCs display a late-time exponential decay behavior. Some studies \cite{Khemani_2018,Rakovszky_2018,cheng2021scrambling} have found late time power law relaxations of the OTOC in systems with U(1) conservation or energy conservation due to diffusion, which scales as $\sim t^{d/2}$, where $d$ is the number of spatial dimensions. However, we are working in $0$ spatial dimensions where the charge is static and therefore the OTOC still exponentially decays in the late-time regime.

\section{Discussion and summary} \label{sec:discussion}
In this work we study the Brownian SYK model with and without charge conservation. We introduce a symmetry-based approach which maps the Brownian SYK dynamics to SU($n$) spin dynamics after taking the disorder average, where $n=2$ for the Majorana model and 4 for the model with complex fermions. This mapping drastically reduces the dimension of the effective dynamical Hilbert space from $\sim e^N$ to $\sim \mathcal{O}(N)$, allowing us to numerically compute the OTOCs exactly for large system size and all time scales. For the non-interacting case ($q_{\syk}=2$), one finds that the SU($n$) algebras are promoted to exact symmetries and the OTOCs can be solved analytically. We also provide a method to connect the Hamiltonian approach used in the formalism to the approach which maps to a stochastic model, via a similarity transformation, and utilize it to derive a hydrodynamical description of the OTOC in the large-$N$ limit. In this limit, we find that the OTOC in the complex model follows the same differential equation as the Majorana model, up to a density dependent overall scaling. 

For the Majorana model, we verify previously known results using the new formalism and compute the OTOC for $N=10000$ fermions. We also utilize the approach to analyze the early and late time exponents as a function of $N$ and demonstrate how they reach their asymptotic values for large $N$. In this context we also discuss the emergence of the positive Lyapunov exponent which arises from the combined effect of the negative eigenvalues of the emergent Hamiltonian.

For the complex model, multiple kinds of OTOCs are discussed which involve different operators. We study the scrambling dynamics of these different operators and provide exact early and late-time values of the OTOC restricted to different charge sectors in a general complex model with charge conservation, which are later verified numerically for the case of the complex Brownian model. We also make use of the formalism to numerically compute the exact OTOC for $N=500$ fermions, and analyze how the OTOC approaches the late time value starting from the initial value. We find that for $q_{\syk}=4$ the Lyapunov exponent has density dependence $\lambda_L^{\rho}= 4 \rho(1-\rho)$ and that the late time dynamics is marked by an exponential decay as well, with an exponent that has a similar functional dependence on the charge density, i.e.\ $\lambda_{late}^{\rho}= 2 \rho(1-\rho)$. Since the formalism provides access to the dynamics for finite $N$, we study how these exponents change and approach their asymptotic values starting from $N=2$ to $N \gtrsim 100$.

The approach used in this work has several directly visible extensions. One direction is to explore higher order correlators which will involve more than $4$ time-contours and therefore for the Brownian model will still have an emergent SU($n$) algebra, albeit with a higher $n$ and more complex symmetry structure when compared with the conventional OTOC. As an example, the complex Brownian fermionic model will give rise to an SU($n$) $\otimes$ U(1) algebra on $n$ time contours, while the Majorana model will display an SO($n$) algebra. Another direction is to start with some Brownian model with a non-abelian symmetry such as SU($n$), instead of U(1), and explore the behavior of the correlators as a function of $n$ to probe the relation between the rank of the continuous symmetry in the model, and scrambling. Furthermore, it would be very interesting to generalize the procedure given in this work to higher dimensions, especially for the charge conserved case where one can derive the hydrodynamic equations to describe the interplay between local Lyapunov growth and ballistic operator spreading~\cite{Xu_2019, keselman2021scrambling} as well as charge diffusion. We expect that the coupled diffusion equation of the charge and the FKPP equation of the operator~\cite{aleiner2016microscopic,Xu_2019} can lead to algebraic decay of OTOC~\cite{Khemani_2018,Rakovszky_2018} in the late time. From the coupled equations between charge and operator, one can also study the charge dependence of the butterfly velocity and the relation between the diffusion constant, butterfly velocity, and the Lyapunov exponent at different charge density. Some other directions for future work are studying related observables such as entanglement entropy, tripartite mutual information and spectral form factors in the presence of the U(1) symmetry.

\section{Acknowledgement}
We thank Brian Swingle, Subhayan Sahu, Shaokai Jian, Christopher  M.  Langlett, Xiao Chen, Andrew Lucas and Tianci Zhou
for helpful discussions and comments on the manuscript. S. Xu acknowledges Subhayan Sahu and Brian Swingle for collaborations on related projects. S. Xu also thanks the hospitality of KITP supported by the National Science Foundation under Grant No. NSF PHY-1748958, and hospitality of Aspen Center for Physics supported by National Science Foundation grant PHY-1607611. Portions of this research  were  conducted  with  the  advanced  computing resources provided by Texas A\&M High Performance Research Computing.  

\bibliography{paper2}
\onecolumngrid
\newpage
\appendix
\begin{appendices}

\section{The emergent Hamiltonian}
\subsection{The Majorana model}
The Emergent Hamiltonian for the Majorana case, for $q_{\syk}=2$ and $q_{\syk}=4$ body couplings in the original Hamiltonian, is given by:
\bea
&\H_{q_{\syk}=2}=\frac{1}{2N}\Bigg(-2 \binom{N}{2}-3 N-\frac{1}{2} \sum_{(\alpha \neq \beta)} (S^{\alpha \beta})^{2}\Bigg)\\
&\H_{q_{\syk}=4}=\frac{3}{N^{3}}\Bigg(-2 \binom{N}{4}+\frac{1}{4 !} \sum_{(\alpha \neq \beta)} (-1)^{\gamma_{\alpha, \beta}}\left[(S^{\alpha\beta})^{4}-(S^{\alpha \beta})^{2}(-6 N+8)+3 N(N-2)\right] \Bigg)
\eea
Where 
\bea
S^{\alpha \beta}= \sum_{i=1}^N \psi^{\alpha}_i \psi^{\beta}_i ; \quad  (-1)^{\gamma_{\alpha,\beta}}=\left\{\begin{array}{ll}
1 & (\alpha,\beta)=(a, b),(a, d),(b, c),(c, d) \\
-1 & (\alpha,\beta)=(a, c),(b, d)
\end{array}\right. \quad 
\eea
The emergent Hamiltonians for larger $q_{\syk}$ can be obtained iteratively, and are always functions of the 6 operators $S^{\alpha\beta}$.

We will devote the rest of the Majorana section to the discussion of the qualitative behavior of the OTOC for larger $q_{\syk}$ in the Brownian SYK model. We will use $q_{\syk}=8$ as an example to build intuition about the behavior for general $q_{\syk}$. The emergent Hamiltonian for $q_{\syk}=8$ in terms of the SU(2) algebra takes the manifestly square symmetric form:
\bea 
\H_{q_{\syk}=8}=\frac{7!}{2N^{7}}\left(-2\left(\begin{array}{c}
N \\
8
\end{array}\right)+\frac{1}{8!} \bigg((H_x)_{q_{\syk}=8} +(H_z)_{q_{\syk}=8} - (H_y)_{q_{\syk}=8} \bigg) \right)
\eea
Where
\bea
(H_{\alpha})_{q_{\syk}=8} = 512 L_{\alpha}^8 - 3584(-4 + N)L_{\alpha}^6 +448(176 + 5 N (-22 + 3 N))L_{\alpha}^4 \\+32(2112 - 7 N (424 + 15 (-10 + N) N))L_{\alpha}^2 + 210 (-6 + N) (-4 + N) (-2 + N) N
 ; && \alpha = x,y,z
\eea
Now we expand the Hamiltonian obtained within the $L=N/2$ irrep, in the infinite-$N$ limit, keeping the input and output states the same as in Sec.~\ref{subsec:Majorana-hydrodynamics}. The Logistic equation in this case is given by:
\bea
\xi'(t) = 2 \xi (\xi^6-1) \, \, ; \, \, \xi(0) = 1-2 \delta \, , \, \delta \ll 1
\eea
with the solution:
\bea
(\F_{N/2}(t))_{q_{\syk}=8} = -\frac{1}{\sqrt[6]{1+ 12 e^{12t} \delta}}.
\eea
This solution has the following early and late-time behavior:
\bea
(\F_{N/2}(t))_{q_{\syk}=8} \sim \left\{\begin{array}{ll}
 -1 + 2 e^{12 t} \delta \quad & t\ll-\frac{1}{12}\operatorname{ln} \delta \\
e^{-2t} \quad & t\gg-\frac{1}{12}\operatorname{ln} \delta
\end{array}\right.
\eea
The Hamiltonian for larger $q_{\syk}$ can be derived iteratively through the equation:
\bea
(H_{\alpha})_{q_{\syk}+1} =  2 i L_{\alpha} (H_{\alpha})_{q_{\syk}} + q_{\syk} (N+1-q_{\syk}) (H_{\alpha})_{q_{\syk}-1} \, \, ; \, \, (H_{\alpha})_{q_{\syk}=0} = 2 \, \, ; \, \, (H_{\alpha})_{q_{\syk}=1} = 2 i L_{\alpha}
\eea
The emergent Hamiltonian for general $q_{\syk}$ can then be written as the square or cubic symmetric function of $(H_{\alpha})_{q_{\syk}}$, depending on whether $q_{\syk}/2$ is even or odd. Based on the results obtained so far, we can conjecture that the logistic equation and OTOC for general $q_{\syk}$ will take the form:
\bea
\xi'(t) = 2 \xi (\xi^{(q_{\syk}-2)}-1) \implies (\F_{N/2}(t))_{q_{\syk}} = -\frac{1}{\sqrt[(q_{\syk}-2)]{1+ 2(q_{\syk}-2)e^{2(q_{\syk}-2)t} \delta}}.
\eea
which implies that the ratio of the early to late exponent is $(\lambda_{L}/ \lambda_{late})_{q_{\syk}} = (q_{\syk}-2)$, for $q_{\syk}>2$. Therefore changing the $q_{\syk}$ leads to a larger Lyapunov exponent but leaves the late-time exponent unchanged ($\lambda_{late}=2$).

\subsection{The complex model}

To derive the emergent Hamiltonian, one has to account for the fact that the Hamiltonian on the replicas $a,c$ looks slightly different from the one on the replicas $b,d$ due to the particle-hole transformation. 
\bea
H^{\alpha} &= \sum_{i,j,k,l} J_{i,j,k,l} \, {\psi_i^{\alpha}}^{\dagger}\,{\psi_j^{\alpha}}^{\dagger} \,\psi_k^{\alpha}\, \psi_l^{\alpha} \quad ; \alpha = a,c \\
H^{\alpha*} &= \sum_{i,j,k,l} J^*_{i,j,k,l} \, \psi_i^{\alpha} \,\psi_j^{\alpha}\, {\psi_k^{\alpha}}^{\dagger}\, {\psi_l^{\alpha}}^{\dagger} \, \quad ; \alpha = b,d.
\eea
Since the couplings are complex, $J$ only couples to $J^{*}$, and when the disorder average is computed, the resultant operator depends on which sites are paired together. As an example, for $q_{\syk}=4$:
\bea
\overline{H^a H^{a}} \propto \sum_{i,j,k,l} {\psi_i^a}^{\dagger}{\psi_j^a}^{\dagger} \psi_k^a \psi_l^a {\psi_l^a}^{\dagger}{\psi_k^a}^{\dagger} \psi_j^a \psi_i^a \, \, ; \, \,
\overline{H^a H^{b*}} \propto \sum_{i,j,k,l} {\psi_i^a}^{\dagger}{\psi_j^a}^{\dagger} \psi_k^a \psi_l^a \psi_i^b \psi_j^b {\psi_k^b}^{\dagger} {\psi_l^b}^{\dagger}.
\eea
The emergent Hamiltonian after disorder average reads
\bea
\H = -\frac{1}{2} &\left(\overline{H^{a}H^{a}} + \overline{H^{b*}H^{b*}} + \overline{H^{c}H^{c}} + \overline{H^{d*}H^{d*}} \right)  +\left(\overline{H^{a}H^{b,*}} + \overline{H^{b,*}H^{c}} + \overline{H^{c}H^{d,*}} + \overline{H^{d,*}H^{a}} \right)\\
    -&\left( \overline{H^{a}H^{c}} + \overline{H^{b,*}H^{d,*}} \right).
\eea
To simplify the notation for $q_{\syk}=4$, we define the operators $\mathcal{P}_{\alpha \beta}$ which sends the index $\alpha \rightarrow \beta$ and the conjugation operator $\mathcal{C}$ which sends $\psi^{a,c} \rightarrow (\psi^{b,d})^{\dagger}$ and therefore $\mathcal{C} S^{aa} \rightarrow N-S^{bb}$ and $\mathcal{C} S^{ac} \rightarrow -S^{db}$. One can then write :
\bea 
\H_{q_{\syk}=4} = -\frac{1}{2}(1+\mathcal{C})(1+\mathcal{P}_{ac})\overline{H^{a} H^{a}}
+(1+\mathcal{P}_{bd})(1+\mathcal{P}_{ac})\overline{H^{a} H^{b, *}} - (1+\mathcal{C})\overline{H^{a} H^{c}}
\eea
Where
\bea
&\overline{H^{a} H^{a}}  = S^{aa}(S^{aa}-1)(N-S^{aa}+2)(N-S^{aa}+1)\\
&\overline{H^{a} H^{b,*}}
= [(S^{ab}S^{ba})^2 + (S^{aa}-S^{bb}-2)S^{ab}S^{ba}]\\
&\overline{H^{a} H^{c}} = [2 (S^{aa}-1)(S^{aa})-(3S^{aa} + S^{cc}-2)S^{ac}S^{ca} + (S^{ac} S^{ca})^2]
\eea
and $S^{\alpha \beta} = \sum_{i}  {\psi_i^{\alpha}}^{\dagger}\psi_i^{\beta}$. It is evident that this Hamiltonian, although more complicated than the $q_{\syk}=2$ case, also preserves the charge profile $(S^{aa}, N-S^{bb}, S^{cc}, N-S^{dd})$ since it commutes with $S^{\alpha \alpha}$, $\alpha = \{ a,b,c,d \}$.

\section{Consistency checks}

We will dedicate this section to independent consistency checks to make sure our results are accurate. 

\subsection{Invariance of the identity operator}
First we will note that within our formalism, we expect operator-states which correspond to symmetries of the Hamiltonian to vanish under the emergent Hamiltonian. One such operator that vanishes identically without depending on the details of the theory, is the identity operator $\ket{I}$. A generalised version of this statement for an operator $O$ can be written as
\bea \label{symmetry}
\big[ H,O \big] = 0 \implies UOU^{\dagger} = O \implies \big(U \otimes U^{*}\big) \ket{O} = \ket{O}
\eea
What is theory dependent however, are the details of how the identity state splits into the representations of the algebra respected by the Hamiltonian. For the complex Brownian SYK model, the identity state splits into different states within the $(0,N,0)$ irrep as 
\bea
\ket{I \otimes I} &= \sum_{l,m=0}^{N} \left| \sum_{\alpha_{1}<\ldots<\alpha_{l}} \bar{n}_{\alpha_{1}} \ldots \bar{n}_{\alpha_{l}} n_{i_{1}} \ldots n_{i_{N-l}} \otimes \sum_{\beta_{1}<\ldots<\beta_{m}} \bar{n}_{\beta_{1}} \ldots \bar{n}_{\beta_{m}} n_{j_{1}} \ldots n_{j_{N-m}}\right\rangle \\
&= \sum_{l,m=0}^{N} (-1)^{N/2}\frac{(S^{dc})^{m}(S^{ba})^{l}(S^{cb})^{N}}{(N-1)! \,l! \, m!} \ket{W_{(0,N,0)}}
\eea
This shows that the identity splits into $(N+1)^2$ different states with different weights. Since the identity cannot have dynamics, it must vanish under the action of the emergent Hamiltonian, i.e. $\H \ket{I \otimes I} = 0$. Because the emergent Hamiltonian conserves weights, it implies that each state in the decomposition of the identity with a unique weight must vanish independently as well. We have already seen that this happens for the free case, i.e. $\H_{q_{\syk}=2}^{(0,N,0)}\ket{I \otimes I} = 0$, in the section \ref{subsec:q=2}, because the operator vanishes for the entire irrep $(0,N,0)$. One can check that the identity indeed does vanish even under the action of the emergent Hamiltonian for $q_{\syk}=4$, which completes our first consistency check.

\subsection{Comparison of initial and final values obtained numerically vs analytically}
\begin{figure}
\includegraphics[scale = 0.4]{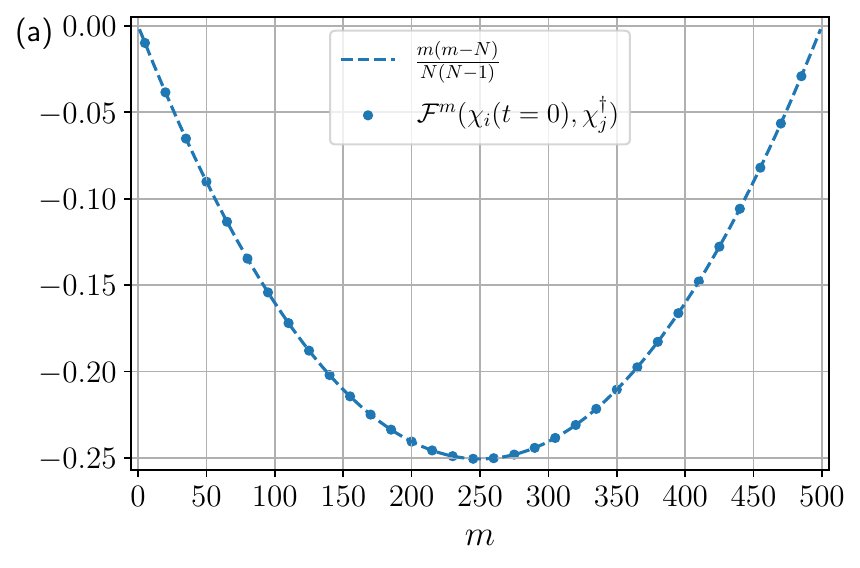}
\includegraphics[scale = 0.4]{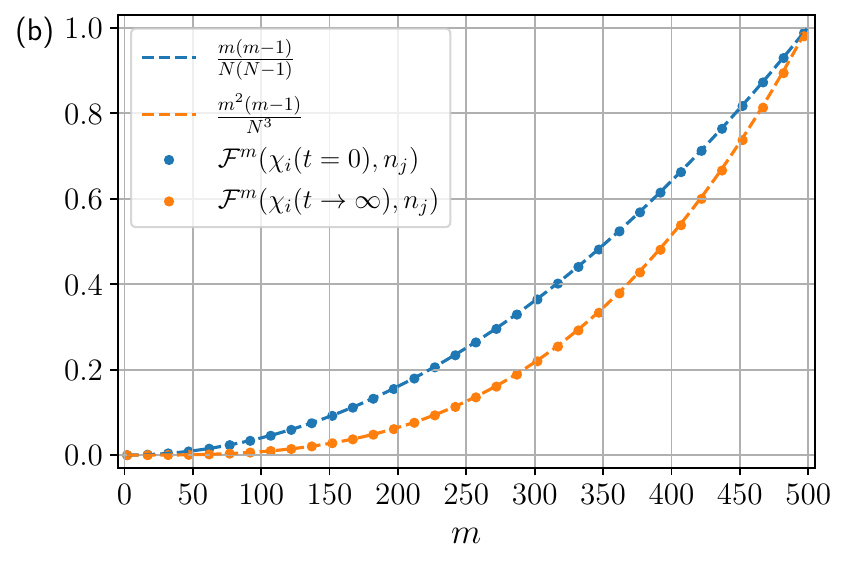}
\includegraphics[scale = 0.4]{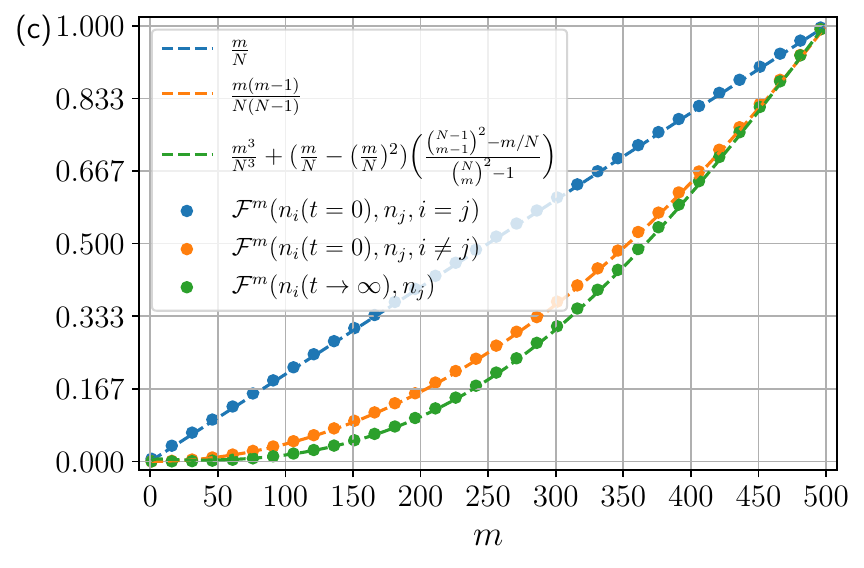}
\caption{ (a) Benchmarking of the initial values of the OTOC $\F^{m}(\chi_i(t),\chi_j^{\dagger})$, for $N=500$ and different $m$. (b) Benchmarking of the initial and final values of the OTOCs $\F^{m}(\chi_i(t),n_j)$, for $N=500$ and different $m$. (c) Benchmarking of the initial and final values of the OTOCs $\F^{m}(n_i(t),n_j)$, for $N=500$ and different $m$. All of the benchmarking has been performed using the numerical values obtained using the Lie Algebra method and the values obtained analytically listed in Table.~\ref{tab:earlylatevalue}.}
\label{Fig:initialfinal}
\end{figure}

The second consistency check comes from comparing the theoretical and numerical values of the initial and final values of the OTOCs. We plot this check in Fig.~\ref{Fig:initialfinal}. The numerical values are obtained using the Lie Algebra method and the theoretical values are listed in Table.~\ref{tab:earlylatevalue}

\subsection{Benchmarking with small systems using explicit random averaging}

\begin{figure}
    \includegraphics[scale = 0.5]{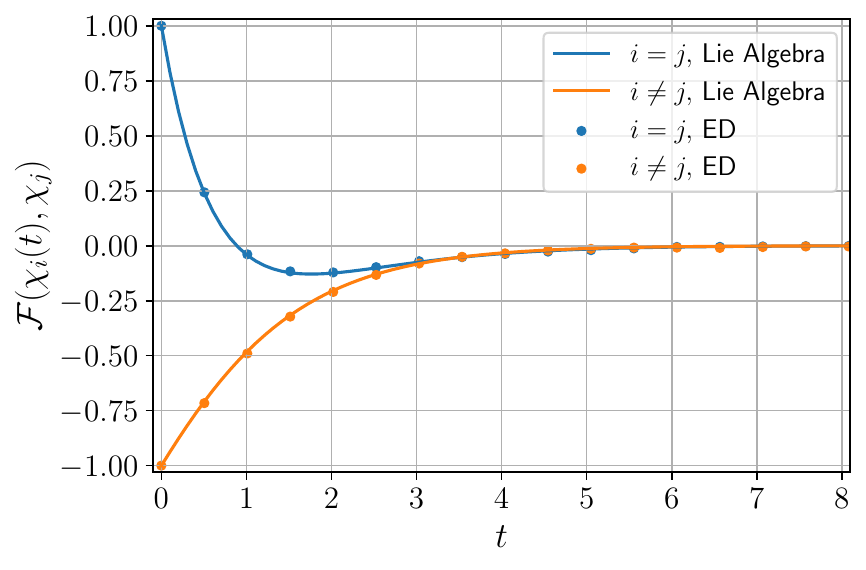}
    \caption{A plot of the OTOC $\F(\chi_i(t),\chi_j)$ in the Majorana model vs time, from both the ED of the original model in Eq.~\eqref{eq:MajoranaHamiltonian}  with explicit disorder average, and Lie Algebra (SU(2)) methods. Here $N=12$, $q_{\syk}=4$ and the numerical ED results are computed using 200 disorder averages.}
    \label{Fig:MajoranaED}
\end{figure}

Another consistency check comes from comparing the prediction of the OTOC obtained using the emergent Lie Algebra in Eq.~\eqref{eq:OverallMajorana}, with the numerical results obtained from directly computing the OTOC via ED. We display the plot in Fig.~\ref{Fig:MajoranaED} for $N=12$ particles in the Majorana model using $200$ disorder averages. 

\begin{figure}
    \includegraphics[scale = 0.35]{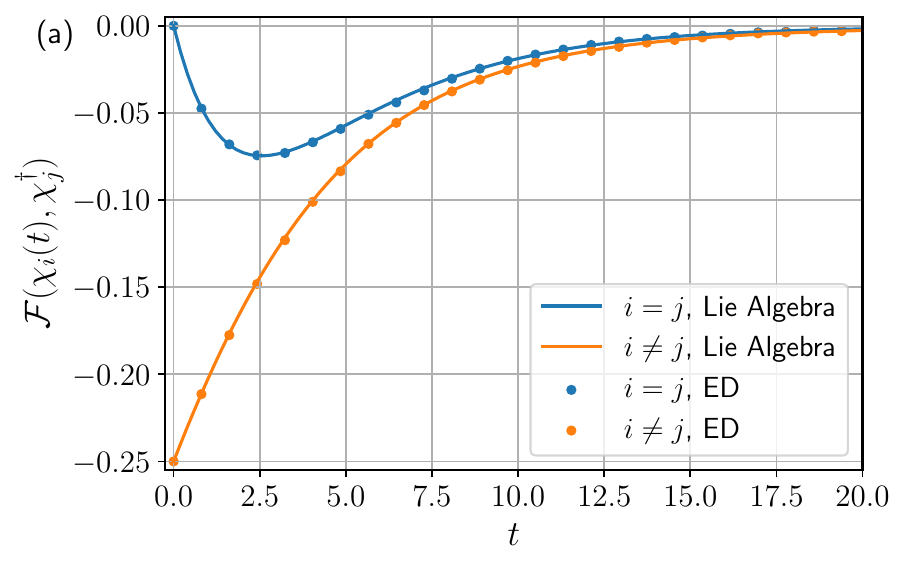}
    \includegraphics[scale = 0.35]{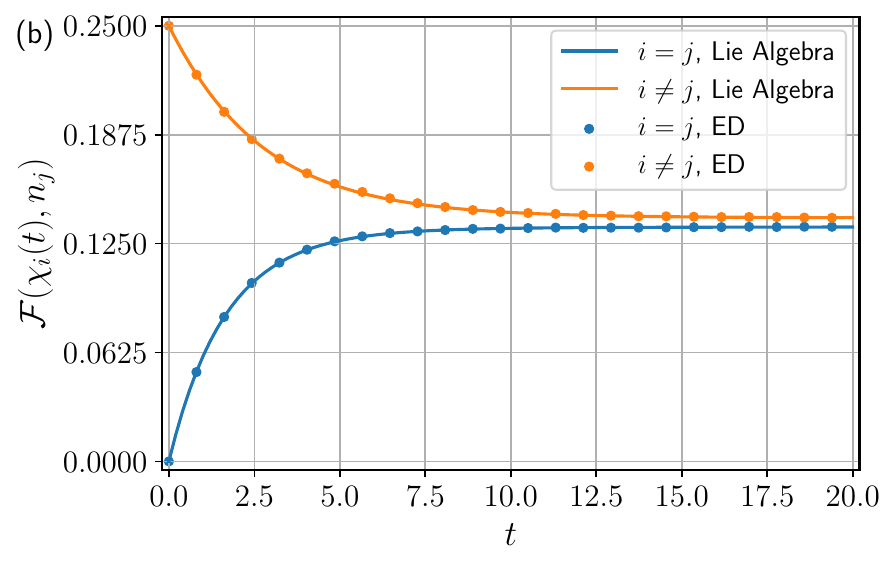}
    \includegraphics[scale = 0.35]{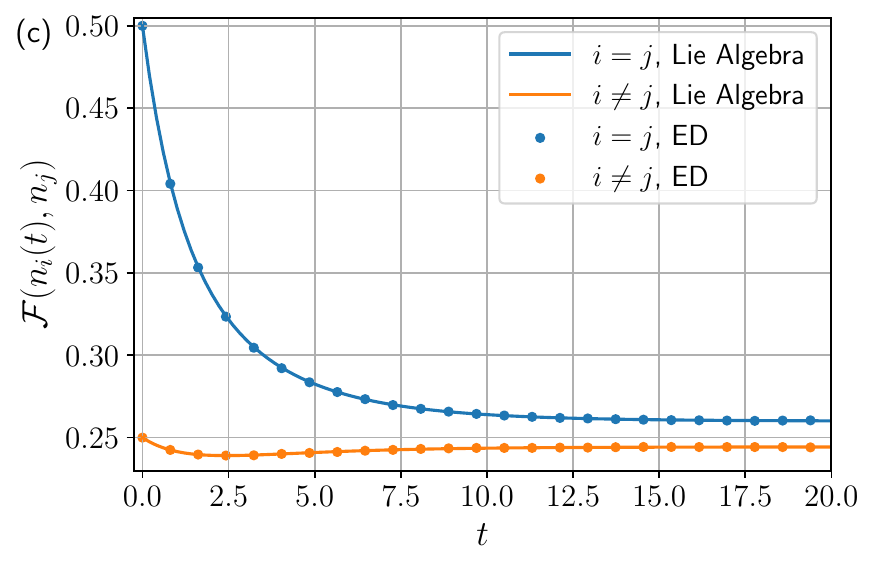}
    \caption{A plot of the OTOCs $\F(\chi_i(t),\chi_j^{\dagger})$ in (a), $\F(\chi_i(t),n_j)$ in (b) and $\F(n_i(t),n_j)$ in (c) vs time, from both the ED of the original model in Eq.~\eqref{eq:ComplexBrownianHamiltonian} and Lie Algebra (SU(4)) methods. Here $N=6$, $q_{\syk}=4$ and the numerical ED results are computed using 500 disorder averages.}
    \label{Fig:ComplexED}
\end{figure}

We also make the same comparison for the complex model in Fig.~\ref{Fig:ComplexED}, for OTOCs involving different local operators. The Lie Algebra method corresponds to computing the OTOC using the contribution from each irrep after evolving the input state and taking the overlap with the output state. As an example, Eq.~\eqref{eq:OverallOTOC} shows how to do this for the OTOC $\F(\chi_i(t),\chi_j^{\dagger})$, and the ED is computed for $N=6$ particles over $500$ disorder averages for all the OTOCs considered.

\section{Gelfand-Tsetlin pattern calculus} \label{appdx.GT}
To label states in the SU(4) irreps and explicitly build the matrices in the group for numerical purposes, we will make use of Gelfand-Tsetlin patterns. In this section we will provide a brief introduction to GT-patterns (a more detailed review can be found in~\cite{Alex_2011}) and demonstrate their utility for the purpose of solving problems  that involve SU($n$) algebras. For SU(4), GT-patterns are labelled by 10 non-negative integers arranged in the triangular pattern depicted below:

	\[ \begin{pmatrix} m_{1,4}&&m_{2,4}&&m_{3,4}&&m_{4,4}\\&m_{1,3}&&m_{2,3}&&m_{3,3}&\\&&m_{1,2}&&m_{2,2}&&\\&&&m_{1,1}&&&\end{pmatrix}\]

 Each entry in the pattern is uniquely labelled by the 2 integers $k,l$ such that the entry $m_{k,l}$ lies along the $k^{th}$ diagonal and in the $l^{th}$ row. A valid GT-pattern has entries that satisfy the constraint $m_{k, l} \geq m_{k, l-1} \geq m_{k+1, l}$. The top-most row of the pattern is determined by the irrep and stays fixed for all states within the irrep. To label the highest-weight state of an irrep, one needs to compute the values of $(S^{aa},S^{bb},S^{cc},S^{dd})$ corresponding to the state, and the values in the highest row of the corresponding pattern $(m_{1,4},m_{2,4},m_{3,4},m_{4,4})$ take these values respectively. The values in the lower rows are given by the maximum values that are allowed in a valid GT-pattern. A subtle point here is that the charge sensitivity of the Hamiltonian that emerges after disorder-averaging makes it important to label the patterns differently for irreps with the same young diagrams but different charges. For example, the states $|{\chi_1}^{\dagger}...{\chi_N}^{\dagger} \otimes {\chi_1}^{\dagger}...{\chi_N}^{\dagger} \rangle$ and  $| \chi_1...\chi_N \otimes \chi_1...\chi_N \rangle$ are both singlets but have charge profiles $(N,N,N,N)$ and $(0,0,0,0)$ respectively and therefore have different eigenvalues with respect to the emergent Hamiltonian. Often this charge is 'gauged' away by requiring that the element $m_{N,N}$ be set to zero, however it is unwise to do so for this problem. The Young-diagrams for the irreps can be read off using the rule that $m_{k,4}$ labels the number of boxes in the $k^{th}$ row of the diagram. For the highest-weight states in Eq.~\eqref{eq:highestweight} (with the correct normalizations) : 
\bea
 N \ket{W_{(0,N,0)}} &&&= \begin{pmatrix} N&&N&&0&&0\\&N&&N&&0&\\&&N&&N&&\\&&&N&&&\end{pmatrix} \in \overbrace{\begin{ytableau}  &  \none[\dots] & \\
&  \none[\dots] &
\end{ytableau}}^{N \text{ boxes}}\\ 
\frac{ \sqrt{N} \ket{W_{(1,N-2,1)}}}{\sqrt{(N-1)}} &&&= 	\begin{pmatrix} N&&N-1&&1&&0\\&N&&N-1&&1&\\&&N&&N-1&&\\&&&N&&&\end{pmatrix} \in \overbrace{\begin{ytableau}  &  \none[\dots] &  &\\
&  \none[\dots] &\\
\;
\end{ytableau}}^{N \text{ boxes}}
\eea
To construct states lower in these irreps, the rule states that the action of the operators $E_{\pm \alpha_l}$ results in adding(subtracting) $M^{k,l}$, the pattern with all entries $0$ except for the position $k,l$ which has value $1$ ($M^{k,l}$ is not a valid GT-pattern by itself). If $|M \rangle$ represents a GT-pattern then the raising(lowering) action is given by: 
\bea
E_{\pm \alpha_l} |M \rangle  = \sum_k c^{\pm}_k |M \pm M^{k,l} \rangle.
\eea
Thus it is possible to act on a single pattern and generate multiple ones, all corresponding to states with the same weight. The coefficients $c^{\pm}_k$ have been determined as: 
\bea
c^-_{k}= \left\langle M-M^{k, l}\left|E_{-\alpha_{l}}\right| M\right\rangle=\left(\begin{array}{c} \frac{\prod_{k^{\prime}=1}^{l+1}\left(m_{k^{\prime}, l+1}-m_{k, l}+k-k^{\prime}+1\right) \prod_{k^{\prime}=1}^{l-1}\left(m_{k^{\prime}, l-1}-m_{k, l}+k-k^{\prime}\right)}{\prod_{k^{\prime}=1 \atop k^{\prime} \neq k}^{l}\left(m_{k^{\prime}, l}-m_{k, l}+k-k^{\prime}+1\right)\left(m_{k^{\prime}, l}-m_{k, l}+k-k^{\prime}\right)} \end{array}\right)^{\frac{1}{2}}\\
c^+_{k} = \left\langle M+M^{k, l}\left|E_{+\alpha_{l}}\right| M\right\rangle=\left(\begin{array}{c} \frac{\prod_{k^{\prime}=1}^{l+1}\left(m_{k^{\prime}, l+1}-m_{k, l}+k-k^{\prime}\right) \prod_{k^{\prime}=1}^{l-1}\left(m_{k^{\prime}, l-1}-m_{k, l}+k-k^{\prime}-1\right)}{\prod_{k^{\prime}=1 \atop k^{\prime} \neq k}^{l}\left(m_{k^{\prime}, l}-m_{k, l}+k-k^{\prime}\right)\left(m_{k^{\prime}, l}-m_{k, l}+k-k^{\prime}-1\right)} \end{array}\right)^{\frac{1}{2}}.
\eea
If $|M \pm M^{k,l} \rangle$ is not a valid pattern, $c^{\pm}_k$ will be $0$. These equations can be used to build the matrix elements of raising/lowering operators in any SU($n$) algebra. The 'traditional' weight of a state, which is composed of the eigenvalues of the Cartan subalgebra, is given by 'z-weights' which are 
\bea
\lambda_{l}^{M}=\sigma_{l}^{M}-\frac{1}{2}\left(\sigma_{l+1}^{M}+\sigma_{l-1}^{M}\right); \quad \bigg(\sigma_{l}^{M}=\sum_{k=1}^{l} m_{k, l}, \quad \sigma_{0}^{M}=0\bigg)
\eea
However one can also use an equivalent formalism called 'p-weights' ($w_{l}^{M}=\sigma_{l}^{M}-\sigma_{l-1}^{M}$) which can be mapped to the z-weights. Patterns with the same sum of individual rows have the same p-weights and therefore the same z-weights and form the subspaces in the irrep with the same weight. For the irrep $(0,N,0)$, one can write down the states appearing in the decomposition of the input state as individual GT-patterns, owing to the symmetrisation of indices in the irrep. The states that occur within the sum in Eq.~\eqref{eq:inputweights} when restricted to the $(0,N,0)$ irrep can be represented as:
\bea \label{eq:inputpattern}
 \frac{(S^{dc})^m (S^{ba})^l}{l!m!} \ket{\chi_1^{\dagger} n \otimes  \chi_1 n}_{(0,N,0)} \propto \begin{pmatrix} N&&N&&0&&0\\&N&&N-m&&0&\\&&N&&1&&\\&&&N-l&&&\end{pmatrix}
\eea
One can take advantage of this feature to gain analytical leverage. Since the Hamiltonian conserves weights, it is important to know the multiplicity of the weights that reside in the input states within each irrep. For the $(0,N,0)$ irrep, this can be calculated by realising that the patterns with the same weight(p-weight) are the ones that differ along the second row but give the same sum. This is because they are the only valid GT-patterns that have the same row sums as Eq.~\eqref{eq:inputpattern}. These patterns look like
\bea
\begin{pmatrix} N&&N&&0&&0\\&N&&N-m&&0&\\&&x&&y&&\\&&&N-l&&&\end{pmatrix} \qquad \begin{matrix} x+y=N+1; \qquad N \geq x \geq N-m ; \\N-m \geq y \geq 0 ; \qquad  x \geq N-l \geq y \end{matrix}
\eea
The number of different non-negative $x,y$ that satisfy these constraints determine the weight multiplicity $I$, and can be computed for both the irreps that contribute to the OTOC
\bea\label{eq:weightmultiplicity}
&I\big((S^{dc})^m (S^{ba})^l \ket{\chi_1^{\dagger} n \otimes  \chi_1 n}_{(0,N,0)}\big)=\min(l+1,m+1,N-l,N-m)\\
&I\big((S^{dc})^m (S^{ba})^l \ket{\chi_1^{\dagger} n \otimes  \chi_1 n}_{(1,N-2,1)}\big) = 2\min(l+1,m+1,N-l,N-m)\\&+min(l+1,m+2,N-l,N-m-1)+\min(l+1,m,N-l,N-m+1)-2.
\eea
This means the subspaces that the Hamiltonian can have dynamics within are at most of size $\mathcal{O}(N)$.

\section{Decomposition of operator states to SU(4) irreps} \label{sec:Operatorstatemap}
In this section, we present more details on how to organize the input and output operator states into irreps of SU(4) algebra. We consider the initial operators that take the form $\ket{O_1 I \otimes O_1^\dagger I}$, where $O_1$ acts on the first fermion. The operator states on the first fermion belongs to the six dimensional $(0,1,0)$ irrep of SU(4). The operator state acting on the remaining fermions is $I\otimes I$ belongs to $(0, N-1, 0)$. As a result, the total states can be decomposed into three irreps,
\bea
(0,1,0)\otimes (0, N-1, 0) = (0,N,0) \oplus (1, N-2, 1) \oplus (0,N-2,0).
\eea
The operator states corresponding to the highest weight states in $(0,N,0)$ and $(1,N-1,1)$ are
\bea
\ket{W_{(0,N,0)}}=\frac{1}{N} \ket{\chi^\dagger\otimes\chi}, \quad
\ket{W_{(1,N-1,1)}} = - \ket{n_1 \chi^\dagger\otimes n_1\chi} +\frac{1}{N} \sum_i \ket{n_i \chi^\dagger \otimes n_i\chi}
\eea

\subsection{$W = n_1$ as the input state}
Now we consider the decomposition of $\ket{n_1 I \otimes n_1 I}$ in the interest of computing correlators involving this state.
We first perform the double rotation in the SU(2)$\otimes$SU(2) subgroup of SU(4) to rotate the state to $\ket{(n_1-\bar{n}_1) n \otimes (n_1-\bar{n}_1)n}$. We study the states in the expansion, which have fixed weights, term by term. The most involved term is $\ket{\bar{n}_1 n \otimes \bar{n}_1 n}$. Based on the weight counting, in the basis of GT patterns, there are two states in $(0,N,0)$, 4 states in $(1,N-1,1)$ and 1 state in $(0,N-2,0)$ contributing to the decomposition. The state $\ket{g}$ in the irrep $(0, N-2, 0)$ is
\bea
\ket{g} &= \sum\limits_{i\neq 1}\left( \ket{(\bn_1 n_i -n_1 \bn_i)\otimes (\bn_1 n_i -n_1 \bn_i)} + \ket{\chi_1^\dagger\chi_i n\otimes \chi_1\chi_i^\dagger n} + \ket{\chi_1\chi_i^\dagger n \otimes \chi_1^\dagger \chi_i n} \right)\\
&-
\frac{1}{2N}\left(\sum\limits_{i,j\neq 1}\ket{(\bn_i n_j -n_i \bn_j)\otimes (\bn_i n_j -n_i \bn_j)} + \ket{\chi_i^\dagger\chi_j n\otimes \chi_i\chi_j^\dagger n} + \ket{\chi_i\chi_j^\dagger n \otimes \chi_i^\dagger \chi_j n} \right).
\eea
The two independent states in the irrep $(0,N,0)$ that are required for the construction are:
\bea
&\ket{a}=\frac{1}{N!}S^{dc}S^{ba}(S^{cb})^N \ket{\chi^\dagger \otimes \chi} \, ; \,
\ket{b} = S^{bc}S^{cb} \ket{a}.
\eea
For the contribution from the $(1,N-2,1)$ irrep, we will begin by defining the state $\ket{X}$
\bea
(-1)^{N/2+1} \frac{(S^{cb})^{N-2}}{(N-2)!} |W \rangle_{1,N-2,1} = |\chi_1^{\dagger} n \otimes \chi_1 n \rangle - \frac{1}{N} \sum_{j} |\chi_j^{\dagger} n \otimes \chi_j n \rangle = \ket{X}.
\eea
Using this state, one can construct the $4$ independent states in $(1,N-2,1)$ irrep required:
\bea
&\ket{k} = S^{cb}S^{dc}S^{ba} \ket{X} \, ; \, 
\ket{l} = S^{dc}S^{cb}S^{ba} \ket{X} \, ; \,
\ket{m} = S^{ba}S^{cb}S^{dc} \ket{X} \, ; \,
\ket{j} = S^{bc}S^{cb} \ket{m} 
\eea
and we use a linear combination of the $4$ states above to build the required state:
\bea 
\implies \ket{g} + \frac{N+2}{N^2} (\ket{b} + \ket{a}) + (1+1/N)(\ket{j} - \ket{k} + 2(\ket{l} + \ket{m}))= (N+3+2/N)|\bn_1 n \otimes \bn_1 n \rangle.
\eea
Hence we have built the desired operator-state using states from all three contributing irreps. The other parts of the state $\ket{(n_1-\bn_1)n \otimes (n_1-\bn_1)n}$ can be built using just two of the three irreps in the young decomposition:

\bea
&\ket{\bn_1 n \otimes n} = (-1)^{\frac{N}{2}+1} \frac{(S^{cb})^{N-1}S^{ba}}{(N-1)!}(S^{cb} \ket{W_{(0,N,0)}} - \ket{W_{(1,N-2,1)}})\\
&\ket{n \otimes \bn_1 n} = (-1)^{\frac{N}{2}+1} \frac{(S^{cb})^{N-1}S^{dc}}{(N-1)!}(S^{cb} \ket{W_{(0,N,0)}} - \ket{W_{(1,N-2,1)}})\\
&\ket{n \otimes n} = (-1)^{N/2} \frac{(S^{cb})^{N}}{(N-1)!} \ket{W_{(0,N,0)}}
\eea
Now the overall state can be rotated back into the desired $z$-basis by applying the following rotation operator:
\bea
\sum_{l,m=0}^{N} \frac{(S^{ba})^l (S^{dc})^m}{l!m!} \ket{(n_1-\bn_1)n \otimes (n_1-\bn_1)n} = \ket{n_1 I \otimes n_1 I}.
\eea
The OTOC $\F(n_1(t),\chi_j^{\dagger})$ in each charge sector, labelled by the charge $m$, has the charge-profile $(m,m,m+1,m+1)$ on the four replicas. This state can therefore be split into different charge-profiles that contribute to the OTOC by applying the appropriate part of the rotation operator. The full output state $\sum_{\S}\ket{\chi_j \S \chi_j^{\dagger} \otimes \S^{\dagger}}$ has been built in the main text, which can also be rotated into the relevant charge profiles to find the output state restricted to each charge sector. Hence the equation above builds the charge resolved input and output states required to compute the OTOC $\F^m(n_1(t),\chi_j^{\dagger})$. This OTOC can also be used to compute the charge resolved OTOC $\F^m(\chi_i(-t),n_1)$ in the following way:
\bea
\tr(P_m \, n_1(t) \, \chi_j \, n_1(t) \, \chi_j^{\dagger}) = \tr(P_{m+1} \, \chi_j^{\dagger}(-t) \, n_1 \, \chi_j(-t) \, n_1)
\eea
Hence the OTOCs can be equated once the charge-sector is shifted by $1$.

\subsection{$V=n_1$ as the output state}
Next, we task ourselves with computing $\ket{n_1 I \otimes n_1 I}$ as the output state in the interest of computing the correlator $\F(n_i(t),n_1)$. Since we know how to build the input state $\ket{n_i I \otimes n_i I}$, we will build the output state using the SU(2) $\otimes$ SU(2) subalgebra formed by the operators $S^{da}$ and $S^{cb}$. Inserting the resolution of the identity into the output state, we get 
\bea
\ket{\o} =  \frac{1}{4^N}\sum_S \ket{n_1 \S^{\dagger} n_1 \otimes \S} = \frac{1}{2^{2N-1}}  \sum_{\S^c} \ket{n_1 {\S^c}^{\dagger} \otimes n_1 \S^c},
\eea
where the sum $S^c$ is over operator-strings on all sites except $1$. This state can be rotated into just four states and then rotated back in the following way:
\bea
&\sum_{\S^c} \ket{n_1 {\S^c}^{\dagger} \otimes n_1 \S^c} 
\\
\quad \quad &= (-1)^{N/2} \sum_{k=0}^{N}\sum_{l=0}^{N-1} (-1)^{l} \frac{(S^{da})^k}{k!} \frac{(S^{cb})^l}{l!} (1+S^{bc})\bigg(\ket{\chi^{\dagger} \otimes \chi} - \ket{\chi_1 \chi^{\dagger} \otimes \chi_1^{\dagger} \chi} + \ket{\bn_1 \chi^{\dagger} \otimes \bn_1 \chi} - \ket{n_1  \chi^{\dagger} \otimes n_1 \chi} \bigg)
\eea
The state $\ket{\chi^{\dagger} \otimes \chi}$ is proportional to the highest weight state $\ket{W_{(0,N,0)}}$. Two of the other states are simple to build using 
\bea
\ket{n_1  \chi^{\dagger} \otimes n_1 \chi} = S^{cb}\ket{W_{(0,N,0)}} - \ket{W_{(1,N-2,1)}} \, \, ; \, \,
\ket{\bn_1  \chi^{\dagger} \otimes \bn_1 \chi} = S^{ba}S^{dc} \ket{n_1  \chi^{\dagger} \otimes n_1 \chi}
\eea
Now we move onto building the fourth state and to build this, we first construct the highest weight state in $(0,N-2,0)$:
\bea
\ket{W_{(0,N-2,0)}} = (N-1)\ket{\chi_1 \chi^{\dagger} \otimes \chi_1^{\dagger} \chi} + \sum_{i \neq 1}( \ket{\bn_1 n_i \chi^{\dagger} \otimes \bn_1 n_i \chi} - \ket{n_1 \bn_i \chi^{\dagger} \otimes \bn_1 n_i \chi} - \ket{\bn_1 n_i \chi^{\dagger} \otimes n_1 \bn_i \chi} + \\ \ket{n_1 \bn_i \chi^{\dagger} \otimes n_1 \bn_i \chi})
+\frac{2}{N}\sum_{i \neq 1}\ket{\chi_i \chi^{\dagger} \otimes \chi_i^{\dagger} \chi} - \frac{1}{N} \sum_{i,j \neq 1} (\ket{\bn_i n_j \chi^{\dagger} \otimes \bn_i n_j \chi} - \ket{\bn_i n_j \chi^{\dagger} \otimes n_i \bn_j \chi})
\eea
The state in the $(0,N,0)$ irrep relevant to the construction is:
\bea
\ket{c} = (S^{cb}S^{ba}S^{dc}S^{cb}-\frac{1}{4}S^{ba}S^{dc}(S^{cb})^2) \ket{W_{(0,N,0)}} 
=\sum_{i} \ket{\chi_i \chi^{\dagger} \otimes \chi_i^{\dagger} \chi}- \frac{1}{2} \sum_{i,j} (\ket{\bn_i n_j \chi^{\dagger} \otimes \bn_i n_j \chi} - \ket{\bn_i n_j \chi^{\dagger} \otimes n_i \bn_j \chi})
\eea
Hence 
\bea
\ket{W}_{(0,N-2,0)}-\frac{2}{N}\ket{c} &= (N-1-2/N) \ket{\chi_1 \chi^{\dagger} \otimes \chi_1^{\dagger} \chi} + (1+1/N)\sum_{i \neq 1}( \ket{\bn_1 n_i \chi^{\dagger} \otimes \bn_1 n_i \chi} \\ &- \ket{n_1 \bn_i \chi^{\dagger} \otimes \bn_1 n_i \chi} - \ket{\bn_1 n_i \chi^{\dagger} \otimes n_1 \bn_i \chi} + \ket{n_1 \bn_i \chi^{\dagger} \otimes n_1 \bn_i \chi})
\eea
Now in the irrep $(1,N-2,1)$, we build the $4$ states:
\bea
&\ket{x}=S^{dc}S^{cb}S^{ba}\ket{W_{(1,N-2,1)}} \, ; \,
\ket{y}=S^{ba}S^{cb}S^{dc}\ket{W_{(1,N-2,1)}} \, ; \, \\
&\ket{z} = S^{cb}S^{ba}S^{dc}\ket{W_{(1,N-2,1)}} \, ; \,
\ket{w} = S^{ba}S^{dc}S^{cb}\ket{W_{(1,N-2,1)}}
\eea
And using these, we can build the fourth state required to rotate into the overall state $\sum_{S^c} \ket{n_1 {S^c}^{\dagger} \otimes n_1 S^c} $
\bea
\ket{W}_{(0,N-2,0)}+ \bigg( \frac{2N+4}{N^2} \bigg) \ket{c}+\bigg(1+\frac{1}{N}\bigg)(2(\ket{x}+\ket{y})-\ket{w}-4\ket{z}) = (N+3+2/N)\ket{\chi_1 \chi^{\dagger} \otimes \chi_1^{\dagger} \chi} 
\eea
For the correlator $\F^m (n_i(t),n_1)$, the values of $S^{\alpha \alpha}$ on each replica corresponding to the overall charge $m (\in [1,N])$ are $(m,N-m,m,N-m)$. Therefore, the input and output state built can be rotated appropriately into these charge sectors to compute the OTOC.

\section{Additional irreps}
\begin{figure}[h]
\includegraphics[scale = 0.5]{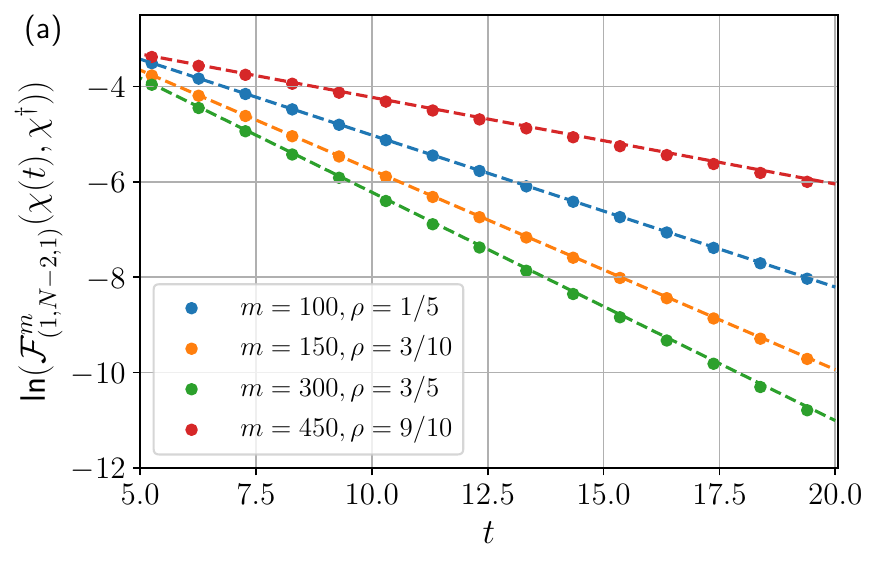}
\includegraphics[scale = 0.5]{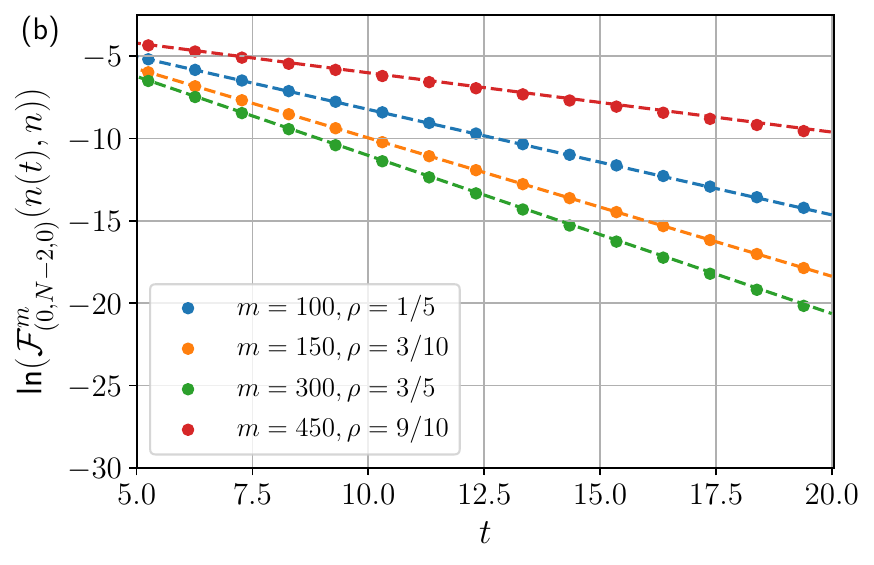}
\caption{(a) The log-behavior of different sectors within the $(1,N-2,1)$ contribution to the $\F^m(\chi_i(t),\chi_j^{\dagger})$ OTOC for $N=500$, compared with straight line fits with slope $-\lambda_{late}^{\rho}$ and intercept ln($\rho(1-\rho)$) for the corresponding $\rho$. (b) The log-behavior of different sectors within the $(0,N-2,0)$ contribution to the $\F^m (n_i(t),n_j)$ OTOC for $N=500$, compared with straight line fits with slope $-\lambda_{L}^{\rho}$ and intercept ln($\rho(1-\rho)$) for the corresponding $\rho$. The irreps in both (a) and (b) exponentially decay in each charge sector and therefore do not contribute to the scrambling dynamics.}
\label{Fig:nonsymbehavior}
\end{figure}

The OTOC $\F^m(\chi_i(t),\chi_j^{\dagger})$ contains contributions from two irreps. The early and late time behavior of sectors in the $(1,N-2,1)$ irrep displays an exponential decay (for large enough $N$) governed by eigenvalues similar to $\lambda_{late}^{\rho}$, shown in Fig.~\ref{fig:lyapcartan}(a), for large $N$
\bea
\F^m_{(1,N-2,1)}(\chi(t),\chi^{\dagger}) \simeq \rho(1-\rho) e^{-\lambda^{\rho}_{late} t}
\eea
where smaller sectors have larger eigenvalues and therefore decay slower. This behavior has been plotted in Fig.~\ref{Fig:nonsymbehavior}(a) and is similar to the behavior of the irrep $L=N/2-1$ in the Majorana case. It should be noted that this irrep shows similar behavior for other OTOCs. The OTOC $\F(n_i(t),n_j)$ is unique amongst the ones discussed in this work because it contains the contribution from the third irrep $(0,N-2,0)$. This irrep also shows an exponential decay as well, with eigenvalues picked from the $\lambda_L^{\rho}$ distribution
\bea
\F_{(0,N-2,0)}^{m}(n(t),n) \simeq \rho(1-\rho) e^{-\lambda_L^{\rho} t}
\eea
We plot this behavior in Fig.~\ref{Fig:nonsymbehavior}(b).

\end{appendices}
\end{document}